\documentclass[letterpaper,english,twocolumn, pra]{revtex4}
\usepackage[T1]{fontenc}
\usepackage[latin9]{luainputenc}
\usepackage{verbatim}
\usepackage{prettyref}
\usepackage{textcomp}
\usepackage{amsmath}
\usepackage{amssymb}
\usepackage{array}
\usepackage{graphicx}
\usepackage{color,soul}
\usepackage[dvipsnames]{xcolor}
\usepackage{makecell}
\usepackage[normalem]{ulem}
\usepackage{subfig}
\usepackage{caption}
\usepackage[bookmarksopen=true]{hyperref}
\usepackage[labelfont=bf]{caption}
 
\makeatletter

\pdfpageheight\paperheight
\pdfpagewidth\paperwidth

\usepackage{graphicx}
\usepackage{graphics}

\makeatother

\usepackage{babel}
\begin{document}
\noindent\begin{minipage}[t]{1\columnwidth}%
\begin{comment}
Defined functions
\end{comment}

\global\long\def\anticommutator#1#2{\left\{  #1,#2\right\}  }

\global\long\def\commutator#1#2{\left[#1,#2\right]}

\global\long\def\braket#1#2{\langle#1|#2\rangle}

\global\long\def\bra#1{\langle#1|}

\global\long\def\ket#1{|#1\rangle}

\global\long\def\Tr{\operatorname{Tr}}

\begin{comment}
Questions:

-Do we need to consider a reorganization energy for the initial excitation?

-460nm vs 665 nm $\lambda$ of chlorophyll a, 
\end{comment}
%
\end{minipage}

\title{Single-photon absorption by single photosynthetic light-harvesting complexes}

\author{Herman C. H. Chan$^{1}$}

\author{Omar E. Gamel$^{1,2}$}

\author{Graham R. Fleming$^{1,2}$}

\author{K. Birgitta Whaley$^{1}$}
%\email{whaley@berkeley.edu}

\affiliation{$^{1}$Department of Chemistry, University of California, Berkeley,
California 94720, USA}

\affiliation{$^{2}$Physical Biosciences Division, Lawrence Berkeley National
Laboratory, Berkeley, California 94720, USA }

\date{\today}
\begin{abstract}
We provide a unified theoretical approach to the quantum dynamics of absorption of single photons and subsequent excitonic energy transfer in photosynthetic light-harvesting complexes. Our analysis combines a continuous mode 
$\langle n\rangle$-photon quantum optical master equation for the chromophoric system with the hierarchy of equations of motion describing excitonic dynamics in presence of non-Markovian coupling to vibrations of the chromophores and surrounding protein.  We apply the approach to simulation of absorption of single-photon coherent states by pigment-protein complexes containing between one and seven chromophores, and compare with results obtained by excitation using a thermal radiation field.  We show that the values of excitation probability obtained under single-photon absorption conditions can be consistently related to bulk absorption cross-sections.
Analysis of the timescale and efficiency of single-photon absorption by light-harvesting systems within this full quantum description of pigment-protein dynamics coupled to a quantum radiation field reveals a non-trivial dependence of the excitation probability and the excited state dynamics induced by exciton-phonon coupling during and subsequent to the pulse, on the bandwidth of the incident photon pulse.  
For bandwidths equal to the spectral bandwidth of Chlorophyll a, our results yield an estimation of an average time of $\sim$0.09 s for a single chlorophyll chromophore to absorb the energy equivalent of one (single-polarization) photon under irradiation by single-photon states at the intensity of sunlight.
\end{abstract}

%\pacs{03.65.Ud, }


\maketitle

\onecolumngrid 
\textbf{Keywords:} photosynthesis, single-photon absorption, quantum dynamics of absorption and energy transfer, \hspace{1cm} non-Markovian exciton-phonon coupling, quantum efficiency
\bigbreak
\bigbreak
\twocolumngrid 

\section{Introduction}
\label{sec:introduction}

\makebox[0.962\linewidth][s]{Photosynthetic light harvesting of plants and bacteria} \textit{in vivo} is characterized by the remarkable capability of \makebox[\linewidth][s]{reaching near unit yield of electron-hole pairs from} electronic excitations produced by absorption of a photon.  Plants usually only operate with such high efficiency under weak illumination conditions and also possess the ability to modify this quantum efficiency in \makebox[\linewidth][s]{real time to respond to changes in environmental} conditions, in particular to reduce it when the intensity of incident light becomes high enough to cause damage to the photosynthetic apparatus. 
The yield of electron-hole pairs from absorbed photons is commonly referred to as the ``quantum efficiency'' of light harvesting and the upper limit of unity is often referred to as implying the transfer of a single photon to a single electron-hole pair, although it is extracted from bulk kinetic measurements of fluorescence.  However this efficiency does not take the effectiveness of light absorption into account and it is thus, strictly speaking, an ``internal'' quantum efficiency.
In contrast, the overall or ``external'' quantum efficiency of light, equal to the number of electron-hole pairs \makebox[\linewidth][s]{produced per incident photon, involves the quantum} dynamics of absorption of light in addition to the dynamics of excitonic energy transport and charge separation that determine the internal quantum efficiency.  The external quantum efficiency, which describes the overall transduction of incident photons to electron-hole pairs, can be significantly lower than the internal quantum efficiency~\cite{Yaghoubi2014}
and in many situations it is the efficiency of this overall transduction process that is important when comparing the effectiveness of complete light-harvesting systems.

Although extensive studies of light harvesting in photosynthetic systems have been carried out, considerable gaps still remain in our understanding of the microscopic \makebox[\linewidth][s]{mechanism and dynamics that leads to this overall} transduction of energy from light to electron-hole pairs. \makebox[\linewidth][s]{In particular, while much is now known about the} microscopic dynamics of excitonic energy transport (EET) and charge separation (CS), relatively little is known about the spatio-temporal interplay of this with the initial process of photoexcitation.  
Understanding of both EET and CS dynamics has increased significantly in recent years due to advances in ultrafast laser techniques \makebox[\linewidth][s]{that have enabled experiments probing the dynamics} of these processes on fs to ns timescales following excitation of light-harvesting complexes by laser pulses~\cite{Berera2009,Ginsberg2009,Schlau2012,Lewis2016,Romero2017}.  In particular, detailed experimental and theoretical studies of EET have revealed a complex quantum dynamical \makebox[\linewidth][s]{process that is enabled and driven by non-Markovian} coupling between excitonic and vibrational modes, and that is characterized by a significant extent of coherence \makebox[\linewidth][s]{over timescales of hundreds of fs.  However, in stark} \makebox[\linewidth][s]{contrast to this situation of extensive microscopic} dynamical understanding for EET and to a lesser extent \makebox[\linewidth][s]{also for CS, we still have very little understanding of} the microscopic dynamics of absorption of individual photons, whether these derive from pulsed lasers or from sunlight. This is an essential missing link for understanding the extent to which the quantum efficiency has a valid microscopic meaning of individual electron-hole pair \makebox[\linewidth][s]{production deriving from absorption of single photons or} whether the meaning of this is restricted to a macroscopic kinetic scheme. 

We note that the absorption of thermal light by photosynthetic systems has been addressed in a recent model of a photosynthetic reaction center as a quantum heat engine driven by absorption from a thermal radiation bath~\cite{Dorfman2013,Nalbach2013}, as well as in studies addressing the role of spatial and/or temporal correlations in thermal light in design of solar energy collection devices~\cite{Mashaal2011,Mendoza2017}.
Several theoretical studies have also addressed the question of whether absorption of incoherent and thermal radiation sources can generate excitonic coherences~\cite{Manvcal2010,Brumer2012,Fassioli2012,Han2013,Olsina2014,Chenu2014,Chenu2016,Shatokhin2016a}.

In order to understand the microscopic dynamics of this critical first step of photosynthesis it is thus essential to undertake a full quantum analysis of the dynamics and efficiency of absorption of individual photons by the chromophores in light-harvesting complexes. Such analysis is further complicated by the need to incorporate the non-Markovian coupling of the electronic excitations of the chromophores to phonon modes, both intramolecular modes and intermolecular modes deriving from exciton-\makebox[\linewidth][s]{photon coupling to vibrational modes of the protein} scaffold in which the chromophores are embedded.
In this work we take the first step in developing such a full quantum analysis, with a study of the dynamics for absorption of a multi-mode single-photon coherent state by a typical pigment-protein complex in a light-harvesting system and comparing this with the conventional master equation description of absorption from a thermal radiation field.  This analysis yields novel microscopic understanding of both the dynamics and energy budget of absorption of single photons, as well as providing the \makebox[\linewidth][s]{first indication of how such single-photon absorption is} integrated with the dynamics of energy transport within a light-harvesting complex. In a subsequent paper we shall address the more complex description of absorption of a single photon from the filtered thermal radiation field of sunlight.  

While we do not specifically analyze the radiation field of sunlight in the current work, the ultraweak nature of sunlight nevertheless provides key additional motivation for the study of single-photon absorption. 
The intensity of solar radiation corresponds to very small photon \makebox[\linewidth][s]{densities, whether expressed per mode ($\sim$10$^{-2}$ for a} single optical mode at 600 - 700 nm, corresponding to the dominant absorptions of chlorophyll and bacteriochlorophyll molecules), per unit volume (at most $\sim$10$^{-5}$ per cubic micron~\cite{Bachor2004}) or per unit time ($\sim$10$^{-3}$ s$^{-1}$ incident on a single chlorophyll molecule~\cite{Blankenship2014}). Probing the dynamics of excitation generation by sunlight thus requires analysis of the absorption of light at the single-photon level. This necessitates a fully quantum treatment of the interaction of realistic pigment-protein complexes with light, which has so far not been carried out. This lack of theoretical analysis is currently mirrored by an absence of experimental studies of light-harvesting systems with ``quantum light'' sources delivering single photons.  

The weakness of the radiation field incident on natural light-harvesting systems is also remarkable in the context of the energetic parameters controlling the dynamics of \makebox[\linewidth][s]{the EET that follows absorption of light to create an} exciton.  The ultraweak nature of sunlight means that coupling of the radiation field to chlorophyll molecules is correspondingly exceedingly weak, of order 10$^{-3}$ cm$^{-1}$ for a transition dipole of magnitude $\sim$6 Debye with the electric field at the earth's surface, $E$ $\sim870$ V cm$^{-1}$. This interaction is markedly smaller than the parameters of order 10 - $10^{2}$ cm$^{-1}$ characterizing the exciton-exciton, exciton-phonon, phonon-phonon and energetic disorder parameters typically measured in light-harvesting complexes~\cite{Novoderezhkin2005,Novoderezhkin2006,Ishizaki2010,Ishizaki2012,Hoyer2014} and is thus also ultraweak in the context of the dynamical parameters controlling light harvesting energetics. Given the apparently highly optimized design of natural light-harvesting systems, we may then \makebox[\linewidth][s]{expect that the strikingly weaker strength of the} coupling of chlorophyll excitons to the radiation field than to both other excitons and vibrations has specific consequences for the spatio-temporal dynamics of excitation generation and transport. 
  
 A key finding of the present work is that the timescale and efficiency of single-photon absorption by light-harvesting systems shows non-trivial dependences on the \makebox[\linewidth][s]{bandwidth of the incident photon pulse, for both the} resulting excitation probability and the coherent excited state dynamics induced by exciton-phonon coupling during and subsequent to the pulse.  We present arguments that the physically relevant value of this bandwidth for natural light-harvesting systems is the inhomogeneous bandwidth of the chromophore absorption band. For the multi-mode single-photon coherent-state pulse this \makebox[\linewidth][s]{results in an optical coherence time of the same order} of magnitude as the timescales associated with the \makebox[\linewidth][s]{excitonic, exciton-phonon, and phonon-phonon couplings.}  
The resulting single-photon excitation probability for chromophores in light-harvesting complexes is exceedingly low, of order 10$^{-6}$ for absorption from a single geometric mode when the excitonic coupling to phonon degrees of freedom is included.  Comparison with calculations for bare chromophores shows that at the physically relevant bandwidth, this low excitation efficiency is primarily due to vibrational dephasing.  We show that when scaled up by the number of geometric modes in the \makebox[\linewidth][s]{coherence volume of sunlight, this single-photon excitation} probability obtained for a single geometric mode yields \makebox[\linewidth][s]{an absorption probability of $\sim$0.147, consistent with} estimates obtained from bulk absorption cross-sections. This allows estimation of the time for absorption of the \makebox[\linewidth][s]{energy equivalent of one (single-polarization) under} sunlight conditions for a single chlorophyll chromophore \makebox[\linewidth][s]{as $\sim$0.09 s. We argue that this implies a timescale of} $\sim$0.1 ms for absorption of this energy equivalent by \makebox[\linewidth][s]{a light-harvesting system such as PSII with $\sim$300} chromophores per reaction center.

Our analysis shows that the excitonic dynamics generated by the pulse are strongly dependent on the pulse bandwidth, with a greater extent of excitonic coherence \makebox[\linewidth][s]{seen in the excitonic dynamics for shorter pulses of} duration 50 fs or less.  In the opposite extreme of pulses longer than 50 ps, phonon relaxation is fast relative to \makebox[\linewidth][s]{the timescale of optical excitation, which effectively} obscures excitonic coherence effects. Comparison of the intensity dependence of absorption from an $\langle n \rangle$-photon coherent pulse with that from a thermal radiation reservoir also shows a strong dependence on the bandwidth of the coherent pulse, with significant differences for the physically relevant bandwidth but similar behavior at the far smaller bandwidths that yield optimal absorption for the isolated bare chromophores.   Detailed analysis of the continuous excitation of a multi-chromophore complex by a thermal radiation reservoir shows that, for a hetero-chromophore complex, excitonic coherence persisting on ns timescales is generated by the combination of excitonic coupling and different intrinsic chromophore excitation rates. These two factors combine to provide a coherent \makebox[\linewidth][s]{drive that becomes established as the excited state} populations increase and that subsequently competes with the incoherent drive of the radiation source. 

\makebox[0.962\linewidth][s]{\smallskip The remainder of the paper proceeds as follows.} Sec.~\ref{sec:Theory} describes the basic theoretical framework for a full quantum description of single-photon absorption and subsequent excitonic energy transfer within a single pigment-protein complex of a typical light-harvesting system.  
The dynamics of interaction with a single-photon radiation field is described by a quantum optical master equation obtained from the quantum stochastic differential equation (QSDE) formulation of equations of motion for a quantum system driven by the quantum noise of a pulsed multi-mode single-photon coherent state~\cite{Baragiola2012}.  To indicate the origin of this approach, we \makebox[\linewidth][s]{shall refer to this as the QSDE master equation.} 
To incorporate the exciton-vibrational coupling and thermal phonon bath dynamics giving rise to homogeneous spectral broadening and dynamic disorder of the excitonic states of chromophores in the native pigment-protein structure of a light-harvesting complex,
this quantum master equation is coupled to a set of hierarchical equations of motion (HEOM) that incorporate the effects of non-Markovian coupling of the pigments to vibrational degrees of freedom~\cite{Ishizaki2009a}. 

\makebox[0.962\linewidth][s]{In this work we do not include the effects of}
inhomogeneous spectral broadening due to a statistical distribution of chromophore energies, focusing on the most elemental event of absorption of single photons by single complexes with inclusion of the homogeneous broadening as described above. We compare the excitation dynamics under absorption from single-photon coherent states with the corresponding dynamics under absorption from a thermal radiation bath, where this is described by the standard reservoir theory of quantum \makebox[\linewidth][s]{optics~\cite{Scully1997}. 
These comparisons will be made under} conditions of similar average photon number for the coherent and thermal radiation fields (see definitions below), for a broad range of pulse bandwidths and photon numbers.   
Sec. \ref{sec:Methods} describes the parameters and gives essential details of the numerical simulations.
In Sec.~\ref{sec:Results} we then present applications of
the theory to monomers, dimers, and a 7-chromophore subcomplex of the light-harvesting complex II (LHCII) of photosystem II (PSII). Sec.~\ref{sec:Discussion} summarizes and discusses the key features of the single-photon absorption dynamics from multi-mode coherent states and the primary differences from excitation by a thermal radiation field.  We conclude by indicating how the present analysis may be generalized to analyze absorption under a partially \makebox[\linewidth][s]{coherent radiation field and briefly discussing implications} for the dynamics of single-photon absorption from the ultraweak radiation field of sunlight.

\section{\label{sec:Theory}Theoretical Model}
\makebox[0.962\linewidth][s]{Analysis of the quantum dynamics of a set of} chromophores interacting with both weak intensity light and vibrational degrees of freedom requires that the chromophore system be coupled to both a quantum radiation field and a quantum phonon field. This results in a complex set of equations of motion for the electronic density matrix.  We present each component of the full simulations here in some
detail, starting from the excitation dynamics of a single chromophore and noting that each additional element may be seen as contributing an additional set
of terms to the density matrix evolution equation. 

\subsection{System Hamiltonian}
\label{subsec:System}

\makebox[0.962\linewidth][s]{We follow the standard approach for single-photon} absorption in which each chromophore $k$ is modeled as a
simple two-level system, setting the ground state energy to zero and the excited state energy to $E_k^{(0)}$.  Unless otherwise stated, this chromophore excited state energy is taken to be the energy of the isolated (gas phase) chromophore, in absence of coupling to phonons, i.e., with no additional reorganization energy due to the protein. For a monomer we have the diagonal 
Hamiltonian
\begin{align}
H_{s}^{(1)} \equiv H_D =\left[\begin{array}{cc}
E^{(0)} & 0\\
0 & 0
\end{array}\right],
\end{align}
while for a dimeric pair of two-level chromophores we add the dipole-dipole coupling $J$ between
the two singly excited states of the dimer.  Without loss of generality, we may write the
energies of the two chromophores as $E_{1}^{(0)}\equiv\hbar\omega_{1}\ge E_{2}^{(0)}\equiv\hbar\omega_{2}$. Defining
the average energy $E^{\prime}\equiv\frac{1}{2}(E_{1}^{(0)}+E_{2}^{(0)})\equiv\hbar\omega^{\prime}$
and the difference energy $\varDelta E\equiv\frac{1}{2}(E_{1}^{(0)}-E_{2}^{(0)})$,
the dimer system Hamiltonian is then given by
\begin{align}
H_{s}^{(2)} & =\left[\begin{array}{cccc}
2E^{\prime} & 0 & 0 & 0\\
0 & E^{\prime} & 0 & 0\\
0 & 0 & E^{\prime} & 0\\
0 & 0 & 0 & 0
\end{array}\right]+\left[\begin{array}{cccc}
0 & 0 & 0 & 0\\
0 & \varDelta E & J & 0\\
0 & J & -\varDelta E & 0\\
0 & 0 & 0 & 0
\end{array}\right]\label{eq:dimerHamiltonian}\\
 & \equiv H_{D}+H_{\delta},\nonumber 
\end{align}
where we have divided the Hamiltonian to two components, a diagonal
one with the average energy, and \makebox[\linewidth][s]{another involving the energy gap and coupling strength.}  

The corresponding matrices for the transition dipole moment operators $M_\mu$ are given as
\[
M_\mu^{(1)}  =\left[\begin{array}{cc}
0 & \mu\\
\mu & 0
\end{array}\right],
\]
and
\begin{align}
M_{\mu}^{(2)} & =\left[\begin{array}{cccc}
0 & \mu_2 & \mu_1 & 0\\
\mu_2 & 0 & 0 & \mu_1\\
\mu_1 & 0 & 0 & \mu_2\\
0 & \mu_1 & \mu_2 & 0
\end{array}\right].
\label{eq:transitiondipolematrix_2site}
\end{align}

Most prior theoretical studies of excitonic energy transfer ignore the initial excitation process and are restricted only to the single
excitation manifold, in which case only the central 2 $\times$ 2 submatrix
of the Hamiltonian is relevant. In this work we always include the fourth state, i.e., the ground state, but in the current study focusing on
linear absorption under single photon conditions we may therefore
exclude
doubly excited states, i.e., consider only the bottom right 3 $\times$ 3
submatrix of the Hamiltonian.
We have verified that including the doubly excited state results in excitation of order $10^{-12}$ in that state, compared with $\sim$10$^{-6}$ in the singly excited states, justifying its neglect.
The division of the Hamiltonian made in Eq.~\prettyref{eq:dimerHamiltonian}
simplifies a rotating frame transformation using the diagonal
term $H_{D}$, which is especially useful when the dynamics 
are restricted to one of the aforementioned submanifolds and which we shall employ below.

\begin{figure}[ht] 
   \centering
 \includegraphics[width=3.4in]{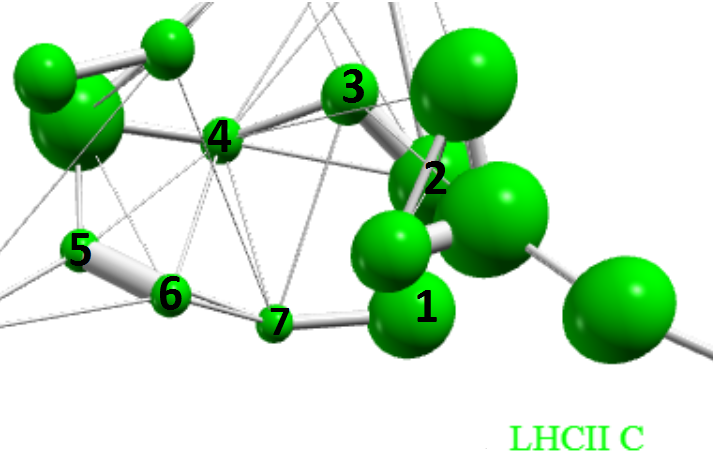} 
 \caption{(Color online) Structural arrangement of the seven chromophores within LHCII in PSII for which single-photon absorption calculations are carried out. See text for specification of  the chromophores labelled 1 - 7.  The sphere size indicates the magnitude of the ground-excited electronic state transition energies and the thickness of the lines between spheres indicates the magnitude of the pigment-pigment electronic interactions. Figure adapted from Ref.~\cite{Roden2016a}. 
 }\label{fig:7mer_structure}
\end{figure}

In the current work we focus on representative monomers and dimers of chlorophyll a (Chla) that are taken from the LHCII antenna complex of the PSII supercomplex.  Specifically, we study the Chla chromophore a603 and the a603-a602 heterodimer.  We also apply our model to a larger pigment-protein system, in particular to the 7-chromophore subcomplex of LHCII that contains Chlb chromophores b608, b609, and Chla chromophores a611, a612, and a610, in addition to the a603-a602 dimer.  This LHCII subcomplex corresponds to chromophores  2 (b608), 3 (b609), 7 (a603), 10 (a602), 11 (a611),  12 (a612), and 13 (a610) in Ref.~\cite{Roden2016a}.  Fig.~\ref{fig:7mer_structure}, adapted from the same reference, shows the structural arrangement of the 7 chromophores in this subcomplex, with the other chromophores that are attached to this sub complex \makebox[\linewidth][s]{de-emphasized. Each chromophore is located at a distinct} site in the three-dimensional structure of the pigment-protein complex. The size of the spheres indicates the relative magnitude of the ground-excited electronic state transition energies and the thickness of the lines connecting the sites is proportional to the electronic interactions between the pigments.  From now on we shall refer to the pigment sites as implying the specified chromophore \makebox[\linewidth][s]{in its corresponding specific location of the protein} environment, unless referring to a property of the bare chromophore molecule. For brevity we shall also refer to the 7-chromophore subcomplex as a 7-mer.

For this 7-mer, we employ a generalization of the Hamiltonian decomposition of Eq.~(\ref{eq:dimerHamiltonian}) that considers only the single-excitation manifold, yielding the 8 $\times$ 8 system Hamiltonian with non-zero matrix elements
\begin{align}
H_s^{(7)} & = H_{D}+H_{\delta}, \nonumber \\
[H_D]_{jk} & = E^{\prime}\delta_{jk}, \,\, j, k = \left \{ 1, 2,...7 \right \}  \nonumber \\
[H_\delta]_{jk} &= (E_{k}^{(0)}-E^{\prime})_{\delta_{jk}} + J_{jk}(1-\delta_{jk}), \,\,
j, k =  \left \{ 1, 2,...7 \right \} \nonumber  \\
\end{align}
where
\begin{align}
E^{\prime} & = \frac{1}{7}\sum_{k=1}^7 E_{k}^{(0)}.
\end{align}
All other matrix elements are zero. The corresponding transition dipole matrix is given by generalization of Eq.~(\ref{eq:transitiondipolematrix_2site}) to the single-excitation manifold for a seven-site complex, resulting in the $8 \times 8$ matrix with non-zero matrix elements
\begin{align}
[M_s^{(7)}]_{jk}  &=  \mu_j, j = \left \{ 1, 2,...7 \right \}, k=8  \nonumber \\
[M_s^{(7)}]_{jk}  &= \mu_k, j=8, k = \left \{ 1, 2,...7  \right \}  \nonumber \\
\label{eq:transitiondipolematrix_7site}
\end{align}
Again all other matrix elements are zero.

In each of the above cases (monomer, dimer, 7-mer), the calculation is simplified if conducted in the rotating frame. The transformation matrix and resulting rotating frame density
matrix are defined by
\begin{align}
U_{r}=e^{\frac{i}{\hbar}H_{D}t},\qquad\rho^{r} & \equiv U_{r}\rho U_{r}^{\dagger},\label{eq:RF_Transform}
\end{align}
with $\rho$ the stationary frame density matrix, which satisfies
the von Neumann equation $\dot{\rho}=-\frac{i}{\hbar}\commutator{H_{s}}{\rho}$.
Applying \makebox[\linewidth][s]{the product rule gives the evolution equation of the}
isolated system in the rotating frame as
\begin{equation}
\dot{\rho}^{r}=-\frac{i}{\hbar}\commutator{H_{\delta}}{\rho^{r}},\label{eq:RF_Evolution}
\end{equation}
where we have used the fact that $H_{\delta}$ commutes with $H_{D}$,
and therefore with $U_{r}$. 

\subsection{Chromophore interaction with radiation field}

In this work we treat the radiation field either as \makebox[\linewidth][s]{a multi-mode pulse of a coherent state with average} photon number $\langle n\rangle=|\alpha|^{2}$, or as a thermal distribution of single-mode Fock states with a thermodynamic average photon number $\overline{n}(\omega)=\ \left[\exp(\beta\hbar\omega)-1 \ \right]^{-1}$. We shall use the rotating wave approximation to describe the interaction of the chromophores with the radiation field.  This is justifiable at the low intensities of the single-photon fields at optical frequencies in sunlight, for which the typical chromophore-radiation interaction is of order $10^{-7}$ times the chromophore absorption frequency~\cite{ultraweakcouplingtosunlight}. A similar \makebox[\linewidth][s]{ratio holds over the range of frequencies of single} excitations in a typical light-harvesting complex and is not significantly changed when the electronic excitations are coupled to phonons.  Given this extremely small \makebox[\linewidth][s]{ratio of coupling strength to transition frequency~\cite{Thimmel1999}, the} contributions of counter-rotating terms can be neglected to a good approximation for both the bare and phonon-coupled chromophores.

In what follows, lowering and raising operators $\sigma_{k},\sigma_{k}^{\dagger}$
apply to the irradiated chromophore at the $k^{\text{th}}$ site. The specific equations given below
relate to photoexcitation dynamics of a photosynthetic dimer ($k=1,2$) in which \makebox[\linewidth][s]{both chromophores have non-zero transition dipole} moments between their ground and first excited states. Extension to the 7-mer and larger chromophore aggregates follows straightforwardly from the definitions given in Sec.~\ref{subsec:System} above.

\subsubsection{Chromophore dynamics under coherent driving by multi-mode pulse}
\label{subsec:methods_coherentdriving}

As described in Refs.~\cite{Loudon2000,Wang2011}, a single-photon pulse of a coherent state with average photon number of one can be represented by a multi-mode wavepacket whose form is controlled by the temporal pulse profile $\xi(t)$. We parameterize this as $\alpha(t)\equiv\alpha\xi(t)$ for
complex $\alpha$ and square-normalized pulse profile $\xi(t)$.
The average number of photons contained in the pulse is equal to
\begin{equation}
\langle n\rangle=\int|\alpha(t)|^{2}dt=|\alpha|^{2},\label{eq:nPhotonsInPulse}
\end{equation}
since $\xi(t)$ is already squared-normalized. 
For a single pulse we may take $\alpha$ to be real, without loss of generality. Note that this average photon number for a pulsed coherent state does not have
the same meaning as the thermodynamic average photon number per mode $\overline{n}$ of a thermal radiation field (see below).
We shall employ a Gaussian pulse profile,
\begin{equation}
\xi(t)\equiv\left(\frac{\Omega^{2}}{2\pi}\right)^{\frac{1}{4}}\exp\left[-\frac{\Omega^{2}}{4}t^{2}\right]\exp\left(-i\omega_0 t\right), \label{eq:GaussianPulse}
\end{equation}
with pulse frequency bandwidth parameter $\Omega$ and carrier frequency $\omega_0$. For this pulse form, the full-width at half-maximum (FWHM) of the pulse amplitude in \makebox[\linewidth][s]{the frequency domain is $\Delta \omega = 2\sqrt{ln 2} \Omega$ in angular} frequency units and $\Delta \nu = (\sqrt{ln2}/\pi) \Omega = \Omega/3.77$ in \makebox[\linewidth][s]{linear frequency units. Other pulse forms have been} studied in~\cite{Wang2011} (see also Sec.~\ref{sec:Discussion}). 
The value of $\Omega$ also defines the coherence time $\tau_{coh}$ of the radiation field. \makebox[\linewidth][s]{Using the simple definition $\tau_{coh}= 1/\Delta \nu$ for this~\cite{Mandel1965}} gives $\tau_{coh} = (\pi/{\sqrt{ln 2}})\Omega = 3.77/\Omega$, which is related to the pulse width by $\tau_{coh}=4.53/ \Delta t$, with $\Delta t$ the FWHM of the pulse Eq.~(\ref{eq:GaussianPulse}) in the time domain.
In the numerical calculations we start the pulse at time $t_0 - 3\sigma$ and end it at $t_0 + 3\sigma$, where $\sigma = \sqrt{2}/\Omega$. The pulse duration for the numerical calculations is thus $6\sqrt{2}/\Omega$.

The quantum optical master equation for a dimeric chromophore system interacting with 
such a pulse can be obtained from the quantum stochastic differential equations (QSDE) for a system interacting with a quantum optical field~\cite{Gardiner2004}.  As shown explicitly in~\cite{Baragiola2012}, this allows generation of quantum master equations for general forms of the optical field.  For an optical field in the form of a coherent state and a single chromophore represented \makebox[\linewidth][s]{by its ground and first excited electronic state, the} resulting system master equation is equivalent to the well-known master equation for a two-level system \makebox[\linewidth][s]{interacting with a coherent state drive~\cite{Wiseman2014}.  Generalizing} this to a dimeric chromophore system yields the excitonic density matrix evolution equation
\begin{align}
\dot{\rho} & =-\frac{i}{\hbar}\commutator{H_{s}}{\rho}-i\sum_{k=1}^{2}\commutator{i\alpha^{*}(t)L_{k}-i\alpha(t)L_{k}^{\dagger}}{\rho}\nonumber \\
 & \hphantom{=}+\sum_{k=1}^{2}\left(L_{k}\rho L_{k}^{\dagger}-\frac{1}{2}\anticommutator{L_{k}^{\dagger}L_{k}}{\rho}\right),\label{eq:CoherentEquation}
\end{align}
where $L_{k}$ are single-chromophore lowering operators given by 
\begin{equation}
L_{k}\equiv\sqrt{\Gamma_{k}}\sigma_{k},\label{eq:LDef}
\end{equation}
\makebox[\linewidth][s]{with $\Gamma_k$ the Weisskopf-Wigner spontaneous emission} decay rate of the $k^{\text{th}}$ chromophore,
\begin{equation}
\Gamma_{k}=\frac{1}{4\pi\epsilon_{0}}\frac{4\omega_{k}^{3}\mu_{k}^{2}}{3\hbar c^{3}}.\label{eq:GammaDef}
\end{equation}
Here $\omega_k$ and $\mu_k$ are the energy and transition dipole moment of the $k^{\text{th}}$ chromophore, respectively. The terms
in Eq. \prettyref{eq:CoherentEquation} involving $\alpha(t)$ correspond to coherent excitation and de-excitation by 
the coherent pulse, while those in the last term correspond to spontaneous emission.

We shall occasionally refer also to the chromophore oscillator strength~\cite{Santori2010}, 
\begin{equation}
f_k = \Gamma_k/3\Gamma_{cl},
\end{equation}
with
\begin{equation}
\Gamma_{cl} =  \omega_k^2 e^2/(6 \pi/\epsilon_0 m_e c^3)
\end{equation}
defined as the dimensionless ratio of the spontaneous emission decay rate to that of a classical oscillator with the corresponding frequency $\omega_k$, which is more
commonly used in molecular spectroscopy.

For ease of numerical simulation, it is more convenient to transform Eq.~(\ref{eq:CoherentEquation}) to the interaction picture, in which case $\alpha(t)$ is replaced by the
slowly varying envelope of the oscillation, $\tilde{\alpha}(t)$:
\begin{align}
\dot{\rho}^{r} & =-\frac{i}{\hbar}\commutator{H_{\delta}}{\rho^{r}}-i\sum_{k=1}^{2}\commutator{i\tilde{\alpha}^{*}(t)L_{k}-i\tilde{\alpha}(t)L_{k}^{\dagger}}{\rho^{r}}\nonumber \\
 & \hphantom{=}+\sum_{k=1}^{2}\left(L_{k}\rho^{r}L_{k}^{\dagger}-\frac{1}{2}\anticommutator{L_{k}^{\dagger}L_{k}}{\rho^{r}}\right),\label{eq:CoherentEquation-RF}
\end{align} 
with
\begin{align}
 \tilde{\alpha}(t) & = \alpha \left(\frac{\Omega^{2}}{2\pi}\right)^{\frac{1}{4}}\exp\left[-\frac{\Omega^{2}}{4}t^{2}\right].
 \end{align}

\subsubsection{Chromophore dynamics under incoherent driving by thermal radiation}
\label{subsec:thermal}

The standard description for interaction of a system of two-level chromophores with a thermal radiation field is obtained with the well-known reservoir theory of quantum optics~\citep{Scully1997} in which the radiation field is treated as a large Markovian reservoir.  For a dimeric chromophore \makebox[\linewidth][s]{system, this yields the following Lindblad form for} evolution of the system density matrix:
\begin{align}
\dot{\rho} & =-\frac{i}{\hbar}\commutator{H_{s}}{\rho}+\sum_{k=1}^{2}\overline{n}(\omega_{k})\left[L_{k}^{\dagger}\rho L_{k}-\frac{1}{2}\anticommutator{L_{k}L_{k}^{\dagger}}{\rho}\right]\nonumber \\
 & \hphantom{=}+\sum_{k=1}^{2}(\overline{n}(\omega_{k})+1)\left[L_{k}\rho L_{k}^{\dagger}-\frac{1}{2}\anticommutator{L_{k}^{\dagger}L_{k}}{\rho}\right],\label{eq:ThermalLindblad}
\end{align}
where
\begin{equation}
\overline{n}(\omega)=\frac{1}{\exp(\beta\hbar\omega)-1}\label{eq:thermalPhotonNumber}
\end{equation}
\makebox[\linewidth][s]{is the thermal average photon number per mode for} energy $E=\hbar\omega$, and
inverse temperature $\beta\equiv 1/{k_{B}T}$.

This description represents the quantum driving of the system by an incoherent radiation field.  In Eq.~\prettyref{eq:ThermalLindblad}, the terms in the square brackets
multiplied by $\overline{n}$ represent stimulated absorption, while
those multiplied by $(\overline{n}+1)$ represent stimulated and
spontaneous emission. 
These \makebox[\linewidth][s]{two terms provide incoherent driving of the system} density matrix, in contrast to the unitary drive provided by the first term $-\frac{i}{\hbar}\left[H_s,\rho\right]$. The latter contains the excitonic couplings between chromophores, which play a key role in the time evolution of off-diagonal elements of the system density matrix for systems larger than a monomer.  We note that inclusion of retardation effects \makebox[\linewidth][s]{can result in additional small coherent light-induced} coupling terms between chromophores~\cite{Shatokhin2016a}.

Applying the rotating frame transformation \prettyref{eq:RF_Transform}
to the thermal equation \prettyref{eq:ThermalLindblad}, 
results in an equation of the same form but with the coherent driving term replaced by $-\frac{i}{\hbar}\left[H_\delta,\rho^r\right]$ and $\rho$ replaced by $\rho^{r}$, i.e., 
\begin{align}
\dot{\rho}^{r} & =-\frac{i}{\hbar}\commutator{H_{\delta}}{{\rho}^{r}}+\sum_{k=1}^{2}\overline{n}(\omega_{k})\left[L_{k}^{\dagger}\rho^{r} L_{k}-\frac{1}{2}\anticommutator{L_{k}L_{k}^{\dagger}}{\rho^{r}}\right]\nonumber \\
 & \hphantom{=}+\sum_{k=1}^{2}(\overline{n}(\omega_{k})+1)\left[L_{k}\rho^{r} L_{k}^{\dagger}-\frac{1}{2}\anticommutator{L_{k}^{\dagger}L_{k}}{\rho^{r}}\right],\label{eq:ThermalLindblad_rf}
\end{align}
\makebox[\linewidth][s]{To see this,
note that in going to the rotating} frame the lowering operators will be transformed as $L_{k}\rightarrow U_{r}L_{k}U_{r}^{\dagger}=e^{-\frac{i}{\hbar}E^{\prime}t}L_{k}$.
Since each term in Eq.~\prettyref{eq:ThermalLindblad} contains both $L_{k}$
and its adjoint $L_{k}^{\dagger}$, the time-dependent phases cancel out. 

We emphasize here that such a thermal radiation bath is characterized by thermodynamic average photon numbers $\overline{n}(\omega_k)$ that are
derived from the temperature of the radiation source.  Unlike a coherent pulse, a thermal bath is not characterized
by finite amount of energy and thus it is not possible to directly compare the energy content of a single coherent pulse with that of a thermal radiation bath.
In fact, the model for a thermal bath model is implicitly assumed
to be of infinite size, and therefore of unlimited energy. 

\subsection{Chromophore interaction with phonon bath}
\label{subsec:HEOM}

Natural light-harvesting complexes are known to be extremely complicated
structures with very large numbers of electronic and vibrational degrees of freedom~\cite{Scholes2011}. 
To model them tractably, the conventional approach has \makebox[\linewidth][s]{been to focus on the reduced density matrix of a few} degrees of freedom of interest.  Generally the vibrational environment of the chromophoric electronic degrees of freedom is traced out and modeled as a phonon bath, \makebox[\linewidth][s]{yielding non-Markovian equations of evolution for the} \makebox[\linewidth][s]{excitonic dynamics of a photoexcited chromophore} system. The most accurate representation of this in \makebox[\linewidth][s]{a time-local formulation convenient for numerical} \makebox[\linewidth][s]{simulations is given by the hierarchy equations of} motion (HEOM), which assume only linear exciton-phonon coupling and a harmonic phonon bath characterized by Gaussian fluctuations \citep{Tanimura1989,Ishizaki2005}. For excitonic \makebox[\linewidth][s]{energy transport in photosynthetic light-harvesting} systems this method reproduces the
results of Redfield theory and F\"{o}rster resonance energy transfer theory in the respective domains of applicability but also allows analysis in the physically relevant domain of similar energy and timescales for exciton-exciton, exciton-phonon, and phonon-phonon couplings~\cite{Ishizaki2009a}.

The HEOM are coupled equations of motion for a set \makebox[\linewidth][s]{of density matrices $\rho^{\boldsymbol{n}}(t)$,
where the superscript is a} vector index $\boldsymbol{n}=(n_{1}, n_{2}, \ldots, n_{N})$
\citep{Ishizaki2005}. Each $n_{k}$ is a non-negative integer and
$N$ is the number of sites at which individual chromophores are located, with $N=1$ for a monomer,
$N=2$ for a dimer, etc. The component density matrix with $\boldsymbol{n} = {\boldsymbol{0}}$, i.e.,  $\rho^{\boldsymbol{0}}(t)$ constitutes the physical reduced density matrix of the
system, while component density matrices with a nonzero vector
index $\boldsymbol{n} \neq 0$ form a hierarchy of auxiliary matrices that represent the non-Markovian bath dynamics generated by the system-bath \makebox[\linewidth][s]{coupling.
The HEOM constitute an infinite series of} coupled differential
equations given by
\begin{align}
\frac{\partial}{\partial t}\rho^{\boldsymbol{n}}(t) & =-\left(i\mathcal{L}_{s}+\sum_{k=1}^{N}n_{k}\gamma_{k}\right)\rho^{\boldsymbol{n}}(t)\nonumber \\
 & \hspace{1em}+\sum_{k=1}^{N}\left[\Phi_{k}\rho^{\boldsymbol{n}_{k+}}(t)+n_{k}\Theta_{k}\rho^{\boldsymbol{n}_{k-}}(t)\right],\label{eq:Hierarchy}
\end{align}
where $\boldsymbol{n}_{k\pm}$ denotes the vector index $\boldsymbol{n}$ resulting from
$n_{k}\rightarrow n_{k}\pm1$, and $\rho^{\boldsymbol{n}_{k-}}$ is
is set to zero if ${n}_{k} = 0$. $\mathcal{L}_{s}=\frac{1}{\hbar}H_{s}^{\times}$ denotes the Liouvillian corresponding to the
system Hamiltonian $H_{s}$ 
and we have defined the following superoperators 
\begin{align}
\Phi_{k} & \equiv iV_{k}^{\times},\nonumber \\
\Theta_{k} & \equiv i\left(\frac{2\lambda_{k}}{\beta\hbar^{2}}V_{k}^{\times}-i\frac{\lambda_{k}}{\hbar}\gamma_{k}V_{k}^{o}\right),\label{eq:HEOMSuperoperators}
\end{align}
where $V_{k}=\ket k\bra k $, the projector onto the excited state
of the chromophore at the $k^{\text{th}}$ site (tensored implicitly with the identity on all other
sites).
Following~\cite{Ishizaki2005,Ishizaki2009a}, we employ the superoperator notation
\begin{align}
O^{\times}f & \equiv\commutator Of,\nonumber \\
O^{o}f & \equiv\anticommutator Of.\label{eq:opCommAntiComm}
\end{align}
for commutators $O^{\times}$ and anticommutators $O^{o}$.

\makebox[0.962\linewidth][s]{The HEOM depend on two parameters that are} specific to the chromophore and/or the site of this.  These are the
reorganization energy $\lambda_{k}$, and the inverse timescale of
phonon relaxation $\gamma_{k}$.  The reorganization
energy $\lambda_{k}$ represents the energy difference between the vibrational configuration reached by a vertical Franck-Condon
transition from the ground electronic state and the equilibrium vibrational configuration in the excited electronic state for a chromophore at the $k^{\text{th}}$ site. The relaxation rate $\gamma_{k}$ characterizes
the timescale of dissipation of the phonon reorganization energy
associated with the chromophore at the $k^{\text{th}}$ site. 
These parameters are determined by the equilibrium response function of the collective energy gap coordinate $\tilde{u}_k$ for the $k^{\text{th}}$ site~\cite{Ishizaki2010}:
\begin{align}
\chi_k(t) \equiv \frac{i}{\hbar} \langle \left[ \tilde{u}_k(t), \tilde{u}_k(0) \right] \rangle.
\end{align}
Specifically, $\lambda_{k}$ is given by the zero time value of the relaxation function
\begin{align}
\tilde{\Gamma}_k(t) & \equiv \int_t^\infty ds \chi_k (s)
\end{align}
according to $\tilde{\Gamma}_k(0) = 2\hbar \lambda_{k}$,
and $\tau_k \equiv \gamma_k^{-1}$ is given by 
\begin{align}
\tau_k \equiv \frac{1}{\tilde{\Gamma}_k(0)} \int_0^\infty dt \tilde{\Gamma}_k(t).
\end{align}
We employ here an exponential decay for the relaxation function, $\tilde{\Gamma}_k(t) = 2\hbar \lambda_k e^{-\gamma_k t} \equiv 2\hbar \lambda_k e^{-t/\tau_k}$, which results in the Ohmic form of spectral density with Drude-Lorentz  regularization describing coupling to an environmental bath of overdamped vibrations,
\begin{align}
 J_k\left[\omega\right] \equiv \text{Im} \left(\chi_k\left[\omega\right]\right) = 2\hbar \lambda_k \gamma_k \omega/(\omega^2 + \gamma_k^2),
\end{align}
with $\chi\left[\omega\right]$ the Fourier-Laplace transform of $\chi(t)$. This Ohmic form with Drude-Lorentz regularization is conventionally employed to represent coupling of chromophores to non-specific environmental vibrational modes that are non-resonant with excitonic energy differences~\cite{Ishizaki2010}.

\makebox[0.962\linewidth][s]{For simulations in the rotating frame, since the} excited state projectors are invariant
under the rotating frame transformation Eq.~\prettyref{eq:RF_Transform}, i.e., 
$U_{r}V_{k}U_{r}^{\dagger}=V_{k}$, the 
\makebox[\linewidth][s]{density matrix $\rho^r$ is represented by a similar set} of hierarchy equations of motion differing only in that the system Hamiltonian
is changed from $H_s$ to $H_\delta$. 

We now take a typical light-harvesting complex that is a pigment-protein complex described as a chromophore system with a vibrational bath, and consider the effect of irradiation by either a coherent pulse (Eq.~\prettyref{eq:CoherentEquation}), or a thermal radiation reservoir (Eq.~\prettyref{eq:ThermalLindblad}). Since the resulting additional terms in the density matrix equation of motion for the excitonic system derive from Markovian couplings to the radiation field, they are time local and hence appear in each level of the hierarchy of equations \makebox[\linewidth][s]{describing the interaction with the phonon field.} (Qualitatively, the interaction with the radiation field may be regarded as modifying the system Liouvillian of the chromophore system interacting with the vibrational bath.) \citep{Kreisbeck2011,Shabani2014}.
Note that the temperature determining the average photon number of the thermal radiation field can be very different from the ambient temperature determining the response of the phonon bath. 

The HEOM form an infinite series which converge on the exact evolution
with increasing accuracy. Depending \makebox[\linewidth][s]{on the system parameters and the accuracy required} for desired observables, one can truncate at different numbers of levels. In simulations shown below in this work we truncate after 5 - 15 levels, depending on the specific system.  We note that depending on the system parameters, truncation of the HEOM can occasionally cause loss of positivity~\cite{Suess2015,Witt2017,Strunz1999} and it is important to check that this is not the case in any specific calculation.  

\makebox[0.962\linewidth][s]{In general, the initial conditions for integration of} the HEOM equations will be the thermal equilibrium density matrix, which may contain correlations between the chromophore system and the vibrational bath.  This may be pre-computed using the approach presented in references \cite{Ishizaki2006,Tanimura2006},
whereby the system density matrix is first initialized to its thermal equilibrium value $\rho^{\boldsymbol{0}}_{th}(t)=
\exp{(-\beta H_s)}/\Tr \exp{(-\beta H_s)}$, and the auxiliary matrices all set to zero, corresponding to a factorized product state of the system and bath, then this factorized initial state evolved under the HEOM using Eq. (\ref{eq:Hierarchy}) until a steady state $\rho_{eq}$ is reached. This equilibrium state is then taken as the true initial state for subsequent dynamics, e.g., under interaction with the radiation field.  In practice, constraints of e.g., vertical photoexcitation in linear response formulations of electronic excitation~\cite{Ishizaki2009a}, or photoexcitation at temperatures low relative to the excited chromophore energy for direct calculation of excitation~\cite{Uchiyama2010}, preclude any significant initial correlation between the system and bath, allowing this pre-equilibration step to be omitted and a factorized initial density matrix to be used.  The effects of initial state correlations in transient linear response of two-level systems has been studied in~\cite{Uchiyama2010}, where it was shown that initial system-bath  correlation has no noticeable effect at temperatures low enough that the initial thermal occupation probability for the excited state is negligible. This is the situation for excitonic states of chromophores at ambient temperatures of $T \sim 300 $ K and therefore we employ the factorized initial condition in this work.

\subsection{Linear response and absorption spectrum} 
\label{subsec:Absorption_spectrum}

It is useful to compare the direct calculations of single-photon absorption described above with the predictions of linear response theories of optical absorption.  Under fields weak enough that the response of the system is \makebox[\linewidth][s]{linear in the applied field, the absorption spectrum is} depended on the imaginary part of a frequency-domain response function that depends only upon dynamical properties of the system at equilibrium \cite{kubo1957,Mukamel1995}. 

\makebox[0.962\linewidth][s]{Within linear response the optical excitation is} represented by the instantaneous action of the dipole operator $M_{\mu}$ on the state density matrix at time $t_{0}=0$, i.e., $\rho_{eq} \rightarrow \frac{i}{\hbar}M_{\mu}^{\times}\rho_{eq}$,
allowing the factorized initial condition to be employed for the system-bath density matrix~\cite{Ishizaki2009a}.  The system then is evolved under the HEOM for a time $t$ and the resulting polarization evaluated by taking the expectation of the dipole operator $M_{\mu}$ under the time-evolved density matrix.  Fourier transforming on the interval $t$ gives the linear susceptibility $\phi\left[\omega\right]$ that determines the linear response absorption spectrum  \cite{Mukamel1995}. 
Operationally, we have the absorbance (in arbitrary units)
\begin{equation}
\label{AbsorptionSpectrumDef}
A(\omega)=\frac{4\pi\omega}{\sqrt{1+4\pi\text{Re}\left(\phi\left[\omega\right]\right)} \  c}  \text{Im}\left(\phi\left[\omega\right]\right) ,
\end{equation}
where 
\begin{align}
\phi\left[\omega\right]=\int e^{-i\omega t}\Tr (M_{\mu}G_{H}(t)\frac{i}{\hbar}M_{\mu}^{\times}\rho_{eq}) dt
\end{align}
is the linear susceptibility, $G_{H}(t)$ denotes evolution \makebox[\linewidth][s]{according to the hierarchy equations, Eq. (\ref{eq:Hierarchy}), for a time} $t$, $M_{\mu}$ is the transition dipole operator (Eqs.~(\ref{eq:transitiondipolematrix_2site}) and (\ref{eq:transitiondipolematrix_7site})), and $c$ is the speed of light. Note that since the optical excitation is represented as an instantaneous event, the initial excited state energy for the evolution under the HEOM is taken as the Franck-Condon energy $E_k + \lambda_k$ (commonly referred to as ``the site energy'') rather than the bare chromophore energy $E_k^{(0)}$.

For a single chromophore-protein complex under weak fields corresponding to single-photon excitation, the linear response spectrum can be expected to provide a good estimate of the homogeneous broadening due to coupling of the excitonic states to the phonon bath and thereby of the energetic range of photons that may be resonantly \makebox[\linewidth][s]{absorbed by chromophores in the presence of this} coupling. This may be used to set the bandwidth of the single-photon coherent pulse, as will be discussed in more detail below (Secs.~\ref{sec:Results} and \ref{sec:Discussion}). 

We note that the linear response formulation assumes \makebox[\linewidth][s]{Markovian bath dynamics. Recent work has shown} additional corrections are required for non-Markovian \makebox[\linewidth][s]{baths (as also when the initial state has system-bath} correlations) \cite{Ban2017,Shen2017}.  We shall neglect such small effects in this work. 

\vfill\null

%Table 1 - monomer and dimer parameters
\begin{table*}[ht]
 \centering
\begin{tabular}{ | c | c | c | c | } 
\hline
 & \thead{Chla monomer} & \thead{LHCII dimer} & \thead{LHCII 7-mer}  \\ 
\hline
  \makecell{System Hamiltonian  \\ parameters } & 
\makecell{$E^{(0)}=15287\text{ cm}^{-1}$ \\ } & 

\makecell{$E_{1}^{(0)}=15287\text{ cm}^{-1}$ \\ $E_{2}^{(0)}=15157\text{ cm}^{-1}$ \\ $J_{12}=38.11 \text{ cm}^{-1}$} &

\makecell{See \hyperref[sec:Appendix]{Appendix}}
 \\

\hline
\ \makecell{Phonon (HEOM) \\ bath parameters} &\makecell{ $\lambda_a=37 \text{ cm}^{-1}$ \\ $\gamma_a=30 \text{ cm}^{-1}$\\ (i.e., $\tau_{a}=177$ fs) \\ $T=300$ K \\$L=5$}  
 & \makecell{ $\lambda_a=37 \text{ cm}^{-1}$ \\ $\gamma_a=30 \text{ cm}^{-1}$ \\ (i.e., $\tau_{a}=177$ fs) \\ $T=300$ K \\$L=5$ \\ ($L=15$ for short-time inserts of Fig.~\ref{fig:Fig5_dimer_coherent})}  
 & \makecell{ $\lambda_a=\lambda_b=37 \text{ cm}^{-1}$ \\ $\gamma_a=30 \text{ cm}^{-1}$, $\gamma_b=48 \text{ cm}^{-1}$ \\ (i.e., $\tau_{a}=177$ fs, $\tau_{b}=111$ fs) \\$L=5$}
 \\ 
\hline
\makecell{Transition dipole \\ moment $\mu_k$} &  $\mu_a= 4$ Debye  & \makecell{$\mu_{1}= \mu_{2}= \mu_a$ } &
\makecell{ $\mu_a =4$ Debye\\ $\mu_b =3.4$ Debye }
\\
\hline
\makecell{Weisskopf-Wigner \\ atomic decay rate $\Gamma_k$}
  &  $\Gamma= 1.79 \times 10^{7} $ Hz   & \makecell{$\Gamma_{1}= 1.79 \times 10^{7}$ Hz \\ $\Gamma_{2}= 1.75 \times 10^{7}$ Hz } 
  &   
    \makecell{See \hyperref[sec:Appendix]{Appendix}}
  \\
\hline

\end{tabular}
\caption{Parameter values employed in this work to model a Chla monomer, LHCII dimer and LHCII 7-mer, taken from \citep{Novoderezhkin2011intra, Bennett2013a}. $T$ is the temperature and $L$ the level of hierarchy employed in the calculations. See Sec.~\ref{sec:Theory} for a full description of all parameters.}\label{tab:chromophoreparameters}
\end{table*}

\section{Parameters and Techniques}
\label{sec:Methods}

\subsection{Parameters}
\label{subsec:Parameters}
\renewcommand\theadalign{cb}
\renewcommand\theadfont{\bfseries}
\renewcommand\theadgape{\Gape[4pt]}
\renewcommand\cellgape{\Gape[4pt]}

Table~\ref{tab:chromophoreparameters} lists the chromophore-protein parameters \makebox[\linewidth][s]{employed in this work to model the Chla monomer,} LHCII a603-a602 dimer, and the LHC 7-mer. For the monomer and dimer we employ the bare chromophore excited state energies $E_k^{(0)}$ determined in ~\cite{Novoderezhkin2011intra} (i.e., no reorganization energy contribution is included in these energies), together with the excitonic coupling between chromophores, $J_{jk}$, together with the values of Chla chromophore reorganization energy $\lambda_k = \lambda_a$ and phonon \makebox[\linewidth][s]{relaxation rate $\gamma_k = \gamma_a$ (corresponding to the relaxation} time constant $\tau_a$) from that work.  For the 7-mer of LHCII containing Chlb chromophores b608, b609, and Chla chromophores a603, a602, a611, a612, and a610, all parameters (bare chromophore excited state energies, transition dipole elements, excitonic couplings and $\lambda_a, \gamma_a$ values for Chla, $\lambda_b, \gamma_b$ values for Chlb) are similarly taken from ~\cite{Novoderezhkin2011intra}.

Table~\ref{tab:lightparameters} lists the parameter values of the coherent \makebox[\linewidth][s]{and thermal radiation fields.  The coherent pulse employs} a Gaussian form with carrier frequency $\omega_0$ given \makebox[\linewidth][s]{by the bare excited state energy for the monomer} chromophore and by the average of the bare excited state energies for the dimer and 7-mer complexes.  The bandwidth parameter $\Omega$ is varied from the optimal value of $\Omega_{opt} = 2.4\Gamma$ for an isolated two-level system~\cite{Wang2011} to the value $\Omega_{sun} = 3.77 \Delta \nu =1.21 \times 10^{15}$ Hz  that ensures the pulse amplitude covers the commonly used bandwidth estimate for the full spectrum of visible light (400 - 700 nm).
%~\cite{saleh1991}.

%Table 2 - light source parameters
\begin{table}[ht]
 \centering
\begin{tabular}{ | c | c |  } 
\hline
 \thead{Light source} & \thead{Parameter}   \\ 
\hline
 Coherent pulse & \makecell{ 1. Frequency bandwidth parameter \\ $\Omega$: $10^7$ - $10^{15}$
  Hz \\ 2. Gaussian pulse: Eq.~(\ref{eq:GaussianPulse}) \\ 3. Average number of photons \\ contained in pulse: $\left \langle n \right \rangle$} \\

\hline
Thermal radiation &\makecell{ Average photon number per mode  \\ 
$\overline{n}$ 
}   \\ 

\hline
 
\end{tabular}
\caption{Parameter values used for the coherent pulse and thermal radiation light sources.}\label{tab:lightparameters}
\end{table} 

\makebox[0.962\linewidth][s]{Although our primary focus is on single-photon} absorption from radiation fields with average photon number $\langle n\rangle=|\alpha|^{2} = 1$ (coherent pulse field) and thermodynamic average photon number $\overline{n}(\omega) = 1$ (thermal field), we shall also carry out coherent absorption calculations with variable values of both $\langle n\rangle$ and $\overline{n}(\omega)$ in order to characterize the dependence on intensity under very different conditions of the radiation fields.  Since we are interested to compare the absorption of single photons from coherent and thermal fields characterized by equal average photon numbers (within the caveat discussed in Sec.~\ref{subsec:thermal}), it should be noted that the thermal calculations for $\overline{n}(\omega) = 1$ will be carried out with radiation deriving implicitly from a reservoir at higher temperature than the sun.

\subsection{Numerical techniques}

The differential equations were propagated using the standard fourth-order
Runge-Kutta method, with the time step reduced until 
the first maximum (or plateau value if there is no pronounced maximum) of the time-dependent excitation probability, $P_{e}^{\text{max}}$, is converged \makebox[\linewidth][s]{to within 3 significant figures. 
For the coherent pulse} calculations, all frequencies within $\pm 3$ standard deviation \makebox[\linewidth][s]{of the carrier frequency of the pulse were included.} All  simulations were carried out with Python Numpy and Scipy packages.  

%Monomer coherent pulse figure
\begin{figure}[ht] 
   \centering
  \includegraphics[width=3.4in]{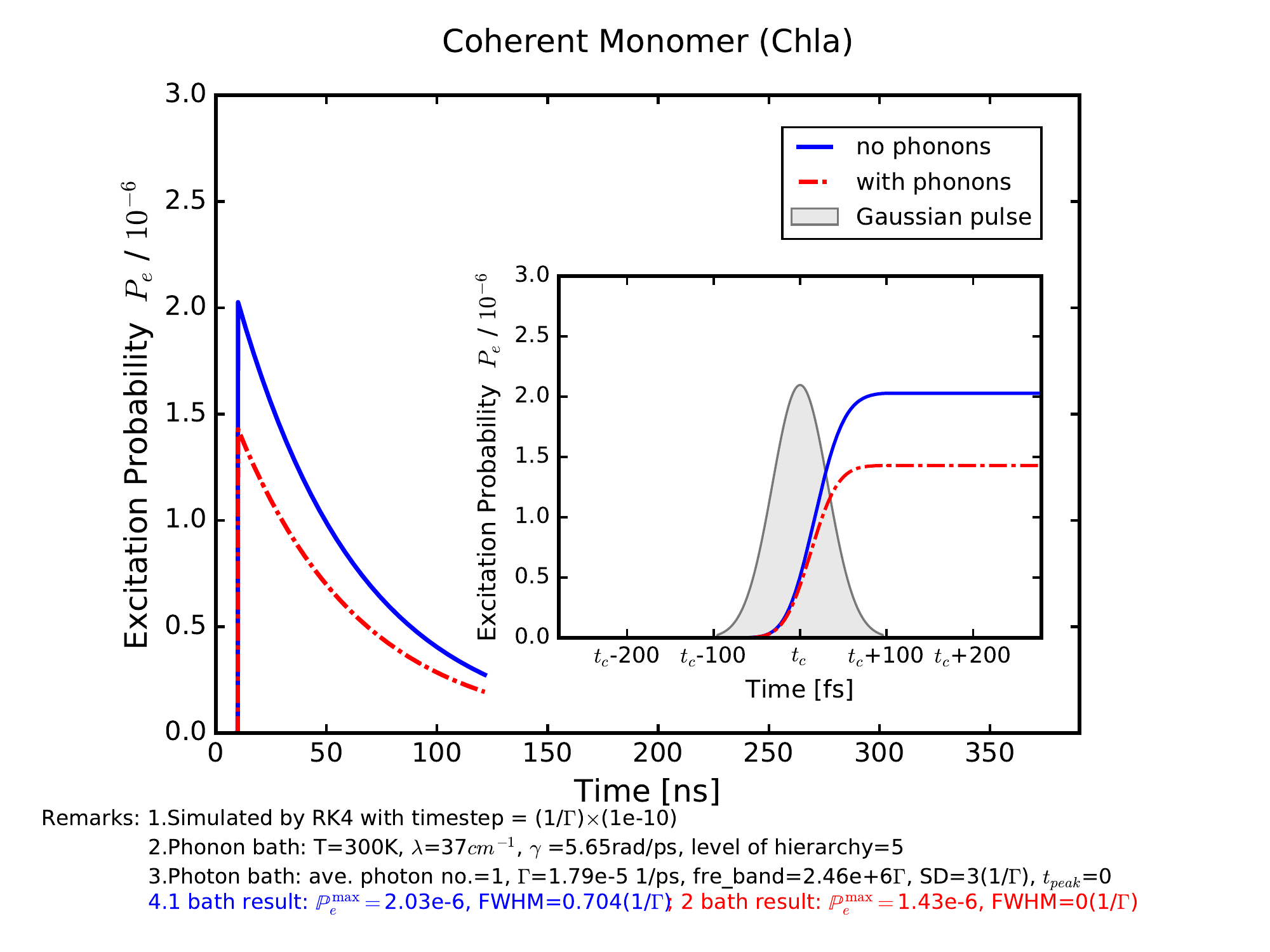} 
 \caption{(Color online) Excitation of the a603 Chla in LHCII by a Gaussian coherent pulse with mean photon number $\langle n\rangle= 1$ and bandwidth parameter $\Omega = 4.41 \times 10^{13}$ Hz, centered at the bare excitation energy of the Chla. Blue line: without phonon bath. Red line: including phonon bath at $T$ = 300 K. The center of the pulse is at $t_c=10$ ns.
}\label{fig:Fig2_monomer_coherent}
\end{figure}

\section{\label{sec:Results}Results}

\subsection{Monomer}
\label{subsec:Monomer}

We first study the dynamics of single-photon absorption by a two-level chromophore monomer.  These calculations serve as
a benchmark against which later results for the dimeric and seven chromophore system are compared.  
We present results here for a monomeric system modeling the a603 chlorophyll (Chla) molecule in LHCII, including the ground and first excited electronic states with absorption in the Q$_y$ band.  The excited electronic state interacts with a vibrational bath at $T = 300$ K via non-Markovian coupling.  For the exciton-vibration \makebox[\linewidth][s]{interaction we employ the usual model of an excited state} dephasing interaction with a Drude-Lorentz, i.e., overdamped Brownian oscillator spectral density, using values of coupling strength  and phonon dissipation energy  for Chla from~\cite{Novoderezhkin2011intra, Bennett2013a}.  All molecular and vibrational bath parameters are given in Table~\ref{tab:chromophoreparameters} of Sec.~\ref{subsec:Parameters} above.

%Monomer absorption spectrum figure
\begin{figure}[ht] 
   \centering
  \includegraphics[width=3.4in]{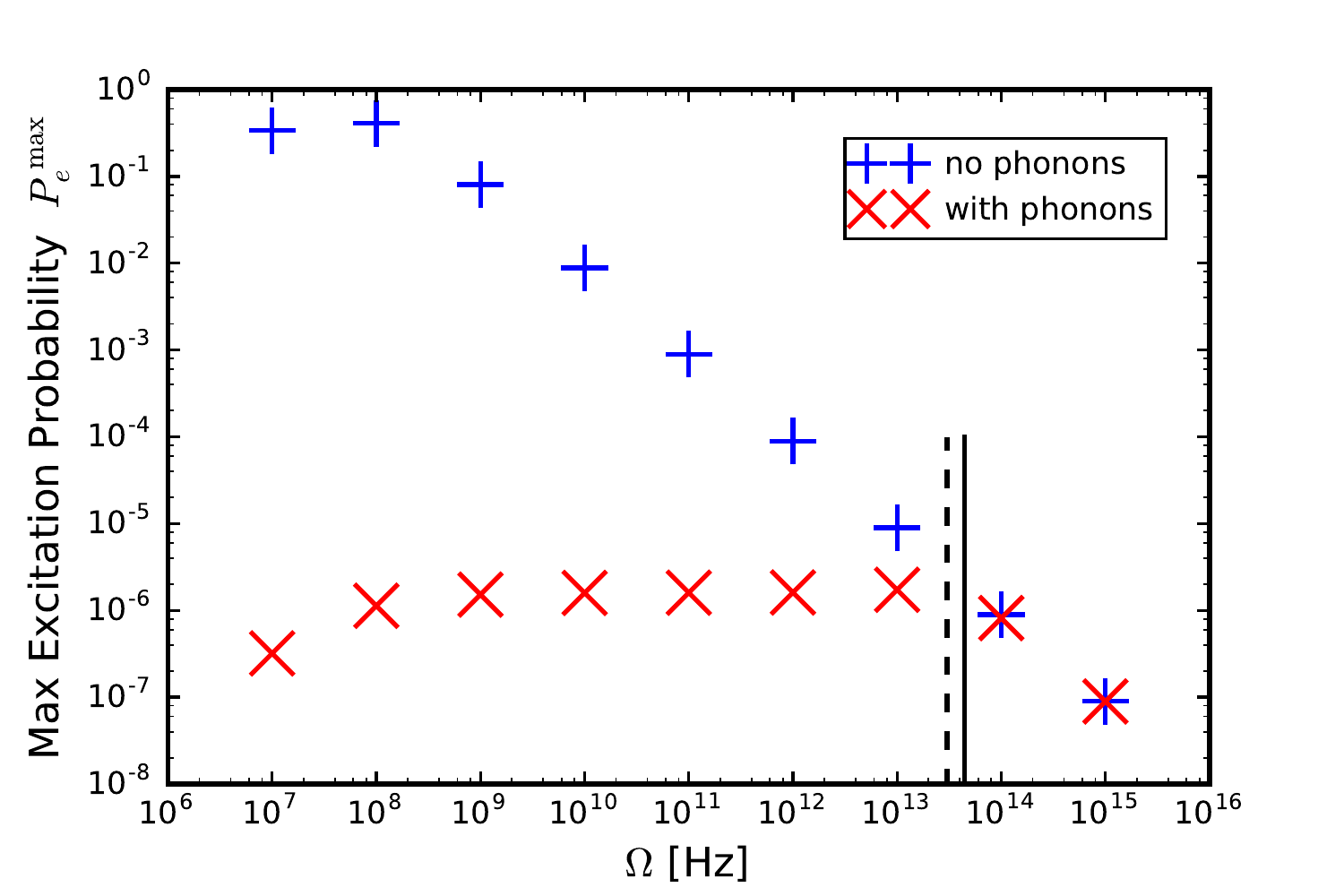} 
 \caption{(Color online) Dependence of maximum excitation probability $P_e^{\text{max}}$ on the coherent pulse bandwidth parameter $\Omega$ for excitation of the Chla monomer. The solid vertical line denotes the value $\Omega=4.41 \times 10^{13}$ Hz corresponding to the typical absorption bandwidth of Chla in solution ($\Delta \nu \sim 390$ cm$^{-1}$~\cite{Blankenship2014}) and the dashed vertical line the value $3.02 \times 10^{13}$ Hz that is derived from the linear response bandwidth (see Fig.~\ref{fig:Fig1_monomer_spectrum}).  Note that both axes are in log scale. Blue symbols `$+$' : without coupling to phonon bath.  Red symbols `$\times$' : with coupling to phonon bath.
}
\label{fig:Fig10_monomercoherent_bandwidth}
\end{figure}

\subsubsection{Monomer excitation with coherent pulse}

Fig.~\ref{fig:Fig2_monomer_coherent} shows the time-dependent excitation probability $P_e$ for excitation of the Chla by a coherent state with an average of one photon, i.e., $\langle n \rangle = 1$.  In this calculation the Gaussian pulse form has carrier frequency $\omega_0$ equal to the monomer bare excited state energy (15287 cm$^{-1}$) 
and the bandwidth parameter $\Omega$ is set to 4.41$\times 10^{13}$ Hz (1471 cm$^{-1}$) to ensure that the pulse amplitude covers the FWHM of a typical Chla absorption in solution
($\Delta \nu \sim 390$ cm$^{-1}$ \cite{Blankenship2014}).
The  insert of Fig.~\ref{fig:Fig2_monomer_coherent}  shows that the 10-90 rise time~\cite{risetimenote} for excitation of the chromophore from its ground state, $t_{rise} = 62.5$ fs, occurs on the timescale of the pulse duration, and that this is followed by a much slower decay due to spontaneous emission on a timescale of $\Gamma^{-1} \sim 56$~ns for this Chla.

One of the most marked features of the Chla excitation by a single-photon coherent pulse in Fig.~\ref{fig:Fig2_monomer_coherent} is the very low value of the excitation probability $P_e$, which is of order $10^{-6}$ for both the bare chromophore and the chromophore coupled to its vibrational bath.  Further analysis based on studying the variation of $P_e^{\text{max}}$ as $\Omega$ is varied reveals two distinct reasons for this reduction, as well as a difference between the excitation probabilities of the bare and phonon-bath coupled chromophore.   
For this value of $\Omega=4.41 \times 10^{13}$ Hz and larger, the value of the excitation probability with and without the phonon bath are quite similar, with both considerably reduced below unity to a value of order $10^{-6}$.  Yet the excitation probabilities show quite different dependences on the coherent pulse bandwidth parameter $\Omega$ when the chromophore is coupled to \makebox[\linewidth][s]{the phonon bath.  Fig.~\ref{fig:Fig10_monomercoherent_bandwidth} summarizes the maximal} values of excitation probability $P_{e}^{\text{max}}$ in both situations, for values of $\Omega$ varying from $10^7$ to $10^{15}$ Hz. 

For the bare Chla monomer without the phonon bath (blue symbols in Fig.~\ref{fig:Fig10_monomercoherent_bandwidth}), the value of $P_e^{\text{max}}$ is seen to depend strongly on the order of magnitude of $\Omega$.   In this case, $P_e^{\text{max}}$ increases from $\sim$10$^{-7}$ to $P_e^{\text{max}} =0.48$  as $\Omega$ is decreased from $\sim$10$^{15}$ Hz to $\Omega_{opt} = 4.30 \times 10^7$ Hz, consistent with the maximal excitation probability for excitation of an isolated two-level system by a pulsed single-photon Guassian coherent state in vacuum with optimal bandwidth parameter $\Omega_{opt} = 2.4\Gamma$~\cite{Wang2011}.   This is due to the increasing contribution of resonant absorption as $\Omega$ decreases to $\Omega_{opt}$.

In contrast, when the chromophore is coupled to the \makebox[\linewidth][s]{phonon bath (red symbols in Fig.~\ref{fig:Fig10_monomercoherent_bandwidth}), $P_{e}^{\text{max}}$ shows} \makebox[\linewidth][s]{a plateau at the value $P_e^{\text{max}} \sim 10^{-6}$ over a broad range} \makebox[\linewidth][s]{of $\Omega$ values.  The bare chromophore absorption (blue} symbols) approaches the absorption calculated with \makebox[\linewidth][s]{phonon coupling at the $\Omega$ value corresponding to} \makebox[\linewidth][s]{the Chla absorption ($\Omega= 4.41 \times 10^{13}$ Hz) and the two} \makebox[\linewidth][s]{values of $P_e^{\text{max}}$ are equal for larger  $\Omega$ values. As $\Omega$} increases beyond this value and approaches $\Omega_{sun}$, both values of $P_{e}^{\text{max}}$ then decrease, continuing the smooth fall with $\Omega$ seen for the bare chromophore absorption for $\Omega > 10^8$ Hz.  

\makebox[0.962\linewidth][s]{This behavior of $P_e^{\text{max}}$ reflects the role of} dephasing fluctuations induced by the thermal phonon bath at  $T=300$ K, which substantially lower the excitation probability, substantially reducing any gain due to greater resonant absorption at smaller values of $\Omega$. The resulting phonon-induced reduction is similar to the differential transmission factors $ \Delta T/T \sim 10^{-6}$ seen in single molecule attenuation measurements at ambient temperatures in solution~\cite{Chong2010,Celebrano2011}, from which empirical absorption cross-sections may be extracted (see below). In contrast, at very high bandwidths the dephasing plays a secondary role to the dominant effect of off-resonance of the chromophore transition energy that results in the fall off of $P_e^{\text{max}}$. The fall off at very low bandwidths is due to the weaker photon density of significantly longer pulses, which effectively reduces the molecule-field coupling~\cite{Wang2011}.

Carrying out the coherent pulse calculations with bandwidth parameter $\Omega = \Omega_{sun}$ (Sec.~\ref{subsec:Parameters}) results in \makebox[\linewidth][s]{excitation evolution profiles that are qualitatively} similar to those shown in Fig.~\ref{fig:Fig2_monomer_coherent}, differing only in a shorter pulse time duration that results in a 10-90 rise time of $t_{rise} = 2.44$ fs. 

%Monomer coherent pulse intensity dependence
\begin{figure}[ht] 
   \centering
      \includegraphics[width=3.4in]{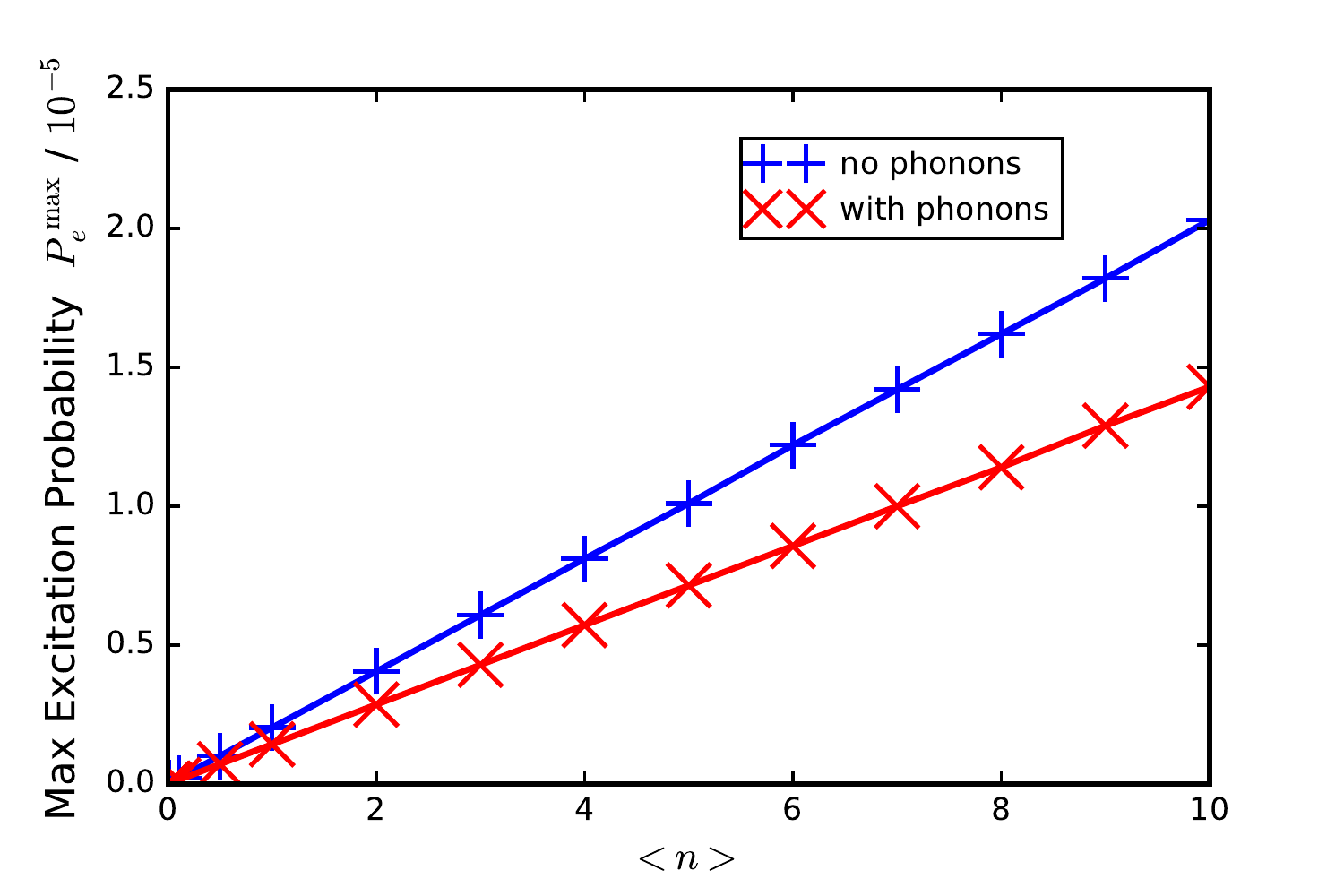}  
 \caption{(Color online) Dependence of maximum excitation probability of Chla on the mean photon number $\langle n\rangle$ under excitation by a coherent  Gaussian pulse with bandwidth parameter $\Omega = 4.41 \times 10^{13}$ Hz. Linear behavior $P_e^{\text{max}} = a \times 10^{-6} \langle n \rangle$ is seen with and without coupling to phonons. Blue line: no coupling to phonon bath, $a = 2.03$. Red line: with coupling to phonon bath, $a = 1.43$.}\label{fig:Fig9_monomercoherent_intensity}
\end{figure}

The broad range of $\Omega$ in Fig.~\ref{fig:Fig10_monomercoherent_bandwidth} corresponds to a very broad range of optical coherence time $\tau_{coh}=1/\Delta \nu$ (Sec.~\ref{subsec:methods_coherentdriving}). The relevant values of $\tau_{coh}$ for the Chla chromophore are 87.7 ns at $\Omega_{opt}=4.30 \times 10^7$ Hz, 85.6 fs at $\Omega = 4.41\times 10^{13}$ Hz, and  3.11 fs at $\Omega_{sun}=1.21 \times 10^{15}$ Hz.  Comparing this timescale with that of the inverse exciton-phonon and phonon-phonon couplings ($\gamma$ and $\lambda$, \makebox[\linewidth][s]{respectively), it is evident that when the pulse} \makebox[\linewidth][s]{bandwidth is selected to cover the Chla absorption} spectrum ($\Omega = 4.41 \times 10^{13}$ Hz), the optical coherence time is of the same order of magnitude as the timescale for phonon relaxation and exciton-phonon energy transfer.  This coincidence is intriguing in that it mirrors the coincidence between the energetic scales of exciton-exciton, exciton-phonon, phonon-phonon couplings and energetic disorder characterizing natural photosynthetic light-harvesting systems (see Sec.~\ref{sec:introduction}). This suggests that \makebox[\linewidth][s]{this coincidence of timescales between the optical} coherence of single photons absorbed by chlorophylls in light-harvesting complexes and the critical timescales of excitonic energy transfer in these systems may constitute a key factor in overcoming the limitations imposed by the ultraweak chromophore-light coupling under illumination by sunlight (see Sec.~\ref{sec:introduction}), as well as playing a role in ensuring the robustness of optimal excitation under sunlight conditions~\cite{Scholes2000,Scholes2017}.    
Further evidence for this derives from additional calculations carried out with randomization of the optical phase at discrete time intervals during the single photon pulse, which showed that the excitation probability decreases as the time duration of the coherent intervals decreases and hence as the effective pulse bandwidth increases beyond its optimal value.  For example, when the optical phase is randomized every 2 fs, the value of $P_e^{\text{max}}$ decreases by two orders of magnitude relative to the value shown in Fig.~\ref{fig:Fig2_monomer_coherent} and also becomes independent of the presence or absence of phonon coupling.

\makebox[0.962\linewidth][s]{With the coherent pulse QSDE approach we may} also readily study the intensity dependence of absorption outside the single photon regime by varying the average photon number $\langle n\rangle=|\alpha|^{2}$ (Eq.~\ref{eq:nPhotonsInPulse}).  This dependence is shown in Fig.~\ref{fig:Fig9_monomercoherent_intensity} for excitation of Chla by the single-photon coherent pulse with the same bandwidth parameter $\Omega=4.41 \times 10^{13}$ Hz, including (red line) and not including (blue line) the chromophore coupling to the phonon bath.  The linear behavior seen for both situations shows that for coherent fields with up to 10 photons at these large bandwidths where the excitation is primarily non-resonant, excitation of the chromophore coupled to the coherent photon field with or without a phonon bath is consistent with a linear response description of absorption of chromophores in pigment-protein complexes, in agreement with conventional analyses of molecular absorption for bulk samples.  Note that even for $\langle n\rangle = 10$, the excitation probability is still exceedingly low, of order $10^{-5}$.  Calculations for larger $\langle n\rangle$ values (not displayed) show that for this pulse bandwidth the linear behavior continues to $\langle n\rangle \sim 10^{4}$, 
after which the excitation starts to saturate, reaching a plateau at a value near unity for $\langle n\rangle \sim10^{13}$, followed by a decrease at even larger $\langle n\rangle$.

\subsubsection{Monomer absorption spectrum}

\begin{figure}[ht] 
   \centering
  \includegraphics[width=3.4in]{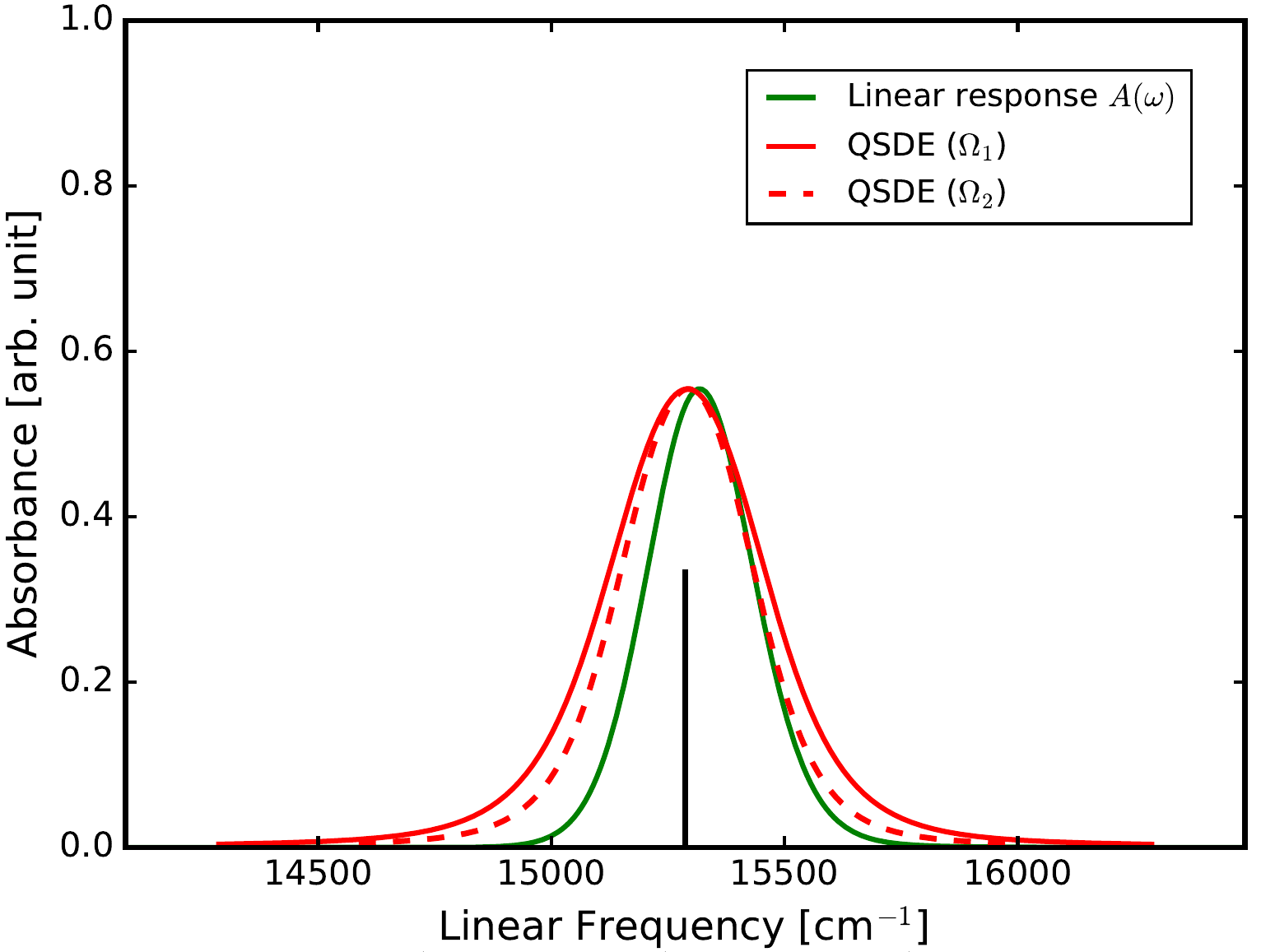} 
 \caption{(Color online) Chla monomer absorption spectrum (all curves are normalized to have the same maximum absorbance). 
 Green line: linear response spectrum. Solid and dashed red lines: direct calculations via QSDE using a coherent-state pulse with two different pulse bandwidth parameters. Solid red line: $\Omega_1 = 4.41 \times 10^{13}$ Hz,  corresponding to a FWHM equal to the experimental value of 390 cm$^{-1}$ for Chla in solution~\cite{Blankenship2014}. Dashed red line: $\Omega_2 = 3.02 \times 10^{13}$ Hz, corresponding to a FWHM covering the linear response spectrum.  The black stick spectrum denotes the excitation energy of the bare chromophore.
 }

\label{fig:Fig1_monomer_spectrum}
\end{figure}

Treating the multi-mode single-photon pulse as a ``broadened single mode pulse'' at the carrier frequency $\omega_0$ further allows us to use these direct calculations of single-photon absorption to estimate a full single-photon absorption spectrum as a function of $\omega_0$ for \makebox[\linewidth][s]{the Chla monomer coupled to the phonon bath} \makebox[\linewidth][s]{without making the usual Franck-Condon assumption} \makebox[\linewidth][s]{of a vertical electronic transition in which the nuclear} \makebox[\linewidth][s]{coordinates of the bath are artificially assumed to be} stationary. This is achieved by scanning the value of \makebox[\linewidth][s]{$\omega_0$ over the frequency range relevant to absorption} \makebox[\linewidth][s]{of the chromophore and recording the  maximum} absorption probability $P_e^{\text{max}}$ at each value, resulting in the absorption spectrum $P_e^{\text{max}}(\omega_0)$. Fig.~\ref{fig:Fig1_monomer_spectrum} shows this \makebox[\linewidth][s]{absorption spectrum calculated using two different values} of the pulse bandwidth parameter $\Omega$, and compares these direct single-photon absorption spectra with the linear \makebox[\linewidth][s]{response prediction calculated using the approach} described in Sec.~\ref{subsec:Absorption_spectrum}, which relies on the Franck-Condon approximation, i.e., assuming that the electronic excitation is instantaneous. The linear response spectrum is shown as the solid green line in Fig.~\ref{fig:Fig1_monomer_spectrum} and has a FWHM of 267 cm$^{-1} = 8.00 \times 10^{12}$ Hz. The solid red line presents direct single photon calculations with bandwidth parameter $\Omega_1 = 4.41 \times 10^{13}$ Hz corresponding to a FWHM equal to the experimental value of 390 cm$^{-1}$ for Chla in solution~\cite{Blankenship2014}, and the dashed red line presents direct calculations with bandwidth parameter $\Omega_2 =  3.02 \times 10^{13}$ \makebox[\linewidth][s]{Hz, corresponding to a FWHM covering the linear} response spectrum. The peak values and FWHM of all three spectra are listed in Table \ref{tab:monomerspectralpeaksFWHM}.

%Monomer absorption spectrum figure

\makebox[0.962\linewidth][s]{The direct single-photon absorption spectrum} calculation is seen to be quite robust to small variations in the bandwidth parameter, as long as this is consistent with the target spectral width. 
For all three spectra, both the two direct QSDE spectra and the linear response spectrum, the values in Table \ref{tab:monomerspectralpeaksFWHM} show that the peak position is blue shifted from the bare excitation energy but by a smaller extent than the reorganization energy, i.e., the peak position is red-shifted from the ``site energy'' of the Chla in the pigment-protein environment, which is conventionally expressed as $E_s = E_1^{(0)} + \lambda$.   This is the usual situation with linear response calculations for chromophores in light harvesting situations at finite temperatures. %Table 3 - Peaks and FHWMs of the spectra in Figure 5
\begin{table}[ht]
 \centering
\begin{tabular}{ | c | c | c | } 
\hline
 & \thead{Peak} & \thead{FHWM} \\ 
\hline

\makecell{Bare chromophore energy $E_1^{(0)}$} & 
\makecell{$15287\text{ cm}^{-1}$ \\ } & 
\makecell{N / A} \\
\hline

\makecell{QSDE ($\Omega_1$)} & 
\makecell{$15292\text{ cm}^{-1}$ \\ } & 
\makecell{$393\text{ cm}^{-1}$} \\
\hline

\makecell{QSDE ($\Omega_2$)} & 
\makecell{$15294\text{ cm}^{-1}$ \\ } & 
\makecell{$334\text{ cm}^{-1}$} \\
\hline

\makecell{Linear response} & 
\makecell{$15314\text{ cm}^{-1}$ \\ } & 
\makecell{$267\text{ cm}^{-1}$} \\
\hline

\end{tabular}
\caption{Peak values and FWHM of the spectra in Fig.~\ref{fig:Fig1_monomer_spectrum}.  See caption of Fig.~\ref{fig:Fig1_monomer_spectrum} for definitions of $\Omega_1$ and $\Omega_2$.
}\label{tab:monomerspectralpeaksFWHM}
\end{table} On decreasing the temperature (not shown here), the spectra show an increasing blue shift from the bare excitation energy, i.e., the red shift from the ``site energy'' decreases, and the width of the spectra decreases.

While there is good overall agreement between the direct single-photon absorption spectrum and the linear response spectrum, as one might expect given the linear dependence of excitation probability on intensity seen in Fig.~\ref{fig:Fig9_monomercoherent_intensity}, there are nevertheless small systematic differences that reflect the lack of any Franck-Condon assumption of vertical transitions in the direct calculation.   At $T=0$ when there is no thermal population of excited vibrations in the ground electronic state, the Franck-Condon principle predicts a peak at $E_s$, but as temperature increases,  the growth of thermal excitations of vibrations in the ground electronic state modifies the \makebox[\linewidth][s]{spectral density $J_s [\omega]$~\cite{Ishizaki2005} and results in a shift of the} absorption peak to the red of $E_s$ as well as increased absorption component to the red of $E_1^{(0)}$. However the differences between the direct and linear response calculations in the wings of the absorption go beyond this Franck-Condon analysis.   In the QSDE continuous time description of coupled radiation-chromophore-phonon dynamics we now have a timescale for photon absorption which is finite on the timescale of relaxation dynamics of phonons, as evidenced by the finite rise times of $P_{e}$ in the direct calculations of Fig.~\ref{fig:Fig2_monomer_coherent} (10-90 rise times of 62.5 fs and 89.3 fs for the blue and red lines, respectively). This results in greater time overlap between the \makebox[\linewidth][s]{phonon and photon dynamics and enables partial} vibrational relaxation to occur during the timescale of photon absorption, modifying the dynamical effect of the finite temperature spectral density.

\makebox[0.962\linewidth][s]{We now address the absolute magnitudes of the} excitation probabilities obtained from these direct calculations with the single photon QSDE approach.  We note first that the conventional linear response evaluation of the absorption spectrum yields only relative magnitudes of absorption and not absolute absorption intensities or cross-sections. Absolute absorption intensities are usually quantified by molecular extinction coefficients or absorption cross-sections, measured in attenuation experiments relying on the Beer-Lambert  law~\cite{Lakowicz2006}.  Recent progress in single molecular spectroscopy has allowed detection of absorption by single molecules using several techniques~\cite{Chong2010,Kukura2010,Gaiduk2010,Celebrano2011}. Of particular note is the  extraction of direct estimates for single molecule cross sections from modulation free transient absorption spectroscopy~\cite{Celebrano2011}.  Assuming similar focusing for a single photon pulse as in ~\cite{Chong2010,Celebrano2011}  leads to an estimated absorption cross-section of $\sigma_{abs} \sim 1$ - 10~\AA$^2$, consistent with typical values for Chla extracted from bulk extinction measurements~\cite{Blankenship2014,Noy2006}.

\makebox[0.962\linewidth][s]{Such an estimate for the absorption cross-section} \makebox[\linewidth][s]{follows the empirical approach of direct absorption} experiments.  We can also make a more fundamental analysis that connects our excitation probability $P_e$ to values of $\sigma_{abs}$ derived from bulk absorption experiments, by a simple estimate based on estimation of the density of geometric modes at each value of the carrier frequency, together with a physically relevant mode volume. Recall that $P_e$ is derived under idealized optical conditions with a single geometric mode (distinct from the multiple frequency modes present in the pulse). Thus, given a density of geometric modes per unit frequency in vacuum (at a given carrier frequency $\omega_0$ and for a single polarization), $ n(\omega_0) \equiv 4\pi\nu^2 /c^3$, with $\nu=\omega_0/2\pi$, and a physically relevant mode volume $V_{m}$,  the probability of excitation by one or more of the possible modes is one minus the probability of not being excited by any modes. This is given by the combinatorial factor
\begin{align}
P=1 - \left( 1 - P_e \right)^{n(\omega_0)V_m d\omega_0},
\end{align}
\makebox[\linewidth][s]{where the differential $d\omega_0$ accounts for the continuous} nature of the carrier frequency.
For $P_e \ll 1$, this reduces to $P_e n(\omega_0)V_{m} d\omega_0$, so that we may then estimate the total absorption probability under conditions of irradiation within a mode volume $V_{m}$ as
\begin{equation}
P_{\text{tot}}= V_{m} \int P_e(\omega_0) n(\omega_0) d\omega_0.
\end{equation}

We may take $P_e^{\text{max}}$ as an estimate of $P_e$, noting that the excited state population at times longer than the pulse duration is not relevant to the absorption.
The relevant mode volume $V_m$ should be the volume in which the irradiating field is coherent.  While this is not easy to \makebox[\linewidth][s]{estimate for chromophores in solution, for natural} photosynthetic systems we may estimate this ``coherence \makebox[\linewidth][s]{volume''~\cite{Mandel1965,Mandel1979} by taking the coherence volume of} sunlight $V_{coh}$ to be given by the product of the Cittert-Zernike coherence area 
$A_{coh} \sim 3\times 10^{-3}$ mm$^2$~\cite{Agarwal2004}  and the coherence length $L_{coh} \sim c/\Delta \nu$~\cite{Kano1962,Mehta1963}, where $\Delta \nu$ is the relevant bandwidth. 
For the Chla monomer, using the bandwidth of the absorption spectrum, which we have argued above is the relevant bandwidth for absorption by the chromophore, results in an integrated absorption probability of $\sim$0.147, in good agreement with values estimated from bulk absorption cross-section values~\cite{Blankenship2014,Noy2006}. 

We can further estimate the photon flux within \makebox[\linewidth][s]{the spectral bandwidth that is incident on the} \makebox[\linewidth][s]{Chla monomer under full sunlight using solar} \makebox[\linewidth][s]{spectral irradiance data~\cite{StandardASTM_2007g173}, which yields photon} \makebox[\linewidth][s]{flux $I \sim$ 122 $\mu$mol m$^{-2}$ s$^{-1}$, equivalent to $\sim$73.7} photons per second incident on a Chla molecule (size $\sim 100$~\AA$^2$~\cite{Blankenship2014}). Combining this with the absorption probability value of $\sim$0.147 results in an estimate of $\sim$0.09 s for the average time required for a single Chla to absorb the full energy equivalent of a single photon. Scaling by the number of Chla per reaction center in the PSII complex ($\sim$270 - 300) and the factor of 3 for \makebox[\linewidth][s]{random polarization, yields an estimated average time of} $\sim$0.1 ms for absorption of the full energy equivalent \makebox[\linewidth][s]{of a single photon by PSII, under the simplest} possible assumption of uncorrelated absorptions and no cooperativity, i.e., no exciton delocalization (this is of \makebox[\linewidth][s]{order 3 - 4 chromophores in PSII~\cite{Amarnath2016}).  Since the timescale} for excitonic energy transport to the reaction center is \makebox[\linewidth][s]{considerably faster, of order ns~\cite{Bennett2013a}, this estimated} photon energy absorption timescale is smaller than, and hence consistent with the measured turnover rate of 200 - 300 electrons per second (i.e., $\sim$3.3 - 5 ms per electron) \makebox[\linewidth][s]{for PSII in its active state~\cite{Lee1989,Chylla1989}.
Note that a smaller} coherence volume will increase the estimated average time for absorption of the full energy equivalent, e.g., a reduction in $V_{coh}$ by a factor of ten increases the time to $\sim$1 ms, which is still consistent with the measured turnover rate for PSII.

%Monomer thermal figure
\begin{figure}[ht] 
   \centering
  \includegraphics[width=3.4in]{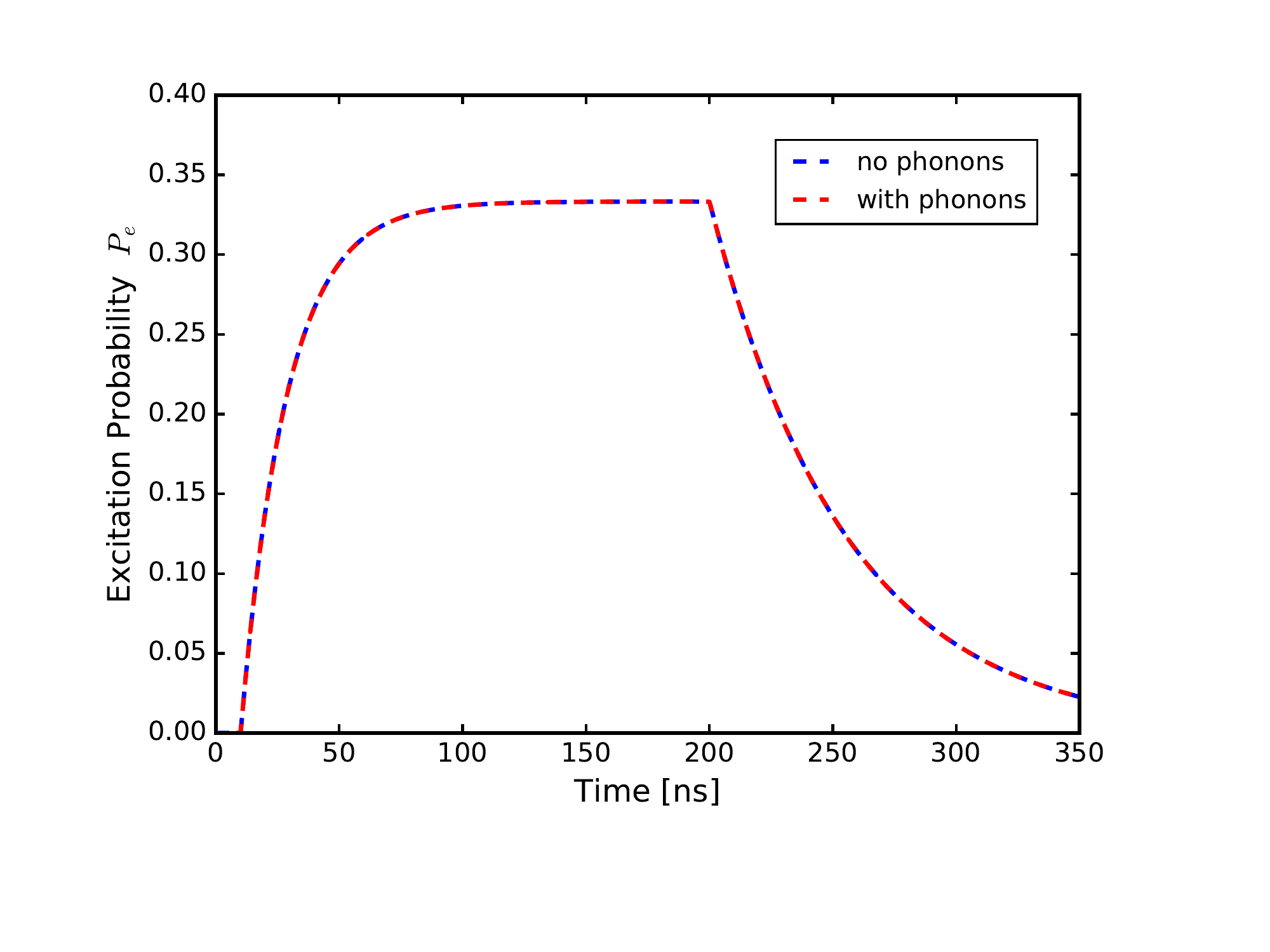} 
 \caption{(Color online) Excitation of Chla by a thermal field with average photon number $\overline{n}$ equal to one. Blue line: without phonon bath. Red line: with phonon bath at $T$ = 300 K.}\label{fig:Fig3_monomer_thermal}
\end{figure}

\vfill\null

\subsubsection{Monomer excitation with thermal radiation}
\label{subsubsec:monomer_thermal}

Fig.~\ref{fig:Fig3_monomer_thermal} shows the time-dependent absorption probability for the Chla monomer under irradiation by the thermal field of Eq.~(\ref{eq:ThermalLindblad}) with a thermal average photon number equal to unity and the field switched on and off at arbitrary times (on at 10 ns and off at 200 ns in this instance).  It is evident that coupling of the excited monomer state to phonons has no effect on the monomer absorption probability.  This results from use of the idealized form of Eq.~(\ref{eq:ThermalLindblad}), which implicitly contains a single mode description of the photons at a single frequency. Consequently the chlorophyll is coupled to the radiation field with the same rate regardless of the energetic fluctuations of the excited state resulting from the phonon-induced dephasing.  The initial rise of the excited state population, its steady-state value, and the subsequent decay after the radiation field is switched off, can all be obtained by analysis of the dynamics in the absence of the phonon coupling, Eq. (\ref{eq:ThermalLindblad_rf}).  For the monomer there is no coherent drive in the rotating frame and the population dynamics given by the diagonal density matrix elements decouple from the off-diagonal matrix elements. The off-diagonal elements (ground-excited state coherences) \makebox[\linewidth][s]{remain zero at all times, while the excited state population}{ satisfies the following master equation for general $\overline{n}$: 
\begin{equation}
\dot{\rho}_e = \Gamma \overline{n} \rho_g - \Gamma (\overline{n}+1) \rho_e.
\label{eq:monomer_mastereq}
\end{equation}
For $\overline{n} = 1$ and initial condition of ground state Chla , i.e., $[\rho^{r}]_{jk} = 1, j=k=1$, $[\rho^{r}]_{jk} = 0$ otherwise, the explicit solution is
\begin{equation}
\rho_e = \frac{N}{3} ( 1 - e^{-3\Gamma t } ), 
\label{eq:monomer_mastereq_solution}
\end{equation}

%P_exe vs n with optimal freq_band figure
\begin{figure}[ht] 
   \centering
  \includegraphics[width=3.4in]{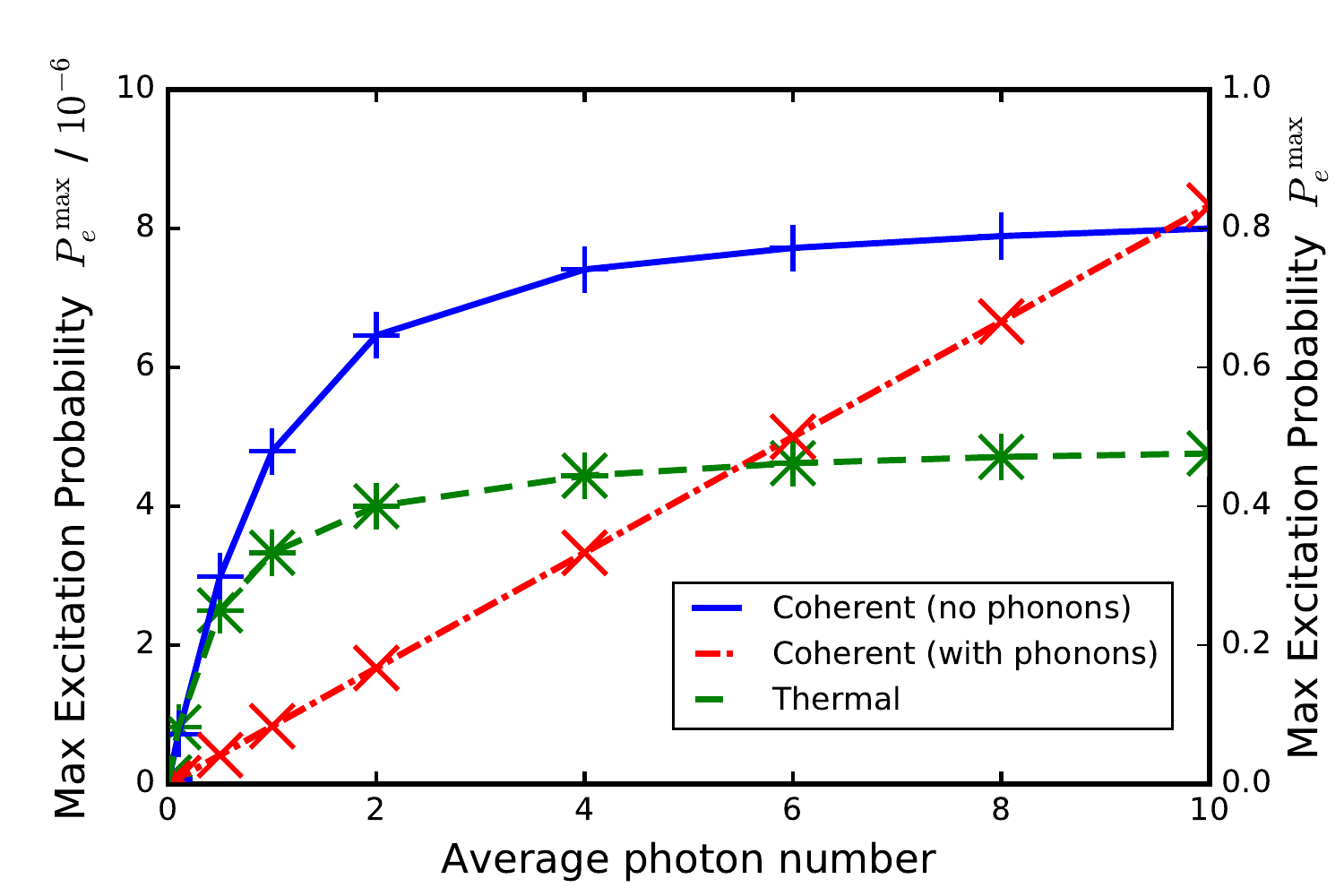} 
 \caption{(Color online) Blue and red lines: $P_{e}^{\text{max}}$ versus average photon number $\langle n \rangle$ for coherent excitation with optimal frequency bandwidth $\Omega_{opt}$ = 4.30 $\times$ 10$^{7}$ Hz. Blue line (right $y$-axis): without phonon coupling. Red line (left $y$-axis): with phonon coupling.  Green line (right $y$-axis): $P_{e}^{\text{max}}$ versus average photon number $\overline{n}$ for thermal excitation. Symbols `$\times$' : with phonons. Symbols `$+$' : without phonons.
}
 \label{fig:Fig7_Pmax_vs_n_with_opt_freq_band}
\end{figure}

which shows an exponential rise of the excited state population $P_e$ with rate $3\Gamma$, yielding a 10-90 rise time of $\sim$2/3$\Gamma$.  For the Chla monomer this yields a 10-90 rise time of 40.9 ns, in excellent agreement with the value $t_{10-90} = 40.9$ ns extracted numerically from Fig. \ref{fig:Fig3_monomer_thermal}. The condition $\dot{\rho}^{r} =0$ further yields a steady-state value of $P_e = 0.333$ in excellent agreement with the plateau value in Fig.~\ref{fig:Fig3_monomer_thermal}. After the field is switched off, the excited state population decays by spontaneous emission.

\makebox[0.962\linewidth][s]{The excitation probabilities with a thermal} radiation field are most similar to those resulting from irradiation with a coherent pulse when the bandwidth of the pulse is very small, corresponding to a long pulse that may in the extreme case be regarded as quasi-adiabatic on the timescale of the phonon motions. 
Fig.~\ref{fig:Fig7_Pmax_vs_n_with_opt_freq_band} shows the dependence of $P_{e}^{\text{max}}$ on the mean photon number $\langle n \rangle$ for coherent excitation with optimal bandwidth parameter $\Omega_{opt} = 2.4 \Gamma$ (blue line - no phonons, red line - with phonons) and compares this with the corresponding dependence of $P_{e}^{\text{max}}$ on the thermal average photon number $\overline{n}$ (green line - with and without phonons). For coherent excitation with the optimal bandwidth and coupling to phonons, the excitation is again primarily off-resonant as a result of the dephasing effect of the phonons, and we find linear behavior as in Fig.~\ref{fig:Fig9_monomercoherent_intensity}, consistent with linear response.  However when the chromophore is now coherently excited without coupling to phonons, the resonant contribution dominates and $P_{e}^{\text{max}}$ reaches its saturation value after $\langle n \rangle \sim$ 8.  This saturation from coherent driving is now more similar to the saturation obtained from classical driving of population with the thermal radiation field (green line), which is independent of the coupling to phonons, as noted above, although the underlying kinetics are quite different.

\subsection{Dimer}
\label{subsec:Dimer}

\makebox[0.962\linewidth][s]{We now consider a dimeric chromophore system,} in particular a heterodimer representing the a603 and a602 Chla chromophores in LHCII of PSII.  This is a strongly coupled dimer in the mid- to low-energy range of the fourteen chlorophylls in LHCII. We employ the chromophore energies and couplings from~\cite{Novoderezhkin2011intra} and the transition dipole moment values taken from~\cite{Bennett2013a}. 

%Dimer absorption spectrum figure
\begin{figure}[ht] 
   \centering
  \includegraphics[width=3.4in]{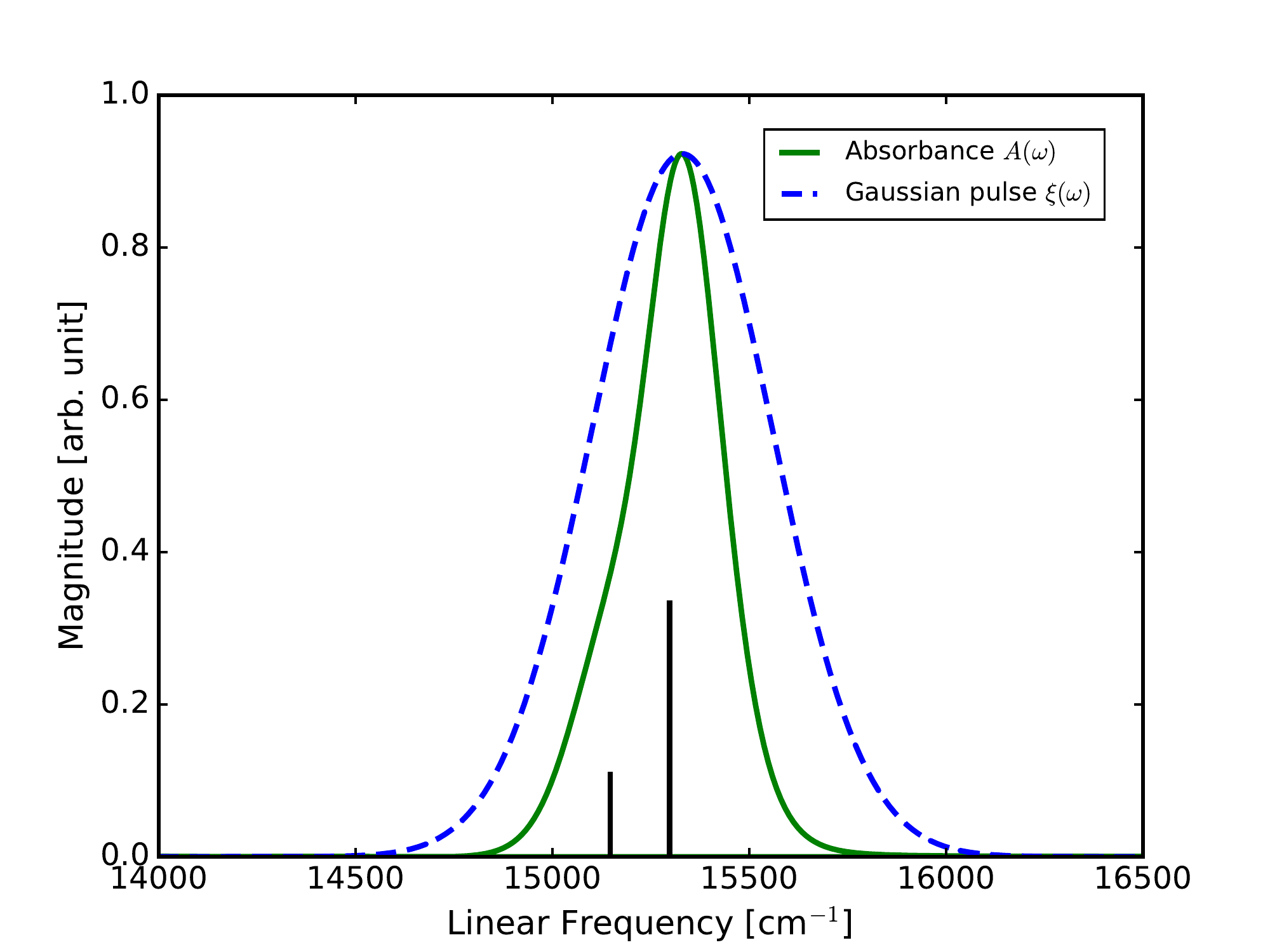} 
 \caption{(Color online) Dimer linear response absorption spectrum for the a603 Chla - a602 Chla dimer in LHCII of PSII. The black stick spectrum denotes the absorption of the bare excitonic states, with the height of each stick denoting the relative magnitude of the excitonic oscillator strengths $f_{i}$.
The Gaussian pulse has bandwidth parameter $\Omega=6.11\times10^{13}$ Hz and is centered here at the maximum of the linear response spectrum for comparison purposes (in the direct absorption calculations (Fig.~\ref{fig:Fig5_dimer_coherent}) it is centered at the average of the two monomer bare excitation energies).
}\label{fig:Fig4_dimer_spectrum}
\end{figure}

\subsubsection{Dimer absorption spectrum}
\makebox[0.962\linewidth][s]{Fig.~\ref{fig:Fig4_dimer_spectrum} shows the linear absorption spectrum} obtained for this heterodimer using the linear response formalism described in Sec.~\ref{subsec:Absorption_spectrum}. The absorptions of the two chromophores overlap to make a single broad asymmetric absorption peak.  Superimposed we show the frequency profile of a multi-mode single-photon Gaussian pulse centered at the maximum of the absorption, with a bandwidth parameter $\Omega = 6.11 \times 10^{13}$ Hz, chosen to cover the dimer absorption band~\cite{dimerbandwidthnote}.

%Dimer coherent pulse figure (2 panels)
\begin{figure*}[ht]   
\centering
\subfloat[]{\includegraphics[width=0.48\textwidth, keepaspectratio]{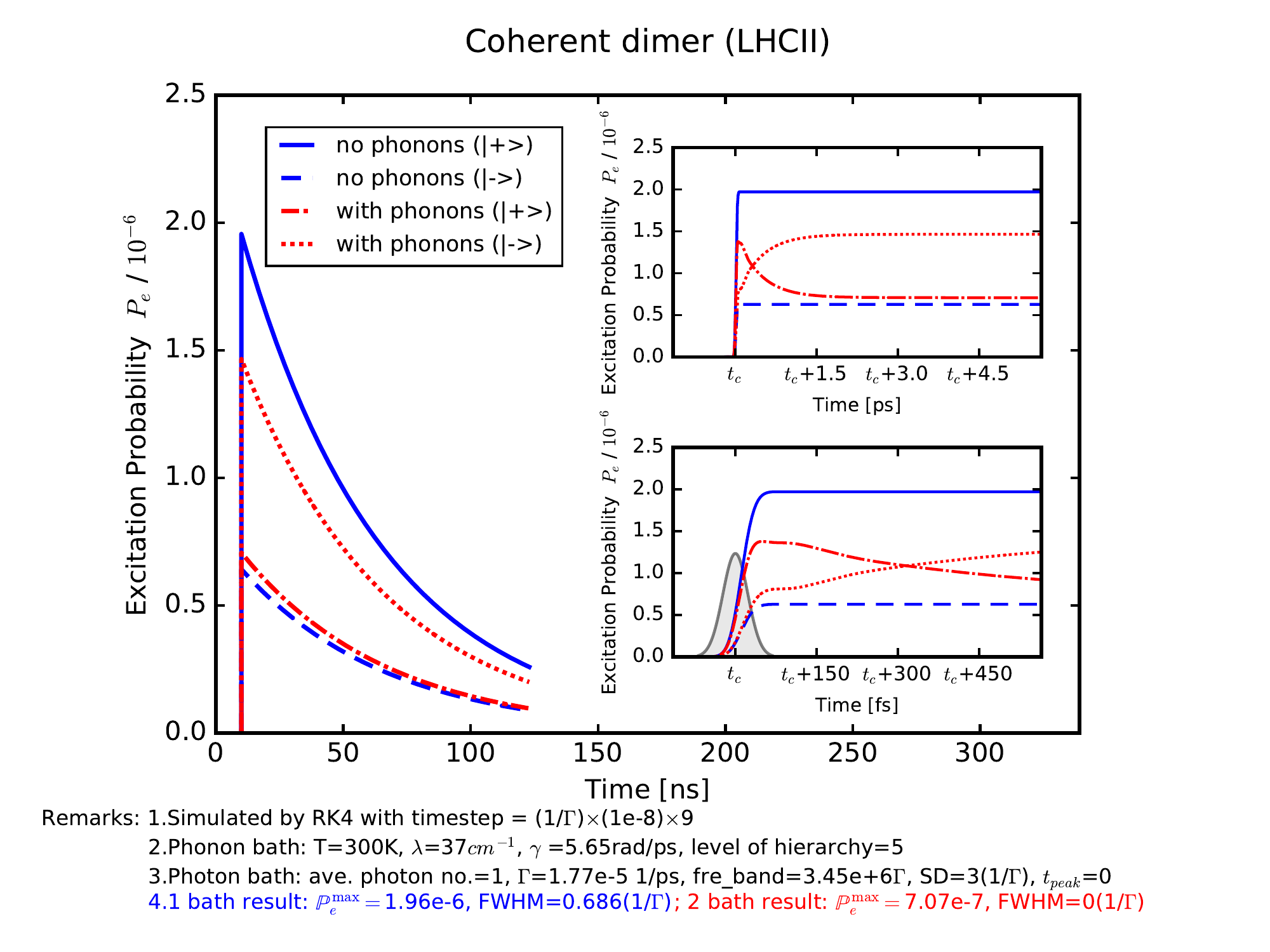}\label{fig:subfig5}}
\subfloat[]{\includegraphics[width=0.53\textwidth, keepaspectratio]{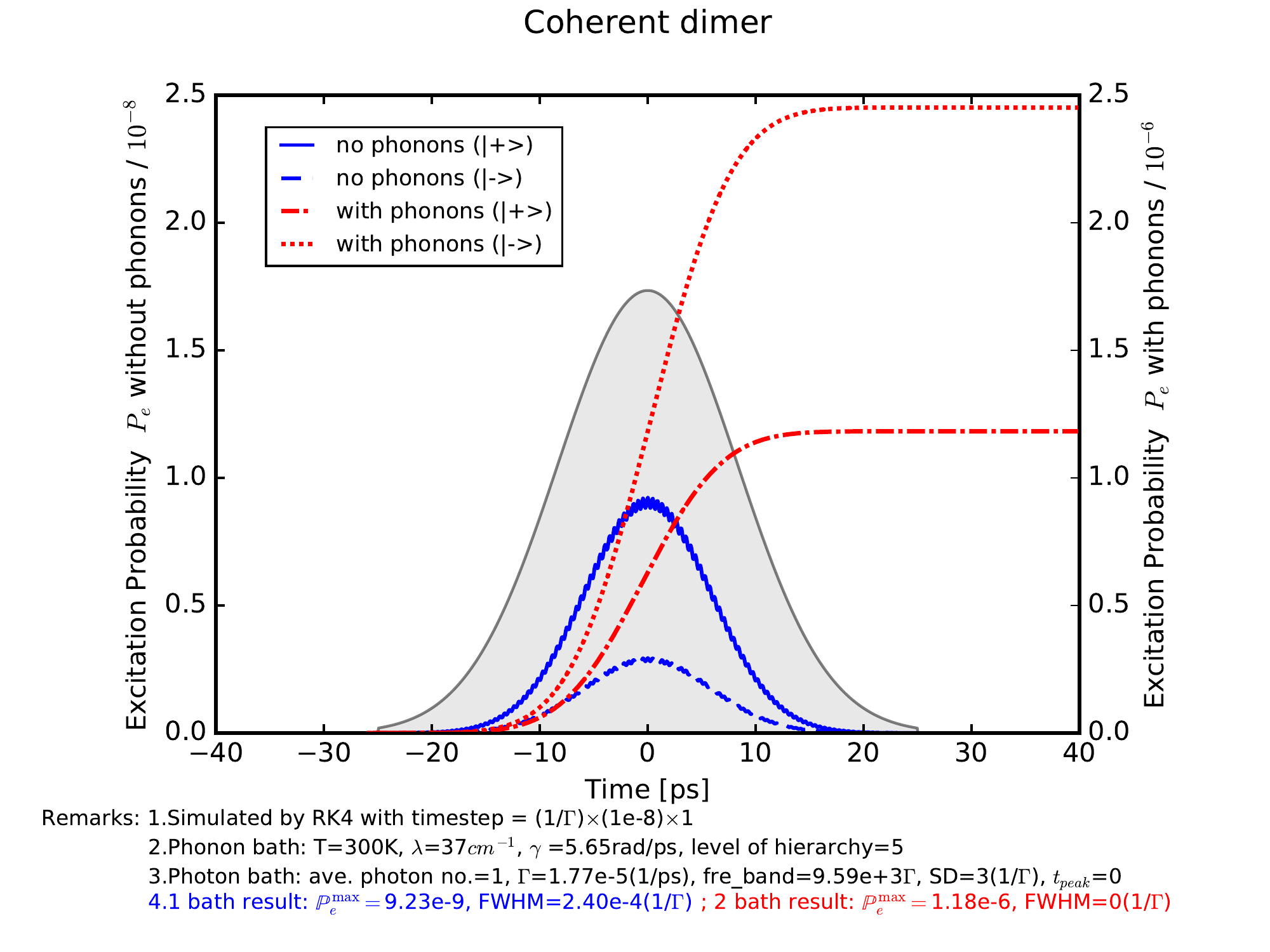}\label{fig:subfig6}}
\caption{(Color online) (a) Excitation of a603-a602 Chla dimer in LHCII by a Gaussian coherent pulse with average photon number  $\langle n \rangle = 1$ and bandwidth parameter $\Omega = 6.11 \times 10^{13}$ Hz, centered at the average of the two bare chromophore excited state energies. $\ket{+}$ and $\ket{-}$ denote the higher and lower energy excitonic states in the single-excitation manifold. Blue lines: populations of excitonic states in the absence of coupling to phonon bath. Red lines: population of excitonic states in the presence of phonon coupling at $T$ = 300 K, and they exhibit a crossover at $t$ = $t{_c}$ + 315 fs. The center of the pulse is at 10 ns, so $t_c=10^4$~ps (upper insert) and $t_c=10^7$~ fs (lower insert).
(b) Coherent excitation of a603-a602 Chla dimer with bandwidth parameter $\Omega = 1.70 \times 10^{11}$ Hz, pulse duration = 50 ps.}
\label{fig:Fig5_dimer_coherent}
\end{figure*}

\subsubsection{Dimer excitation with coherent pulse}

\makebox[0.962\linewidth][s]{Fig.~\ref{fig:Fig5_dimer_coherent}(a) shows the excitation probability $P_e$ as a} function of time for single-photon coherent excitation of the two excitonic states of a heterodimer representing the a603 and a602 Chla chromophores in LHCII of PSII. The excitation is made here with a multi-mode single-photon Gaussian pulse centered at the average of the bare excitation energies of the two chromophores. The result in Fig.~\ref{fig:Fig5_dimer_coherent} is obtained with the bandwidth parameter $\Omega=6.11\times10^{13}$ Hz which excites both chromophores~\cite{dimerbandwidthnote}.
The probabilities of exciting the two excitonic eigenstates are denoted here by $\ket{+}$ and $\ket{-}$, with $\ket{+}$ the higher energy state and $\ket{-}$ the lower energy state.

Analysis of the dependence of the dimer spectra on bandwidth parameter $\Omega$ (not shown here) reveals that the behavior with coupling to the phonon bath is very similar to that seen for the monomer spectra in Fig.~\ref{fig:Fig10_monomercoherent_bandwidth}, namely \makebox[ \linewidth][s]{a plateau value of $P_e^{\text{max}} \sim 5\times 10^{-6} -$ over the range} $\Omega \sim 10^7$ - $10^{17}$ Hz, with a fall off for $\Omega < 10^8$ Hz and for $\Omega > 10^{14}$ Hz.  In the absence of coupling to the bath, the dimer dependence on $\Omega$ resembles the monomer \makebox[ \linewidth][s]{dependence for $\Omega > 10^{13}$ Hz, but drops off sharply as $\Omega$} decreases below this value, in contrast to the increase in $P_e^{\text{max}}$ that is seen in Fig.~\ref{fig:Fig10_monomercoherent_bandwidth} as $\Omega$ is decreased to the optimal vacuum value and the monomer absorption becomes increasingly resonant. In both cases, the behavior at high $\Omega$ values results from  the increasingly dominant role of the off-resonant nature of absorption, just as for the monomer. However when the FHWM of the pulse approaches and then decreases below the energy separation of the two bare chromophore excitonic energies (i.e., $\Omega$ approaches and then decreases below $\sim$1.70 $\times$ 10$^{13}$ Hz), the dimer absorption now becomes increasingly off-resonant again. In the absence of phonons, this results in a sharp drop in $P_e^{\text{max}}$, while the dephasing induced by the phonons will still broaden the absorptive regime of each chromophore and hence that of the full dimer.

In the absence of coupling to the phonon bath, the rise time of the excitation (47.2 fs) and the long timescale (ns) decrease of excitation after the pulse has passed that is due to spontaneous emission (with average rate $\sim$57 ns for these two Chla), are both similar to that seen for the monomer in Fig.~\ref{fig:Fig2_monomer_coherent}.

The dimer behavior at sub ps timescales shows several interesting additional features when the chromophores are coupled to the bath (red lines). The two inserts show that just after the turnover there are two additional timescales. One is a fast timescale of tens of fs over which weak oscillations corresponding to excitonic coherence are seen (lower insert). These oscillations become more pronounced when the bandwidth parameter $\Omega$ is further increased and the pulse length shortens (see below).  The second is a slower timescale of $\sim$3 ps that corresponds to the time to reach the quasi-equilibrium excitonic \makebox[\linewidth][s]{distribution in the presence of phonon coupling.  Note that} without coupling to phonons, the higher energy excitonic state, $\ket{+}$, which has a larger transition dipole moment, shows a larger excitation probability at all times.  When the coupling to phonons is added, the population of $\ket{+}$ is still larger at short (fs) times but the dissipative effect of the non-Markovian coupling causes relaxation to the lower $\ket{-}$ state, resulting in a crossover of the excitation probabilities at approximately $t_{c} $ $+$ 315 fs, after which the populations approach their steady-state values on a ps timescale.

\makebox[0.962\linewidth][s]{As noted above, when the bandwidth parameter $\Omega$} is increased, e.g., to the value of $\Omega_{sun}$ (Sec.~\ref{subsec:Parameters}), with a correspondingly shorter pulse duration, the fs timescale excitonic oscillations 
induced by coupling to the phonons are more pronounced. Conversely, for significantly longer pulses, i.e., pulses of duration 50 ps or greater that correspond to bandwidths less than $\sim$10$^{11}$ Hz, the exciton populations rise smoothly to their steady state values with a very early crossover of the populations so that the lower energy state $\ket{-}$ always appears larger than that of the higher energy state $\ket{+}$ on the timescale resolution of Fig.~\ref{fig:Fig5_dimer_coherent}(b),  and do not show any apparent oscillatory relaxation dynamics (red lines in Fig.~\ref{fig:Fig5_dimer_coherent}(b)). This opposite situation can be understood as an inversion of the conventional hierarchy of timescales for photon absorption from short pulses, resulting instead in a fast, and in the extreme case, an ``instantaneous'' phonon relaxation during slow photon absorption dynamics. In Fig.~\ref{fig:Fig5_dimer_coherent}(b) the electric field is changing on a much slower timescale than that of the phonon relaxation and the exciton relaxation dynamics therefore occurs on a much faster timescale than this.  The result is a very slow rate of interaction of the pulse with the chlorophyll complex, at each point of which excitonic relaxation may be regarded as essentially instantaneous. Note that in the absence of coupling to phonons (blue lines in Fig.~\ref{fig:Fig5_dimer_coherent}(b), the exciton populations now show oscillatory dynamics during the pulse. They also do not saturate at a steady-state value, decaying instead with the pulse. These features can also be rationalized in terms of the slow rate of change of the electric field relative to the excitonic dynamics: since the monomers have different transition amplitudes $\sqrt{\Gamma_i}$, the state reached by the slowly changing field is a superposition of excitonic states with a characteristic oscillation time 
given by the inverse of twice the energy difference between the excitons, which is now much faster than the rate of change of the electric field. Such oscillatory behavior at twice the excitonic frequency difference also results from a perturbative semiclassical analysis for interaction of the dimer with an oscillating field of constant amplitude, which is consistent with the picture of fast excitonic dynamics following an adiabatically changing electric field. Under these conditions the subsequent decay of the pulse results in de-excitation of the excited states, the dynamics of which essentially follow the pulse adiabatically throughout its duration.

This range of behavior can be summarized by the change in the relationship between the optical coherence time $\tau_{coh}$ and the characteristic timescales of exciton and vibrational  dynamics, 
$\tau_{exc} = 1/J, \tau_{vib} =1/\gamma$. 
For the pulse excitation conditions used in Fig.~\ref{fig:Fig5_dimer_coherent}, $\tau_{coh} = 44.2$ fs and we have $\tau_{coh} < \tau_{vib}, \tau_{exc}$.  In this case, during the \makebox[\linewidth][s]{coherent excitation by the pulse the excitonic and} bath degrees of freedom do not have time to relax and we are therefore exciting superpositions of excitonic and vibrational states that show subsequent relaxation \makebox[\linewidth][s]{dynamics.   For the pulse durations longer than 1 ps and} corresponding coherence times (recall $\tau_{coh} =4.53 \Delta t$),  the timescales now satisfy $\tau_{coh} \gg \tau_{vib}, \tau_{exc}$ which ensures \makebox[\linewidth][s]{rapid equilibration of both excitonic and vibrational} degrees of freedom during the optical excitation.

%Dimer thermal figure (5 panels)
\begin{figure*}[!]    %[ht]
\centering
\subfloat[]{\includegraphics[width=0.3258\textwidth, keepaspectratio]{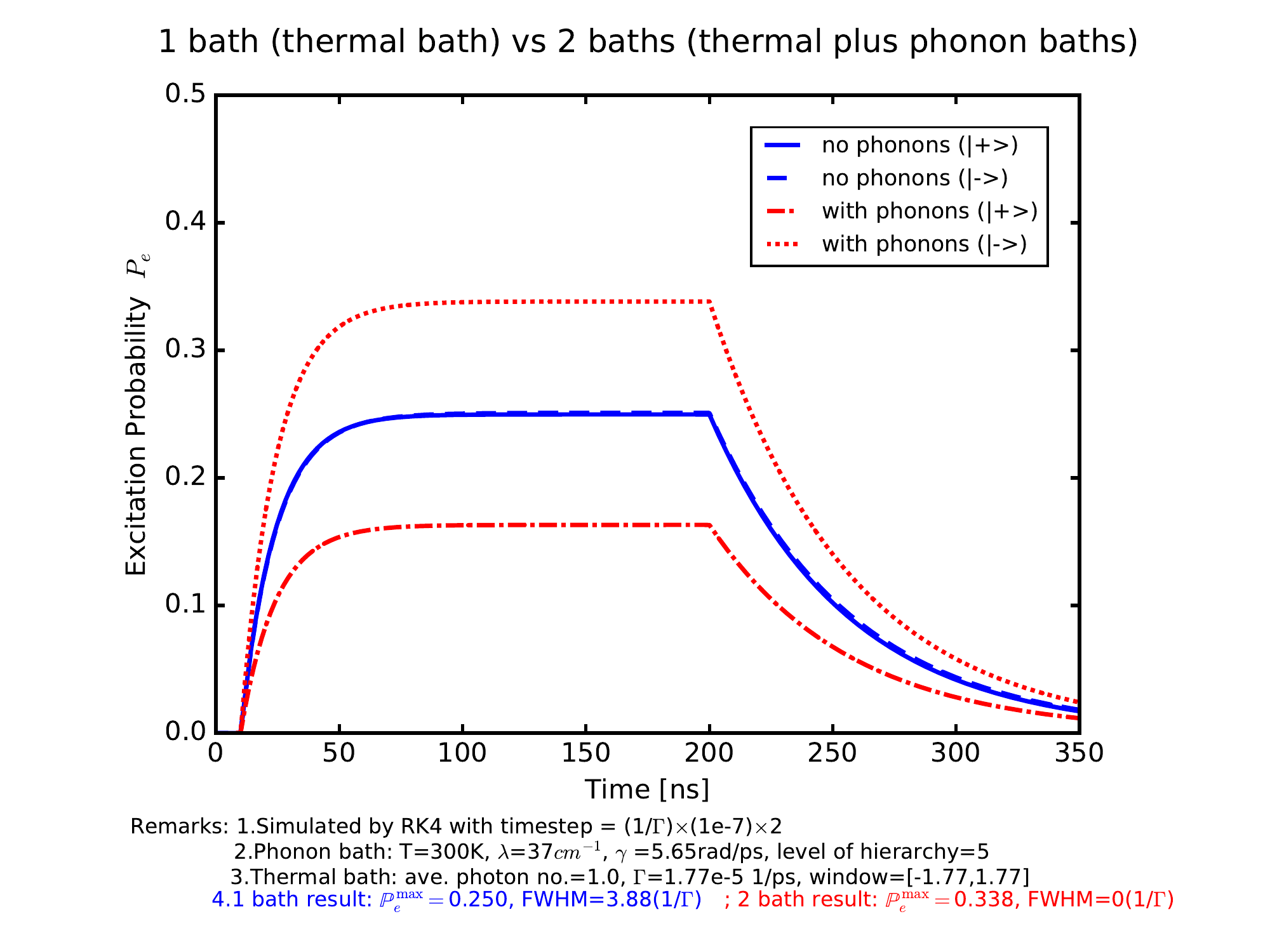}\label{fig:subfig5}} 
\subfloat[]{\includegraphics[width=0.3313\textwidth, keepaspectratio]{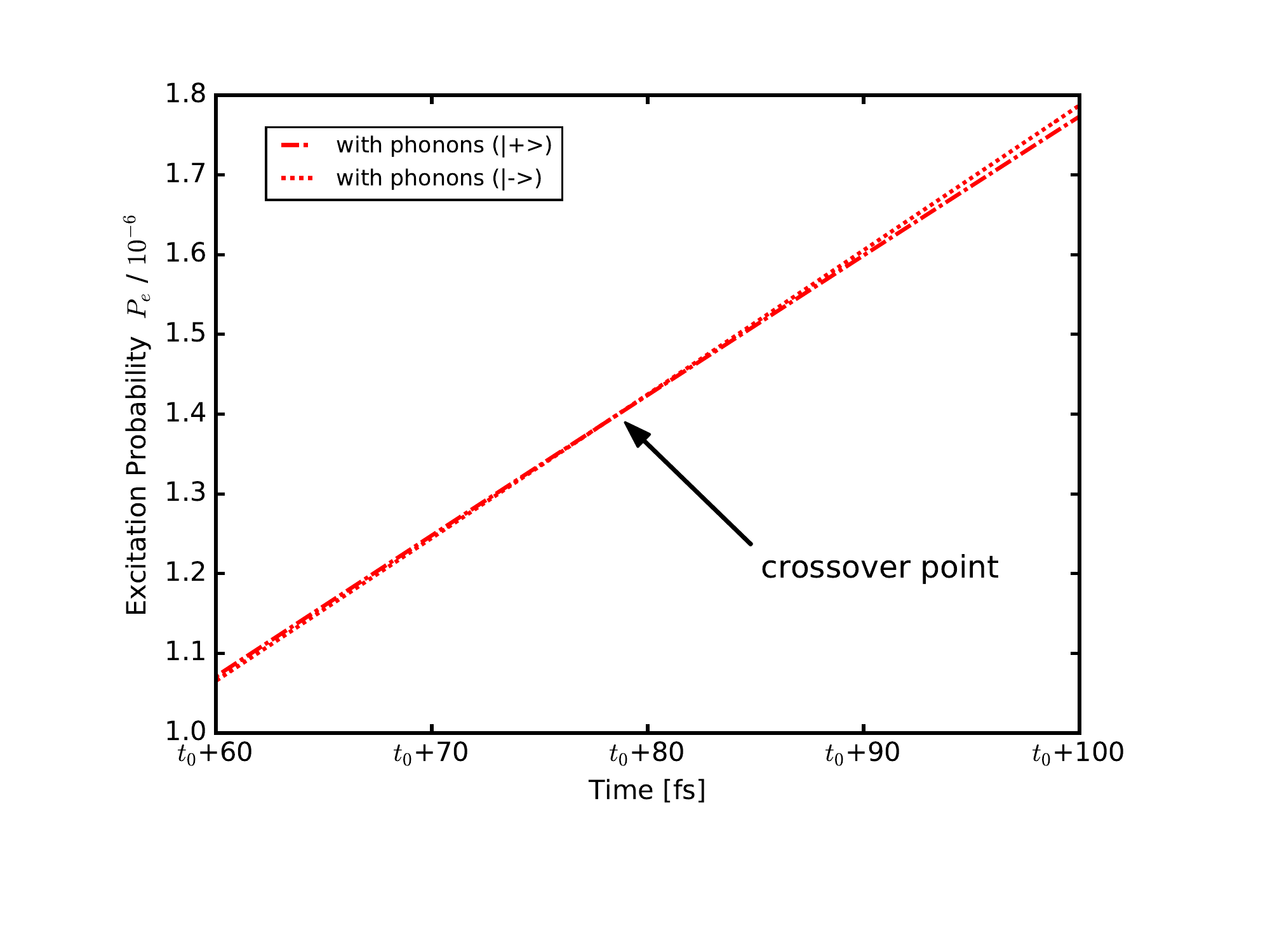}\label{fig:subfig6}}
\subfloat[]{\includegraphics[width=0.33255\textwidth, keepaspectratio]{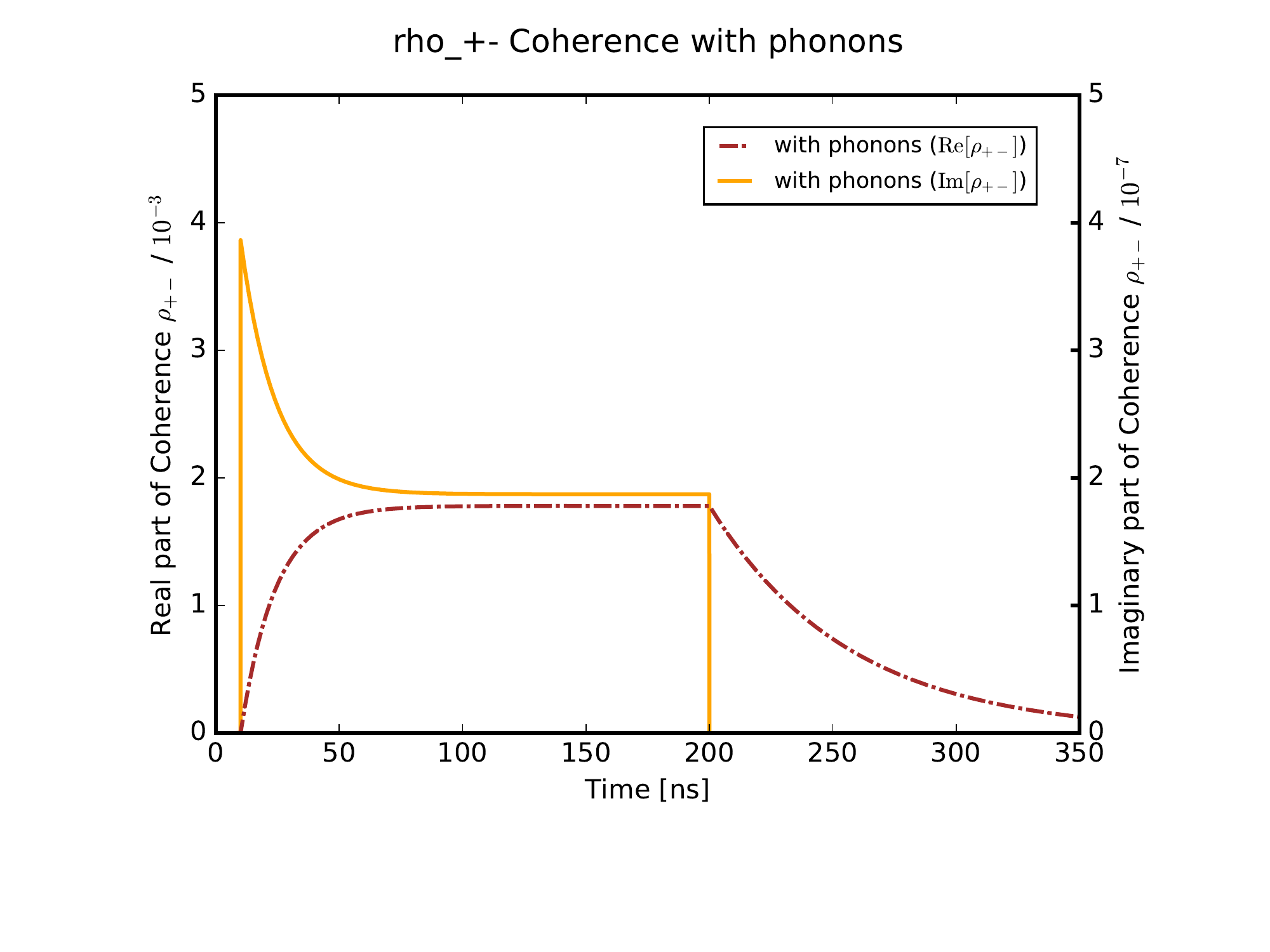}\label{fig:subfig7}}

\subfloat[]{\includegraphics[width=0.4\textwidth, keepaspectratio]{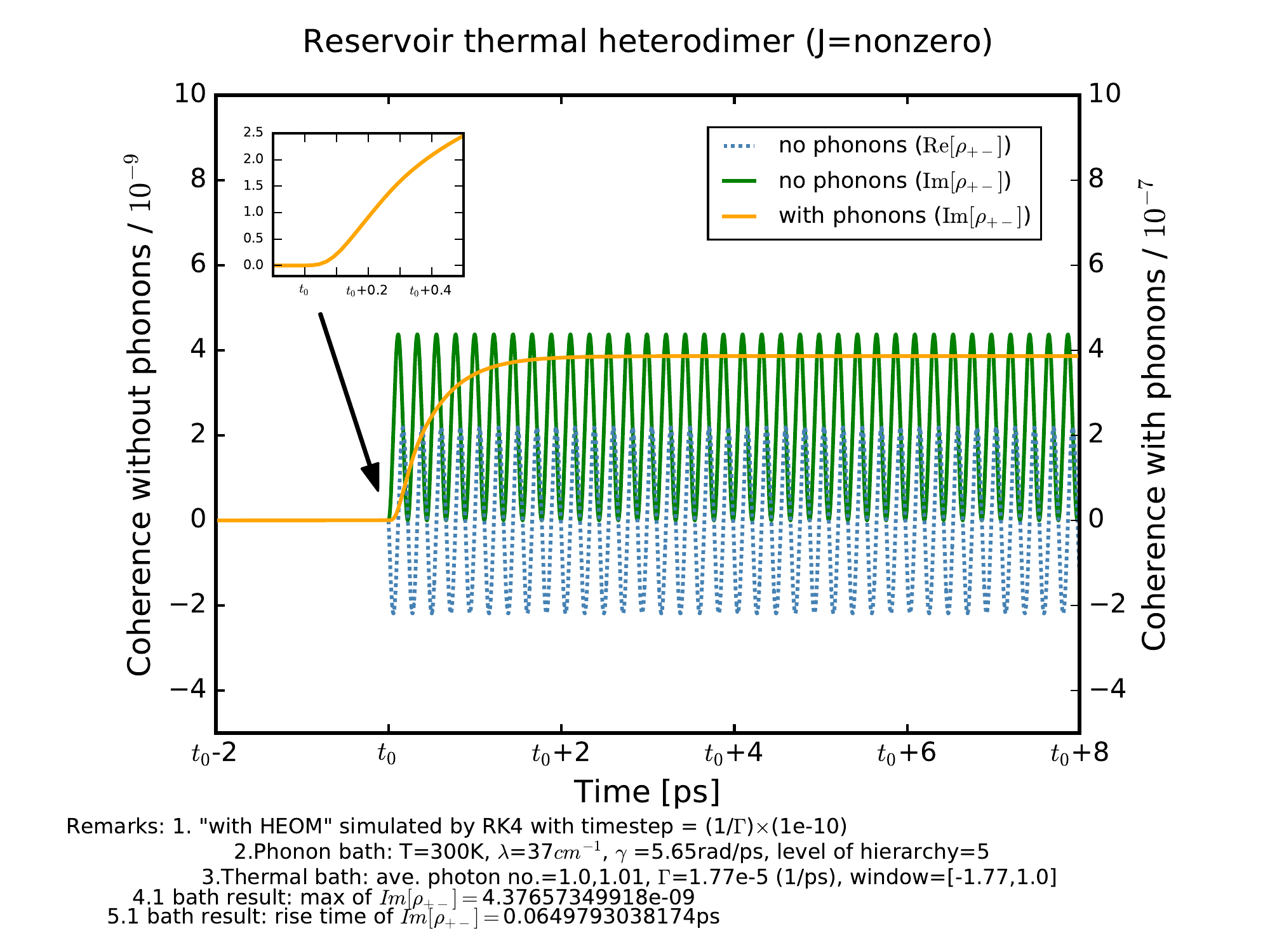}\label{fig:subfig7}}\hspace{2cm}
\subfloat[]{\includegraphics[width=0.4\textwidth, keepaspectratio]{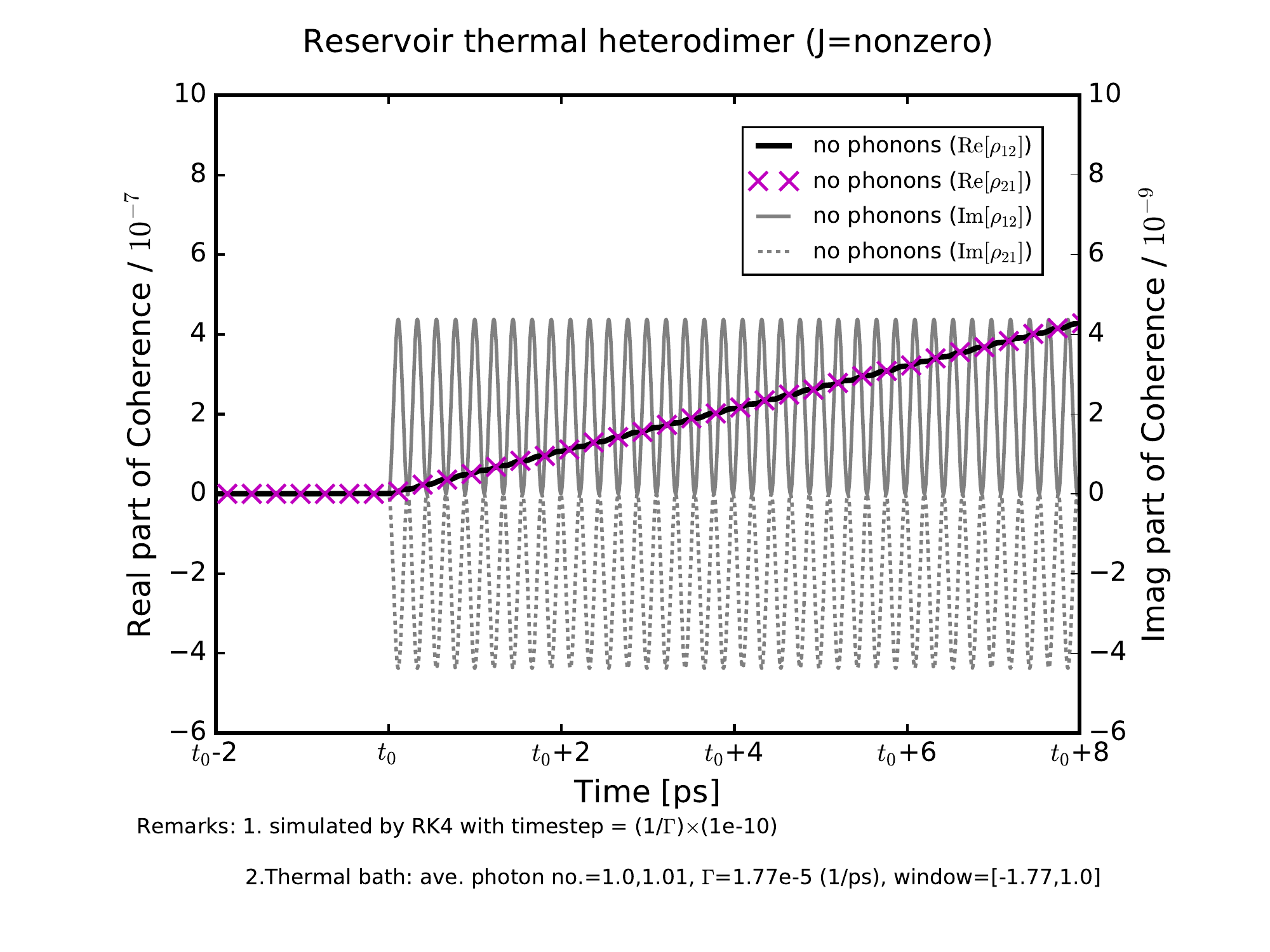}\label{fig:subfig7}} 
\caption{(Color online) Excitation of LHCII dimer by a thermal radiation field with thermal average photon number $\overline{n}=$1.00 and 1.01 for the a603 and a602 Chla chromophores respectively. (a)-(b) Blue lines: without phonon bath. Red lines: including coupling to phonon bath at $T$ = 300 K. The field is switched on at $t_0=$ 10 ns  and the crossover point (panel (b)) is at  $t = t_0+78.5$ fs.
(c) Real and imaginary parts of excitonic state coherence $\rho_{+-}$ in the presence of coupling to phonon bath (d) Real and imaginary parts of excitonic state coherence $\rho_{+-}$ in the absence of phonon coupling (blue and green lines) compared with the non-oscillatory imaginary part of $\rho_{+-}$ in the presence of phonon coupling (yellow line). (e) Real and imaginary parts of the excited state coherences in the site basis, $\rho_{12}$ and $\rho_{21}$, in the absence of phonon coupling. Note that the black and purple hatched lines are coincident.}
\label{fig:Fig6_dimer_thermal}
\end{figure*}

\subsubsection{Dimer Excitation with thermal radiation}

\makebox[0.962\linewidth][s]{Fig.~\ref{fig:Fig6_dimer_thermal} shows the dimer excitation probabilities} obtained with a thermal radiation field, Eq.~(\ref{eq:ThermalLindblad}), where the thermal occupation numbers are now slightly different for the a603 and a602 ($\overline{n} =$ 1.00 and 1.01, respectively) due to their different bare excited state energies $E_k^{(0)}$.  As in the monomer calculation (Fig.~\ref{fig:Fig3_monomer_thermal}), the field 
is switched on at 10 ns and off at 200 ns. As in the comparison for the monomer absorption above, we employ a thermal average photon number of one here.  The overall behavior of both excitation probabilities (Fig.~\ref{fig:Fig6_dimer_thermal}(a), plotted here in the excitonic basis) is similar to that of the monomer (Fig.~\ref{fig:Fig3_monomer_thermal}), showing a rise on ns timescales, with 10-90 rise times $\sim$30 ns, to a steady-state saturation value.   In the absence of phonons (blue lines), both excitonic states are near equally populated, as expected from Eq.~(\ref{eq:ThermalLindblad}) since for this dimer both the stimulated absorption rates $\Gamma_k$   
(Table~\ref{tab:chromophoreparameters}) and the values of $\overline{n}$ (see above) are very similar for the two monomers. However, when the phonon coupling is present (red lines) we now see that the lower energy excitonic state $\ket{-}$ has higher steady-state excitation $P_e$, and that the excitonic probabilities show a crossover of $P_e$ for the $\ket{+}$ and  $\ket{-}$ states at short times.  For the a603-a602 Chla dimer, this crossover occurs at $t_{0}$ + 78.5 fs (Fig.~\ref{fig:Fig6_dimer_thermal}(b)) after which the $\ket{-}$ state rises to a larger steady-state value on account of the relaxation between the excited states that is induced by the non-Markovian dephasing of the chromophore excited states. This relaxation is now fast on the ns timescale of light absorption from the thermal radiation field so that the crossover between the populations of $\ket{+}$ and $\ket{-}$ occurs relatively early on in the initial rise of the populations.

For this heterodimer the time evolution of the excitonic populations is mirrored by a similar exponential rise to a saturation value of off-diagonal excitonic density matrix coherences $\rho_{+-}$ (Figs.~\ref{fig:Fig6_dimer_thermal}(c)-(d)). Such finite off-diagonal excitonic coherences are the result of the non-equivalence of the monomer energies (and hence of their stimulated absorption rates $\Gamma_k$) and are absent for a homodimer as well as for uncoupled ($J$ = 0) chromophores.   The ground-excited state coherences are in all cases zero, as for the monomer in Sec.~\ref{subsubsec:monomer_thermal}.  

Just as for the monomer under thermal irradiation \makebox[\linewidth][s]{(Sec.~\ref{subsubsec:monomer_thermal}), 
these features may be understood by} \makebox[\linewidth][s]{detailed analysis of the dynamics of diagonal and} off-diagonal terms in Eq.~(\ref{eq:ThermalLindblad_rf}).  For a dimeric system, this analysis is now complicated by the presence of the coherent drive term $-\frac{i}{\hbar}[H_{\delta}, \rho^r]$.  In the site basis this term now couples the diagonal population terms to the \makebox[\linewidth][s]{off-diagonal coherences. 
Thus a constant drive by} incoherent thermal radiation will provide a coherent drive of the excited state coherences via the excitonic coupling between chromophores. 

Detailed analysis reveals that the rise times of both populations and coherences are similar to those of the monomer, consistent with the initial absorption being \makebox[\linewidth][s]{essentially defined by the rates $\Gamma_i$ for each excitonic state} $\ket{i}$, analogous to the analysis in Sec.~\ref{subsubsec:monomer_thermal}.  However, the additional coherent drive term introduces several interesting and distinct features into the dynamics.
Explicit analysis of the coupled equations of motion for the dimer density matrix elements, Eq.~(\ref{eq:ThermalLindblad_rf}), shows that
the excited state coherence $\rho_{12}$ between excited states of monomers 1 and 2, and hence also the excitonic coherence $\rho_{+-}$, is initially driven by the coherent term $-\frac{iJ}{\hbar}(\rho_{22} - \rho_{11})$, with \makebox[\linewidth][s]{subsequent additional contributions from stimulated} absorption and stimulated plus spontaneous emission \makebox[\linewidth][s]{terms proportional to $\rho_{12}$, once the value of this} \makebox[\linewidth][s]{coherence has risen from its initial zero value.  This} \makebox[\linewidth][s]{role of the coherent drive is manifested by the slight} concavity evident in the initial growth of $\text{Im}(\rho_{+-})$ (see insert of Fig.~\ref{fig:Fig6_dimer_thermal}(d)). 
For a heterodimer, the} incoherent driving ensures different excited state populations of the two chromophores, so that provided there is some non-zero excitonic coupling $J$, the coherent drive term will then generate non-zero excited state coherences.  These are shown in the exciton basis in Figs.~\ref{fig:Fig6_dimer_thermal}(c)-(d), and in the site basis for the case of no phonon coupling in Fig.~\ref{fig:Fig6_dimer_thermal}(e).
In contrast, for a homodimer the site \makebox[\linewidth][s]{populations $\rho_{11}$ and $\rho_{22}$ will always be equal, and then no} excited state coherence is generated by absorption from the thermal radiation field,
regardless of both the value of $J$ and of whether or not there is coupling to phonons. We note that this analysis applies to the current model of a single light-harvesting complex, i.e., without inhomogeneous broadening effects, and uncorrelated phonon coupling at each chromophore site. Relaxing either of these constraints can potentially give rise to homodimer coherences, as discussed further below.

\makebox[0.962\linewidth][s]{The calculations without phonon coupling reveal} a further interesting feature of these incoherently \makebox[\linewidth][s]{initiated coherences, namely the presence of oscillations} on fs timescales, shown in Fig.~\ref{fig:Fig6_dimer_thermal}(d) for times up to 8 ps. These oscillations decay over a ns timescale of order $\left({\Gamma_1+\Gamma_2}\right)^{-1} \sim 28$ ns. In the absence of phonon coupling we can analytically solve for the steady-state solutions of all density matrix elements in Eq.~(\ref{eq:ThermalLindblad_rf}). This yields a steady-state solution with small but finite values of the coherences. Specifically, the magnitude of the imaginary parts of $\rho_{-+}$ and $\rho_{+-}$  are $\pm \sim$10$^{-10}$ and that of the real parts is  $\sim$10$^{-16}$ in both cases.  The steady-state populations are in agreement with the numerical simulation plateau values that are established on ns timescales (Fig.~\ref{fig:Fig6_dimer_thermal}(a)). To gain a physical understanding of these fast oscillatory coherences, it is useful to analyze them in the site basis, which can be more directly related to a probability current analysis~\cite{Roden2016}. Fig.~\ref{fig:Fig6_dimer_thermal}(e) shows the behavior of $\rho_{12}$ and $\rho_{21}$ over the same ps timescale.  It is evident that the imaginary parts of $\rho_{12}$ and $\rho_{21}$ are of opposite sign but equal magnitude. The unitary contribution to the excited state probability current within the dimer, which is constituted of the sum of the imaginary components of the coherence, is therefore zero. This eventually results in a zero overall probability current, as required in a closed system~\cite{Roden2016}.

\makebox[0.962\linewidth][s]{In the presence of exciton-phonon coupling, the} coherence oscillations are suppressed and die out near the relaxation crossover point. The suppression increases with the coupling to phonons and is not visible for the value of $\lambda$ used in Fig.~\ref{fig:Fig6_dimer_thermal} (Table I). Nevertheless it is evident from Fig.~\ref{fig:Fig6_dimer_thermal}(c) that the magnitude of the steady-state values of the coherence are now significantly larger than the corresponding values in the case of no coupling to phonons.  Thus the phonon bath is assisting in maintaining the coherence. It is possible that this is due to the non-Markovian nature of the coupling, but further analysis is required to characterize the mechanism of this enhancement.
It should be noted that coupling to phonons that are correlated can further enhance the coherence, due to suppression of the relative fluctuations in the excited state energies~\cite{Sarovar2011,Schlau2012}.  Another factor that can provide additional enhancement of the coherences in ensemble systems is inhomogeneous broadening, which can lead to accidental degeneracies that result in long-lived coherences in photon echo experiments~\cite{Dong2014,Butkus2016}.

%7-mer absorption spectrum figure} \label{fig:Fig11_7mer_abs_spectrum}
\begin{figure}[ht] 
   \centering
  \includegraphics[width=3.4in]{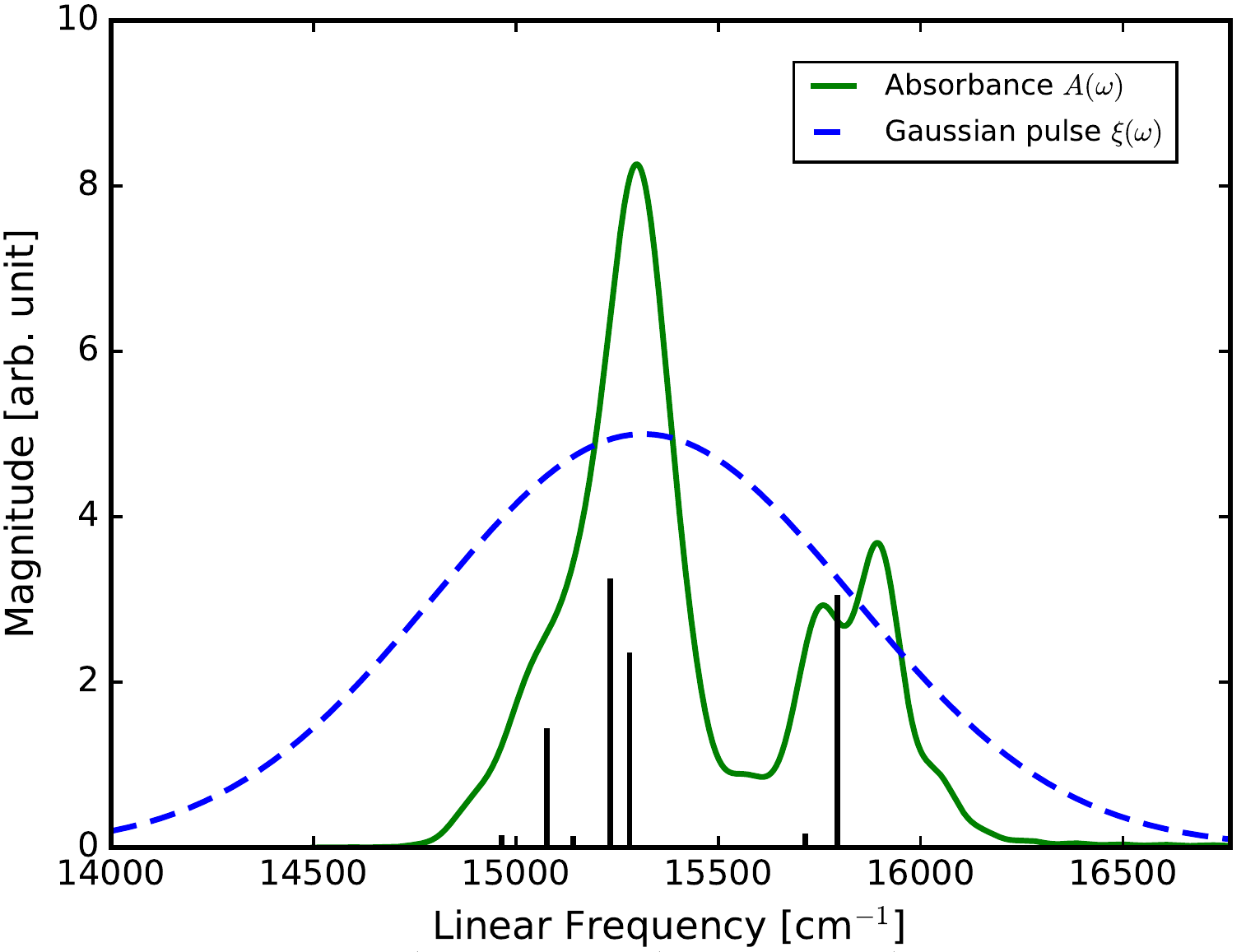} 
 \caption{(Color online) Linear response absorption spectrum for the 7-site chromophore subcomplex of LCHII (green line). The dashed blue line shows a Gaussian pulse centered at the average value of the bare chromophore excited state energies, with bandwidth parameter $\Omega= 1.38 \times 10^{14}$ Hz designed to ensure coverage of the full absorption spectrum~\cite{7merpulsechoice}. The black stick spectrum denotes the absorption of the bare excitonic states, with the height of each stick denoting the relative magnitude of the excitonic oscillator strengths $f_{i}$.}

\label{fig:Fig11_7mer_abs_spectrum}
\end{figure}

\makebox[0.962\linewidth][s]{The above analysis shows that a continuous drive} by an incoherent radiation reservoir will generate excitonic coherences in a heterochromophore system as a result of the combination of excitonic coupling $J$ and different excitation rates $\Gamma_i$ (or equivalently, different oscillator strengths $f_i $), both in absence and presence of coupling to a phonon bath. 
Light-harvesting systems thus constitute a natural example of an open quantum system that can generate steady-state coherences from incoherent driving (see also Ref.~\cite{Karasik2008}).  The steady-state coherences seen here cannot be derived from a classical Einstein rate equation analysis of the excited state populations, which will also predict incorrect population dynamics, since the coherences also enter into the equations of motion for the diagonal terms of Eq.~(\ref{eq:ThermalLindblad_rf}).  We further note that the excitonic coherences generated here by interaction of a dimeric chromophore system from a thermal radiation field are distinct from those observed in Refs.~\cite{Shatokhin2016a} and~\cite{Olsina2014}.   Ref.~\cite{Shatokhin2016a}  reports transient excited state coherences due to additional retarded light-induced inter-chromophore coupling terms in the Hamiltonian, while Ref.~\cite{Olsina2014} reports small oscillatory excited state coherences for an uncoupled dimer ($J=0$) that are due to a collective coherent (superradiant) interaction of the dimer with the field.  In contrast, no collective coherent interaction of the field is assumed in the thermal radiation reservoir analysis employed here.

\subsection{Seven-site chromophore subcomplex of LHCII}
\label{subsec:7Mer}

\makebox[0.962\linewidth][s]{The linear response absorption spectrum for the} 7-mer of LHCII described in Sec. ~\ref{sec:Methods} is shown in Fig.~\ref{fig:Fig11_7mer_abs_spectrum} (green line).  The dashed blue line shows a Gaussian%7-mer coherent pulse figure
\begin{figure}[ht] 
   \centering
  \includegraphics[width=3.4in]{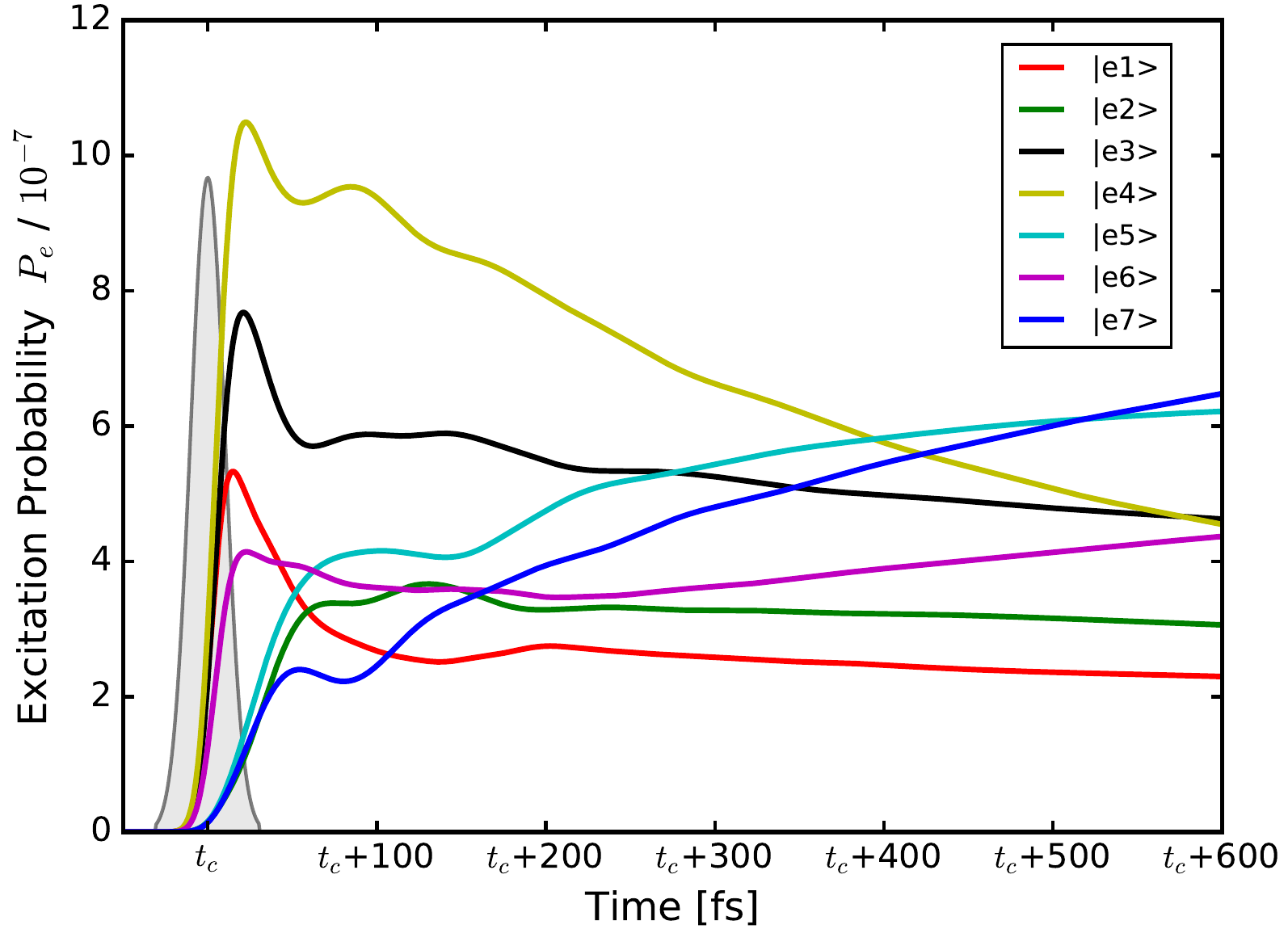} 
 \caption{(Color online) Time dependence of excitation probabilities for the excitonic states (listed in order of decreasing energy) of the 7-site chlorophyll subcomplex in LHCII that results from a Gaussian coherent pulse with mean photon number $\langle n \rangle = 1$ and bandwidth parameter $\Omega= 1.38 \times 10^{14}$ Hz, centered at the average value of the bare chromophore excited state energies.  We include a phonon bath at $T$ = 300 K via the HEOM. See Sec.~\ref{subsec:Parameters} and references therein for details of the 7-mer parameters. The center of the pulse is at $t_c$. 
 }\label{fig:Fig8_7mer_coherent}
\end{figure} \makebox[\linewidth][s]{pulse entered at the average value of the bare} chromophore excited state energies of the 7-mer,
with bandwidth parameter $\Omega= 1.38 \times 10^{14}$ Hz designed to ensure coverage of the full absorption spectrum~\cite{7merpulsechoice}.

\makebox[0.962\linewidth][s]{Fig.~\ref{fig:Fig8_7mer_coherent} shows the time dependence of excitation} probability of each of the seven excitonic states of the 7-mer as a function of time over a period of $\sim$630 fs during and following a single-photon Gaussian coherent pulse with the form shown in Fig.~\ref{fig:Fig11_7mer_abs_spectrum}.  Both the dynamics of photon absorption and excitonic time evolution are now considerably more complex than the corresponding dynamics for the a603-a602 Chla dimer presented earlier.  We see several significant new features. The first is that there is more marked oscillatory behavior in the short time dynamics ($t < t_c+200$~fs). This derives from the greater potential for excitonic coherence in a 7-level system, i.e., more possible superpositions of excitonic states. Secondly, there is a complex dynamical interchange of energy between the seven excitonic states at longer times as the system relaxes (relaxation is clearly not yet complete at 600 fs).  This reflects the greater spatial extent of the 7-mer and the resulting emergence of spatial transport within this subcomplex.  Analysis of the coherences and the probability currents in the site basis can lead to information on the spatial pathways for energy transport~\cite{Roden2016,Roden2016a}.  In addition, we see that for this larger complex there are now two distinct rates of initial absorption for the single photon, with excitons 2, 5, and 7 absorbing at a slower rate than the other four excitons. Detailed analysis shows that these rates of initial absorption correlate with the stimulated absorption rates $\Gamma_i$ (or equivalently, with the oscillator strengths $f_i$) for the excitonic states $\ket{i}$, which depend on both the transition dipole moments $\mu_i$ and the excited state energies $\omega_i$ of the excitonic states. These excitonic quantities can be obtained from the chromophore values $\mu_k$ and $\omega_k$.  However, since for this 7-mer we now have both Chla and Chlb chromophores, the rates $\Gamma_i$ for the excitonic states do not bear a simple relation with the transition dipole moments of the individual chromophores.

\section{\label{sec:Discussion}Summary discussion and outlook}

\makebox[0.962\linewidth][s]{We have presented a full quantum description of} the dynamics of photon absorption by pigment-protein complexes driven by multi-mode pulses of coherent states. 
A quantum optical master equation derived from QSDE equations of motion for chromophore excitations driven by quantum noise originating from the continuum of modes associated with the multi-mode pulses~\cite{Baragiola2012}, was combined with a hierarchical set of equations of motion for the electronic excitations coupled to the phonon environment derived from the pigment and protein vibrational degrees of freedom~\cite{Ishizaki2009a}, to make simulations of photon absorption under realistic pigment-protein conditions that cover the full range of timescales from fs to ns.
Applications to models of monomers, dimers and sub-complexes of chlorophyll pigments in LHCII have revealed several key dynamical features that depend on both the parameters defining the radiation coupling to the chromophores and the parameters defining the coupling of the vibrational environment to the chromophores. 

\makebox[0.962\linewidth][s]{Under single photon conditions corresponding to} the ultraweak intensities of sunlight and correspondingly weak interaction with the pigment-protein complexes, relative to the exciton-phonon and phonon relaxation energy scales, the key radiation parameter determining the quantum dynamics of absorption from a coherent pulse is the magnitude of the pulse bandwidth parameter $\Omega$, or more precisely, the ratio of $\Omega/\Gamma_k$, where $\Gamma_k$ are the underlying spontaneous emission rates of the individual pigments.  Since the latter depends on the bare pigment excited state energies $E_k^{(0)}$ and their transition dipole moments $\mu_k$, the critical radiation-matter parameters defining the dynamics of absorption are thus pulse bandwidth and pigment energy and transition dipole moment. 

Our results show that the rise time of the excitation probability occurs on the timescale of the pulse duration, which is on the order of the coherence time of the pulse, $\tau_{coh}= \pi/(\sqrt{ln 2}\Omega) = 3.77/\Omega$.
 We have argued that the relevant frequency bandwidth for the pulse is that of the chromophore spectral absorption, rather than the bandwidth of visible light.  This is seen to yield excitation rise times on order of tens of fs for Chla, somewhat longer than than the timescale of a few fs that is often predicted using the bandwidth of visible light.   

\makebox[0.962\linewidth][s]{The frequency bandwidths relevant to Chla} spectral absorption and visible light are of order 10$^{13}$ Hz and 10$^{14}$ Hz, respectively, and correspond to ratios $\Omega/\Gamma$ that are several orders of magnitude greater than the optimal bandwidth for absorption by two-level \makebox[\linewidth][s]{systems~\cite{Wang2011} in vacuum. This means that the contribution} of resonant absorption is severely suppressed, independent of whether the chromophores are coupled to the vibrational environment or not. This bandwidth induced resonant suppression is the dominant factor determining the magnitude of the maximum excitation probability for the pigment-protein complexes.  At smaller pulse bandwidths, the coupling to vibrations plays a greater role in differentiating the efficiency of absorption by the chromophore when actively coupled to the vibrational environment from that by a bare chromophore.  We find that even for the optimal bandwidth ratio for a bare single chromophore, $\Omega/\Gamma = 2.4$, in the presence of coupling to the vibrations the maximum excitation probability is of only of order 10$^{-5}$, reflecting the critical role of vibrational dephasing. 

\makebox[0.962\linewidth][s]{For a monomeric chromophore system, this} dephasing is the only effect of the coupling to the \makebox[\linewidth][s]{vibrational environment and after the peak excitation} is reached as the pulse nears completion, the excitation probability slowly decays on the ns timescale of the spontaneous emission.  However for complexes with multiple chromophores, the coupling to vibrations plays a key role in determining the form of the chromophore excitation dynamics subsequent to optical excitation.  For the dimer and 7-mer in LHCII, the full quantum calculations reveal a smooth evolution of the population of the excitonic states from the initial distribution reached during the pulse, which is determined by the relative values of $\Gamma_k$ for each chromophore and its spectral location within the pulse envelope, to a time-dependent set of exciton populations that relaxes coherently within the manifold of singly excited states of the pigment-protein complex over a timescale of ps.  
As is well established in the recent light-harvesting literature~\cite{Ishizaki2010,Ishizaki2012}, this relaxation is controlled by the non-Markovian coupling of the excitonic states to the vibrational environment and is characterized by coherent oscillations on timescales of hundreds of fs that are determined by the energy difference between the excitonic states. The combined QSDE/HEOM calculations show that these oscillations are more pronounced for the shorter pulses deriving from larger bandwidths.  

\makebox[0.962\linewidth][s]{This behavior is seen for the large pulse bandwidth} parameter values $\Omega$ relevant to natural conditions, whether these are determined by the homogenous spectral absorption bandwidths of the pigment complexes or by the even larger conventional bandwidth estimate of sunlight (covering the visible range, 400 - 700 nm). Such pulses have temporal FWHM durations of at most tens of fs.  With significantly longer pulses, e.g., of duration of 10 ps or longer, no oscillatory relaxation dynamics are seen, reflecting an inversion of the conventional timescales in which there is now fast ``instantaneous'' phonon relaxation during  slow photon absorption dynamics.  This constitutes an ``anti Franck-Condon'' regime of single-photon absorption.

\makebox[0.962\linewidth][s]{Within the QSDE/HEOM approach with a} coherent state photon pulse, it is also possible to \makebox[\linewidth][s]{calculate the absorption of coherent states with larger} values of average photon number 
$\langle n \rangle$. Analysis of the absorption as a function of average photon number $\langle n \rangle$ for the large $\Omega$ values relevant to absorption in natural conditions shows a linear response behavior, both with and without phonons, as a result of the dominant non-resonant nature of the absorption.  Comparing the evolution of excitation probability under these coherent conditions with that derived under excitation by a \makebox[\linewidth][s]{thermal reservoir, according to the conventional treatment} of thermal radiation sources, the greatest similarity was found when the bandwidth parameter $\Omega$ is small,  where the absorption without phonon coupling now shows a saturation with increasing $\langle n \rangle$. However when coupling to phonons is added, the absorption intensity is again linear in $\langle n \rangle$ for absorption from a coherent pulse.

\makebox[0.962\linewidth][s]{We showed that the QSDE/HEOM calculations of} single-photon absorption can be used both to make a direct calculation of the absorption spectrum without making a Franck-Condon assumption of instantaneous electronic excitation, and to estimate the absolute absorption probability for absorption from sunlight.  The latter estimate shows that when scaled by the number of geometric modes in the coherence volume of sunlight, the very low value of excitation probability $\sim$10$^{-6}$ for a single chromophore irradiated by a single-photon pulse in a single geometric mode yields absorption probabilities of order $10^{-1}$ that are consistent with estimates based on bulk measurements of extinction coefficients. We presented a simple analysis to estimate from this the average time needed to absorb the full energy equivalent of a single photon by a single Chla molecule under full sunlight, using the radiation incident on the molecule that is within the spectral bandwidth. This yields a value of $\sim$0.09 s for the time required for a single Chla chomophore to absorb the energy equivalent of one (single-polarization) photon.  We argued that this further implies an average time of $\sim$0.1 ms for harvesting the full excitation energy of a single photon and getting this to a reaction center in PSII, assuming the simplest model of uncorrelated single-photon absorptions by individual chlorophyll molecules and $\sim$270 - 300 chlorophyll molecules per reaction center.  This average time is consistent with the experimentally measured turnover rate for PSII, i.e., the rate of production of electrons under active conditions~\cite{Lee1989,Chylla1989}.

This raises two important questions for a microscopic understanding of the quantum efficiency of natural light harvesting. First, a key parameter in this estimate is the \makebox[\linewidth][s]{finite coherence volume of sunlight, raising the intriguing} question of to what extent such single-photon coherent states might be relevant in absorption under natural sunlight conditions.  Additional motivation for investigating this question comes from our observations that shorter single-photon pulses give rise to more manifestly coherent excitonic dynamics, showing oscillatory relaxation dynamics on the fs - ps timescales following a short pulse. Such direct manifestations of exciton dynamics are absent under single-photon absorption from significantly longer pulses (on timescales of tens of ps) because the phonon relaxation is now fast relative to the absorption dynamics. Any contribution of single-photon coherent pulses to absorption under natural conditions will also have implications for design of artificial light-harvesting devices optimized to take advantage of spatial and/or temporal coherence of sunlight~\cite{Mashaal2011,Mendoza2017}.  Second, since in any given single photon absorption, on average the \makebox[\linewidth][s]{energy equivalent of only $\sim$10$^{-1}$ photons is absorbed by} the light-harvesting complex, it is important to further understand the spatiotemporal nature of the sequence of $\sim$270 - 300 single-photon absorption events needed to generate the full energetic equivalent of a single optical photon that is required to generate an electron in the reaction center.  Whether the full energy of a photon can be absorbed instead in a single ``quantum jump'' is a further, challenging, question for future research. Experiments with single-photon sources, as recently carried out for photoabsorption studies of retinal rod cells of \textit{Xenopus leavis} toads~\cite{Phan2014} will be very useful for such investigations.

\makebox[0.962\linewidth][s]{The quantum optical master equation derived from} the QSDE approach can be generalized to radiation fields consisting of an incoherent sum of coherent terms, e.g., coherent states or Fock states~\cite{Baragiola2012}.  Such a description also possesses the benefit of allowing a comparison of absorption from a coherent pulse of well-defined energy content to absorption from an incoherent source composed of pulses with similarly well-defined energy content.  This approach is thus well suited to analysis of absorption under natural sunlight, which is often described as consisting of short bursts of light occurring at infrequent times. Extension of the present study of single-photon absorption from coherent states to a study of single photon absorption under natural conditions from a radiation field modeling sunlight as an incoherent sum of coherent terms using the quantum optical master equation derived from the QSDE approach will constitute the focus of future work.

\clearpage
\onecolumngrid 

\section{\label{sec:Appendix}Appendix}
The system Hamiltonian parameters of the 7-mer (in cm$^{-1}$) taken from Ref.~\cite{Novoderezhkin2011intra}:

\begin{align} H_{s}^{(7)} 
& =\left[\begin{array}{cccccccc}
15761 & 36.07 & 6.72 & -5.84 & 4.35 & -1.08 & 61.97 & 0\\
36.07 & 15721 & 96.66&	-19.25&	4.30&	-2.57&	3.86   & 0\\
6.72 & 96.66 & 15287 & 38.11 &	-2.70&	-0.76 & 12.97 & 0\\
-5.84 & -19.25 & 38.11 & 15157 & 9.69 &	15.83	&-11.39 & 0\\
4.35 & 4.30 & -2.70 &	9.69&	15112 &	126.92&	-24.96 & 0\\
-1.08 & -2.57 & -0.76 &	15.83	& 126.92 &	15094&	23.10 & 0\\
61.97 & 3.86 & 12.97 &	-11.39&	 -24.96	& 23.10	&15073 & 0\\
0 & 0 & 0 & 0 & 0 & 0 & 0 & 0
\end{array}\right]\label{eq:7merHamiltonian}\\
\nonumber 
\end{align}

Following the indexes in Fig.~\ref{fig:7mer_structure}, the chromophores in Eq. (\ref{eq:7merHamiltonian}) are arranged in descending order of their bare choromophore excited state energies and listed together with the corresponding Weisskopf-Wigner decay rates calculated by Eq. (\ref{eq:GammaDef}) in Table~\ref{tab:7merparameters} below.
%Table 1 - monomer and dimer parameters
\begin{table*}[ht]
 \centering
\begin{tabular}{ | c | c | c | c | c | c | c | c |} 
\hline
\makecell{Index in Fig.~\ref{fig:7mer_structure}}
 &  1  &  2  &  3 & 4 & 5 & 6 & 7 \\ 
\hline
  \makecell{Chromophore name \\ in Ref.~\cite{Novoderezhkin2011intra}} & 
\makecell{b608} & 
\makecell{b609} &
\makecell{a603} &
\makecell{a602} &
\makecell{a611} &
\makecell{a612} &
\makecell{a610}
 \\
\hline
\ \makecell{Weisskopf-Wigner \\ atomic decay rate $\Gamma_k$ \\ ($\times 10^{7}$ Hz)}  
 & \makecell{1.42}  
 & \makecell{1.41}
 & \makecell{1.79}
 & \makecell{1.75}
 & \makecell{1.73}
 & \makecell{1.73}
 & \makecell{1.72}
 \\ 
\hline

\end{tabular}
\caption{Indexes and names of the choromophores employed in this work to model a LHCII 7-mer.}\label{tab:7merparameters}
\end{table*}

\twocolumngrid

\begin{acknowledgments}
We thank Joshua Combes, Akihito Ishizaki, Nick Lewis and Daniele Monahan for helpful discussions. 
O. E. G. and G. R. F. were are supported by the Director, Office of Science,
Office of Basic Energy Sciences, of the USA Department of Energy under
Contract No. DE-AC02-05CH11231 and the Division of Chemical Sciences,
Geosciences and Biosciences Division, Office of Basic Energy Sciences
through Grant No. DE-AC03-76F000098 (at LBNL and UC Berkeley).
\end{acknowledgments}

\bibliography{Coherent_Energy_Transfer,Optics,Mathematics,QuantumInfo,Spectroscopy,Photosystem2_BWlatexversionNotes}

\begin{thebibliography}{79}
\expandafter\ifx\csname natexlab\endcsname\relax\def\natexlab#1{#1}\fi
\expandafter\ifx\csname bibnamefont\endcsname\relax
  \def\bibnamefont#1{#1}\fi
\expandafter\ifx\csname bibfnamefont\endcsname\relax
  \def\bibfnamefont#1{#1}\fi
\expandafter\ifx\csname citenamefont\endcsname\relax
  \def\citenamefont#1{#1}\fi
\expandafter\ifx\csname url\endcsname\relax
  \def\url#1{\texttt{#1}}\fi
\expandafter\ifx\csname urlprefix\endcsname\relax\def\urlprefix{URL }\fi
\providecommand{\bibinfo}[2]{#2}
\providecommand{\eprint}[2][]{\url{#2}}

\bibitem[{\citenamefont{Yaghoubi et~al.}(2014)\citenamefont{Yaghoubi, Li, Jun,
  Lafalce, Jiang, Schlaf, Beatty, and Takshi}}]{Yaghoubi2014}
\bibinfo{author}{\bibfnamefont{H.}~\bibnamefont{Yaghoubi}},
  \bibinfo{author}{\bibfnamefont{Z.}~\bibnamefont{Li}},
  \bibinfo{author}{\bibfnamefont{D.}~\bibnamefont{Jun}},
  \bibinfo{author}{\bibfnamefont{E.}~\bibnamefont{Lafalce}},
  \bibinfo{author}{\bibfnamefont{X.}~\bibnamefont{Jiang}},
  \bibinfo{author}{\bibfnamefont{R.}~\bibnamefont{Schlaf}},
  \bibinfo{author}{\bibfnamefont{J.~T.} \bibnamefont{Beatty}},
  \bibnamefont{and} \bibinfo{author}{\bibfnamefont{A.}~\bibnamefont{Takshi}},
  \bibinfo{journal}{J. Phys. Chem. C} \textbf{\bibinfo{volume}{118}},
  \bibinfo{pages}{23509} (\bibinfo{year}{2014}).

\bibitem[{\citenamefont{Berera et~al.}(2009)\citenamefont{Berera, van
  Grondelle, and Kennis}}]{Berera2009}
\bibinfo{author}{\bibfnamefont{R.}~\bibnamefont{Berera}},
  \bibinfo{author}{\bibfnamefont{R.}~\bibnamefont{van Grondelle}},
  \bibnamefont{and} \bibinfo{author}{\bibfnamefont{J.~T.}
  \bibnamefont{Kennis}}, \bibinfo{journal}{Photosynth. Res.}
  \textbf{\bibinfo{volume}{101}}, \bibinfo{pages}{105} (\bibinfo{year}{2009}).

\bibitem[{\citenamefont{Ginsberg et~al.}(2009)\citenamefont{Ginsberg, Cheng,
  and Fleming}}]{Ginsberg2009}
\bibinfo{author}{\bibfnamefont{N.~S.} \bibnamefont{Ginsberg}},
  \bibinfo{author}{\bibfnamefont{Y.-C.} \bibnamefont{Cheng}}, \bibnamefont{and}
  \bibinfo{author}{\bibfnamefont{G.~R.} \bibnamefont{Fleming}},
  \bibinfo{journal}{Acc. Chem. Res.} \textbf{\bibinfo{volume}{42}},
  \bibinfo{pages}{1352} (\bibinfo{year}{2009}).

\bibitem[{\citenamefont{Schlau-Cohen et~al.}(2012)\citenamefont{Schlau-Cohen,
  Dawlaty, and Fleming}}]{Schlau2012}
\bibinfo{author}{\bibfnamefont{G.~S.} \bibnamefont{Schlau-Cohen}},
  \bibinfo{author}{\bibfnamefont{J.~M.} \bibnamefont{Dawlaty}},
  \bibnamefont{and} \bibinfo{author}{\bibfnamefont{G.~R.}
  \bibnamefont{Fleming}}, \bibinfo{journal}{IEEE J. Sel. Top. Quantum
  Electron.} \textbf{\bibinfo{volume}{18}}, \bibinfo{pages}{283}
  (\bibinfo{year}{2012}).

\bibitem[{\citenamefont{Lewis et~al.}(2016)\citenamefont{Lewis, Gruenke,
  Oliver, Ballottari, Bassi, and Fleming}}]{Lewis2016}
\bibinfo{author}{\bibfnamefont{N.~H.} \bibnamefont{Lewis}},
  \bibinfo{author}{\bibfnamefont{N.~L.} \bibnamefont{Gruenke}},
  \bibinfo{author}{\bibfnamefont{T.~A.} \bibnamefont{Oliver}},
  \bibinfo{author}{\bibfnamefont{M.}~\bibnamefont{Ballottari}},
  \bibinfo{author}{\bibfnamefont{R.}~\bibnamefont{Bassi}}, \bibnamefont{and}
  \bibinfo{author}{\bibfnamefont{G.~R.} \bibnamefont{Fleming}},
  \bibinfo{journal}{J. Phys. Chem. Lett.} \textbf{\bibinfo{volume}{7}},
  \bibinfo{pages}{4197} (\bibinfo{year}{2016}).

\bibitem[{\citenamefont{Romero et~al.}(2017)\citenamefont{Romero,
  Novoderezhkin, and van Grondelle}}]{Romero2017}
\bibinfo{author}{\bibfnamefont{E.}~\bibnamefont{Romero}},
  \bibinfo{author}{\bibfnamefont{V.~I.} \bibnamefont{Novoderezhkin}},
  \bibnamefont{and} \bibinfo{author}{\bibfnamefont{R.}~\bibnamefont{van
  Grondelle}}, \bibinfo{journal}{Nature} \textbf{\bibinfo{volume}{543}},
  \bibinfo{pages}{355} (\bibinfo{year}{2017}).

\bibitem[{\citenamefont{Dorfman et~al.}(2013)\citenamefont{Dorfman, Voronine,
  Mukamel, and Scully}}]{Dorfman2013}
\bibinfo{author}{\bibfnamefont{K.~E.} \bibnamefont{Dorfman}},
  \bibinfo{author}{\bibfnamefont{D.~V.} \bibnamefont{Voronine}},
  \bibinfo{author}{\bibfnamefont{S.}~\bibnamefont{Mukamel}}, \bibnamefont{and}
  \bibinfo{author}{\bibfnamefont{M.~O.} \bibnamefont{Scully}},
  \bibinfo{journal}{Proc. Natl. Acad. Sci.} \textbf{\bibinfo{volume}{110}},
  \bibinfo{pages}{2746} (\bibinfo{year}{2013}).

\bibitem[{\citenamefont{Nalbach and Thorwart}(2013)}]{Nalbach2013}
\bibinfo{author}{\bibfnamefont{P.}~\bibnamefont{Nalbach}} \bibnamefont{and}
  \bibinfo{author}{\bibfnamefont{M.}~\bibnamefont{Thorwart}},
  \bibinfo{journal}{Proc. Natl. Acad. Sci.} \textbf{\bibinfo{volume}{110}},
  \bibinfo{pages}{2693} (\bibinfo{year}{2013}).

\bibitem[{\citenamefont{Mashaal and Gordon}(2011)}]{Mashaal2011}
\bibinfo{author}{\bibfnamefont{H.}~\bibnamefont{Mashaal}} \bibnamefont{and}
  \bibinfo{author}{\bibfnamefont{J.~M.} \bibnamefont{Gordon}},
  \bibinfo{journal}{Opt. Lett.} \textbf{\bibinfo{volume}{36}},
  \bibinfo{pages}{900} (\bibinfo{year}{2011}).

\bibitem[{\citenamefont{De~Mendoza et~al.}(2017)\citenamefont{De~Mendoza,
  Caycedo-Soler, Manrique, Quiroga, Rodriguez, and Johnson}}]{Mendoza2017}
\bibinfo{author}{\bibfnamefont{A.~M.} \bibnamefont{De~Mendoza}},
  \bibinfo{author}{\bibfnamefont{F.}~\bibnamefont{Caycedo-Soler}},
  \bibinfo{author}{\bibfnamefont{P.}~\bibnamefont{Manrique}},
  \bibinfo{author}{\bibfnamefont{L.}~\bibnamefont{Quiroga}},
  \bibinfo{author}{\bibfnamefont{F.~J.} \bibnamefont{Rodriguez}},
  \bibnamefont{and} \bibinfo{author}{\bibfnamefont{N.~F.}
  \bibnamefont{Johnson}}, \bibinfo{journal}{J. Phys. B: At. Mol. Opt. Phys.}
  \textbf{\bibinfo{volume}{50}}, \bibinfo{pages}{124002}
  (\bibinfo{year}{2017}).

\bibitem[{\citenamefont{Man{\v{c}}al and Valkunas}(2010)}]{Manvcal2010}
\bibinfo{author}{\bibfnamefont{T.}~\bibnamefont{Man{\v{c}}al}}
  \bibnamefont{and} \bibinfo{author}{\bibfnamefont{L.}~\bibnamefont{Valkunas}},
  \bibinfo{journal}{New J. Phys.} \textbf{\bibinfo{volume}{12}},
  \bibinfo{pages}{065044} (\bibinfo{year}{2010}).

\bibitem[{\citenamefont{Brumer and Shapiro}(2012)}]{Brumer2012}
\bibinfo{author}{\bibfnamefont{P.}~\bibnamefont{Brumer}} \bibnamefont{and}
  \bibinfo{author}{\bibfnamefont{M.}~\bibnamefont{Shapiro}},
  \bibinfo{journal}{Proc. Natl. Acad. Sci.} \textbf{\bibinfo{volume}{109}},
  \bibinfo{pages}{19575} (\bibinfo{year}{2012}).

\bibitem[{\citenamefont{Fassioli et~al.}(2012)\citenamefont{Fassioli,
  Olaya-Castro, and Scholes}}]{Fassioli2012}
\bibinfo{author}{\bibfnamefont{F.}~\bibnamefont{Fassioli}},
  \bibinfo{author}{\bibfnamefont{A.}~\bibnamefont{Olaya-Castro}},
  \bibnamefont{and} \bibinfo{author}{\bibfnamefont{G.~D.}
  \bibnamefont{Scholes}}, \bibinfo{journal}{J. Phys. Chem. Lett.}
  \textbf{\bibinfo{volume}{3}}, \bibinfo{pages}{3136} (\bibinfo{year}{2012}).

\bibitem[{\citenamefont{Han et~al.}(2013)\citenamefont{Han, Shapiro, and
  Brumer}}]{Han2013}
\bibinfo{author}{\bibfnamefont{A.~C.} \bibnamefont{Han}},
  \bibinfo{author}{\bibfnamefont{M.}~\bibnamefont{Shapiro}}, \bibnamefont{and}
  \bibinfo{author}{\bibfnamefont{P.}~\bibnamefont{Brumer}},
  \bibinfo{journal}{J. Phys. Chem. A} \textbf{\bibinfo{volume}{117}},
  \bibinfo{pages}{8199} (\bibinfo{year}{2013}).

\bibitem[{\citenamefont{Ol{\v{s}}ina et~al.}(2014)\citenamefont{Ol{\v{s}}ina,
  Dijkstra, Wang, and Cao}}]{Olsina2014}
\bibinfo{author}{\bibfnamefont{J.}~\bibnamefont{Ol{\v{s}}ina}},
  \bibinfo{author}{\bibfnamefont{A.~G.} \bibnamefont{Dijkstra}},
  \bibinfo{author}{\bibfnamefont{C.}~\bibnamefont{Wang}}, \bibnamefont{and}
  \bibinfo{author}{\bibfnamefont{J.}~\bibnamefont{Cao}},
  \bibinfo{journal}{arXiv:1408.5385v1}  (\bibinfo{year}{2014}).

\bibitem[{\citenamefont{Chenu et~al.}(2014)\citenamefont{Chenu, Mal{\`y}, and
  Man{\v{c}}al}}]{Chenu2014}
\bibinfo{author}{\bibfnamefont{A.}~\bibnamefont{Chenu}},
  \bibinfo{author}{\bibfnamefont{P.}~\bibnamefont{Mal{\`y}}}, \bibnamefont{and}
  \bibinfo{author}{\bibfnamefont{T.}~\bibnamefont{Man{\v{c}}al}},
  \bibinfo{journal}{Chem. Phys.} \textbf{\bibinfo{volume}{439}},
  \bibinfo{pages}{100} (\bibinfo{year}{2014}).

\bibitem[{\citenamefont{Chenu and Brumer}(2016)}]{Chenu2016}
\bibinfo{author}{\bibfnamefont{A.}~\bibnamefont{Chenu}} \bibnamefont{and}
  \bibinfo{author}{\bibfnamefont{P.}~\bibnamefont{Brumer}},
  \bibinfo{journal}{J. Chem. Phys.} \textbf{\bibinfo{volume}{144}},
  \bibinfo{pages}{044103} (\bibinfo{year}{2016}).

\bibitem[{\citenamefont{Shatokhin et~al.}(2016)\citenamefont{Shatokhin,
  Walschaers, Schlawin, and Buchleitner}}]{Shatokhin2016a}
\bibinfo{author}{\bibfnamefont{V.~N.} \bibnamefont{Shatokhin}},
  \bibinfo{author}{\bibfnamefont{M.}~\bibnamefont{Walschaers}},
  \bibinfo{author}{\bibfnamefont{F.}~\bibnamefont{Schlawin}}, \bibnamefont{and}
  \bibinfo{author}{\bibfnamefont{A.}~\bibnamefont{Buchleitner}},
  \bibinfo{journal}{arXiv:1602.7878v3}  (\bibinfo{year}{2016}).

\bibitem[{\citenamefont{Bachor et~al.}(2004)\citenamefont{Bachor, Ralph, Lucia,
  and Ralph}}]{Bachor2004}
\bibinfo{author}{\bibfnamefont{H.-A.} \bibnamefont{Bachor}},
  \bibinfo{author}{\bibfnamefont{T.~C.} \bibnamefont{Ralph}},
  \bibinfo{author}{\bibfnamefont{S.}~\bibnamefont{Lucia}}, \bibnamefont{and}
  \bibinfo{author}{\bibfnamefont{T.~C.} \bibnamefont{Ralph}},
  \emph{\bibinfo{title}{A Guide to Experiments in Quantum Optics}},
  vol.~\bibinfo{volume}{1} (\bibinfo{publisher}{Weinheim: Wiley-VCH},
  \bibinfo{year}{2004}), \bibinfo{edition}{2nd} ed.

\bibitem[{\citenamefont{Blankenship}(2014)}]{Blankenship2014}
\bibinfo{author}{\bibfnamefont{R.~E.} \bibnamefont{Blankenship}},
  \emph{\bibinfo{title}{{Molecular Mechanisms of Photosynthesis}}}
  (\bibinfo{publisher}{Chichester, UK: Wiley-Blackwell}, \bibinfo{year}{2014}),
  \bibinfo{edition}{2nd} ed.

\bibitem[{\citenamefont{Novoderezhkin et~al.}(2005)\citenamefont{Novoderezhkin,
  Palacios, Van~Amerongen, and Van~Grondelle}}]{Novoderezhkin2005}
\bibinfo{author}{\bibfnamefont{V.~I.} \bibnamefont{Novoderezhkin}},
  \bibinfo{author}{\bibfnamefont{M.~A.} \bibnamefont{Palacios}},
  \bibinfo{author}{\bibfnamefont{H.}~\bibnamefont{Van~Amerongen}},
  \bibnamefont{and}
  \bibinfo{author}{\bibfnamefont{R.}~\bibnamefont{Van~Grondelle}},
  \bibinfo{journal}{J. Phys. Chem. B} \textbf{\bibinfo{volume}{109}},
  \bibinfo{pages}{10493} (\bibinfo{year}{2005}).

\bibitem[{\citenamefont{Novoderezhkin et~al.}(2006)\citenamefont{Novoderezhkin,
  Rutkauskas, and van Grondelle}}]{Novoderezhkin2006}
\bibinfo{author}{\bibfnamefont{V.~I.} \bibnamefont{Novoderezhkin}},
  \bibinfo{author}{\bibfnamefont{D.}~\bibnamefont{Rutkauskas}},
  \bibnamefont{and} \bibinfo{author}{\bibfnamefont{R.}~\bibnamefont{van
  Grondelle}}, \bibinfo{journal}{Biophys. J.} \textbf{\bibinfo{volume}{90}},
  \bibinfo{pages}{2890} (\bibinfo{year}{2006}).

\bibitem[{\citenamefont{Ishizaki et~al.}(2010)\citenamefont{Ishizaki, Calhoun,
  Schlau-Cohen, and Fleming}}]{Ishizaki2010}
\bibinfo{author}{\bibfnamefont{A.}~\bibnamefont{Ishizaki}},
  \bibinfo{author}{\bibfnamefont{T.~R.} \bibnamefont{Calhoun}},
  \bibinfo{author}{\bibfnamefont{G.~S.} \bibnamefont{Schlau-Cohen}},
  \bibnamefont{and} \bibinfo{author}{\bibfnamefont{G.~R.}
  \bibnamefont{Fleming}}, \bibinfo{journal}{Phys. Chem. Chem. Phys.}
  \textbf{\bibinfo{volume}{12}}, \bibinfo{pages}{7319} (\bibinfo{year}{2010}).

\bibitem[{\citenamefont{Ishizaki and Fleming}(2012)}]{Ishizaki2012}
\bibinfo{author}{\bibfnamefont{A.}~\bibnamefont{Ishizaki}} \bibnamefont{and}
  \bibinfo{author}{\bibfnamefont{G.~R.} \bibnamefont{Fleming}},
  \bibinfo{journal}{Annu. Rev. Condens. Matter Phys.}
  \textbf{\bibinfo{volume}{3}}, \bibinfo{pages}{333} (\bibinfo{year}{2012}).

\bibitem[{\citenamefont{Hoyer et~al.}(2014)\citenamefont{Hoyer, Caruso,
  Montangero, Sarovar, Calarco, Plenio, and Whaley}}]{Hoyer2014}
\bibinfo{author}{\bibfnamefont{S.}~\bibnamefont{Hoyer}},
  \bibinfo{author}{\bibfnamefont{F.}~\bibnamefont{Caruso}},
  \bibinfo{author}{\bibfnamefont{S.}~\bibnamefont{Montangero}},
  \bibinfo{author}{\bibfnamefont{M.}~\bibnamefont{Sarovar}},
  \bibinfo{author}{\bibfnamefont{T.}~\bibnamefont{Calarco}},
  \bibinfo{author}{\bibfnamefont{M.~B.} \bibnamefont{Plenio}},
  \bibnamefont{and} \bibinfo{author}{\bibfnamefont{K.~B.}
  \bibnamefont{Whaley}}, \bibinfo{journal}{New J. Phys.}
  \textbf{\bibinfo{volume}{16}}, \bibinfo{pages}{045007}
  (\bibinfo{year}{2014}).

\bibitem[{\citenamefont{Baragiola et~al.}(2012)\citenamefont{Baragiola, Cook,
  Branczyk, and Combes}}]{Baragiola2012}
\bibinfo{author}{\bibfnamefont{B.~Q.} \bibnamefont{Baragiola}},
  \bibinfo{author}{\bibfnamefont{R.~L.} \bibnamefont{Cook}},
  \bibinfo{author}{\bibfnamefont{A.~M.} \bibnamefont{Branczyk}},
  \bibnamefont{and} \bibinfo{author}{\bibfnamefont{J.}~\bibnamefont{Combes}},
  \bibinfo{journal}{Phys. Rev. A} \textbf{\bibinfo{volume}{86}},
  \bibinfo{pages}{56} (\bibinfo{year}{2012}).

\bibitem[{\citenamefont{Ishizaki and Fleming}(2009)}]{Ishizaki2009a}
\bibinfo{author}{\bibfnamefont{A.}~\bibnamefont{Ishizaki}} \bibnamefont{and}
  \bibinfo{author}{\bibfnamefont{G.~R.} \bibnamefont{Fleming}},
  \bibinfo{journal}{J. Chem. Phys.} \textbf{\bibinfo{volume}{130}},
  \bibinfo{pages}{234111} (\bibinfo{year}{2009}).

\bibitem[{\citenamefont{Scully and Zubairy}(1997)}]{Scully1997}
\bibinfo{author}{\bibfnamefont{M.~O.} \bibnamefont{Scully}} \bibnamefont{and}
  \bibinfo{author}{\bibfnamefont{M.~S.} \bibnamefont{Zubairy}},
  \emph{\bibinfo{title}{{Quantum Optics}}} (\bibinfo{publisher}{Cambridge:
  Cambridge University Press}, \bibinfo{year}{1997}).

\bibitem[{\citenamefont{Roden et~al.}(2016)\citenamefont{Roden, Bennett, and
  Whaley}}]{Roden2016a}
\bibinfo{author}{\bibfnamefont{J.~J.} \bibnamefont{Roden}},
  \bibinfo{author}{\bibfnamefont{D.~I.} \bibnamefont{Bennett}},
  \bibnamefont{and} \bibinfo{author}{\bibfnamefont{K.~B.}
  \bibnamefont{Whaley}}, \bibinfo{journal}{J. Chem. Phys.}
  \textbf{\bibinfo{volume}{144}}, \bibinfo{pages}{245101}
  (\bibinfo{year}{2016}).

\bibitem[{ult()}]{ultraweakcouplingtosunlight}
\bibinfo{note}{Taking a typical chromophore electronic dipole moment $\mu = 7$
  Debye and the earth's electric field as 870 V cm$^{-1}$ yields an interaction
  strength $\mu E = 10^{-3}$ cm$^{-1}$, seven orders of magnitude smaller than
  typical chromophore transition energies in the visible regime, which are of
  order 10$^{4}$ cm$^{-1}$.}

\bibitem[{\citenamefont{Thimmel et~al.}(1999)\citenamefont{Thimmel, Nalbach,
  and Terzidis}}]{Thimmel1999}
\bibinfo{author}{\bibfnamefont{B.}~\bibnamefont{Thimmel}},
  \bibinfo{author}{\bibfnamefont{P.}~\bibnamefont{Nalbach}}, \bibnamefont{and}
  \bibinfo{author}{\bibfnamefont{O.}~\bibnamefont{Terzidis}},
  \bibinfo{journal}{Eur. Phys. J. B} \textbf{\bibinfo{volume}{9}},
  \bibinfo{pages}{207} (\bibinfo{year}{1999}).

\bibitem[{\citenamefont{Loudon}(2000)}]{Loudon2000}
\bibinfo{author}{\bibfnamefont{R.}~\bibnamefont{Loudon}},
  \emph{\bibinfo{title}{The Quantum Theory of Light}}
  (\bibinfo{publisher}{Oxford: Oxford University Press}, \bibinfo{year}{2000}).

\bibitem[{\citenamefont{Wang et~al.}(2011)\citenamefont{Wang, Minar, Sheridan,
  and Scarani}}]{Wang2011}
\bibinfo{author}{\bibfnamefont{Y.}~\bibnamefont{Wang}},
  \bibinfo{author}{\bibfnamefont{J.}~\bibnamefont{Minar}},
  \bibinfo{author}{\bibfnamefont{L.}~\bibnamefont{Sheridan}}, \bibnamefont{and}
  \bibinfo{author}{\bibfnamefont{V.}~\bibnamefont{Scarani}},
  \bibinfo{journal}{Phys. Rev. A} \textbf{\bibinfo{volume}{83}},
  \bibinfo{pages}{1} (\bibinfo{year}{2011}), \eprint{1010.4661}.

\bibitem[{\citenamefont{Mandel and Wolf}(1965)}]{Mandel1965}
\bibinfo{author}{\bibfnamefont{L.}~\bibnamefont{Mandel}} \bibnamefont{and}
  \bibinfo{author}{\bibfnamefont{E.}~\bibnamefont{Wolf}},
  \bibinfo{journal}{Rev. Mod. Phys.} \textbf{\bibinfo{volume}{37}},
  \bibinfo{pages}{231} (\bibinfo{year}{1965}).

\bibitem[{\citenamefont{Gardiner and Zoller}(2004)}]{Gardiner2004}
\bibinfo{author}{\bibfnamefont{C.}~\bibnamefont{Gardiner}} \bibnamefont{and}
  \bibinfo{author}{\bibfnamefont{P.}~\bibnamefont{Zoller}},
  \emph{\bibinfo{title}{Quantum Noise}}, vol.~\bibinfo{volume}{56}
  (\bibinfo{publisher}{Berlin: Springer}, \bibinfo{year}{2004}).

\bibitem[{\citenamefont{Wiseman and Milburn}(2014)}]{Wiseman2014}
\bibinfo{author}{\bibfnamefont{H.~M.} \bibnamefont{Wiseman}} \bibnamefont{and}
  \bibinfo{author}{\bibfnamefont{G.~J.} \bibnamefont{Milburn}},
  \emph{\bibinfo{title}{Quantum Measurement and Control}}
  (\bibinfo{publisher}{Cambridge: Cambridge University Press},
  \bibinfo{address}{Cambridge}, \bibinfo{year}{2014}).

\bibitem[{\citenamefont{Santori et~al.}(2010)\citenamefont{Santori, Fattal, and
  Yamamoto}}]{Santori2010}
\bibinfo{author}{\bibfnamefont{C.}~\bibnamefont{Santori}},
  \bibinfo{author}{\bibfnamefont{D.}~\bibnamefont{Fattal}}, \bibnamefont{and}
  \bibinfo{author}{\bibfnamefont{Y.}~\bibnamefont{Yamamoto}},
  \emph{\bibinfo{title}{Single-Photon Devices and Applications}}
  (\bibinfo{publisher}{New York: Wiley}, \bibinfo{year}{2010}).

\bibitem[{\citenamefont{Scholes et~al.}(2011)\citenamefont{Scholes, Fleming,
  Olaya-Castro, and van Grondelle}}]{Scholes2011}
\bibinfo{author}{\bibfnamefont{G.~D.} \bibnamefont{Scholes}},
  \bibinfo{author}{\bibfnamefont{G.~R.} \bibnamefont{Fleming}},
  \bibinfo{author}{\bibfnamefont{A.}~\bibnamefont{Olaya-Castro}},
  \bibnamefont{and} \bibinfo{author}{\bibfnamefont{R.}~\bibnamefont{van
  Grondelle}}, \bibinfo{journal}{Nat. Chem.} \textbf{\bibinfo{volume}{3}},
  \bibinfo{pages}{763} (\bibinfo{year}{2011}).

\bibitem[{\citenamefont{Tanimura and Kubo}(1989)}]{Tanimura1989}
\bibinfo{author}{\bibfnamefont{Y.}~\bibnamefont{Tanimura}} \bibnamefont{and}
  \bibinfo{author}{\bibfnamefont{R.}~\bibnamefont{Kubo}}, \bibinfo{journal}{J.
  Phys. Soc. Japan} \textbf{\bibinfo{volume}{58}}, \bibinfo{pages}{1850}
  (\bibinfo{year}{1989}), ISSN \bibinfo{issn}{00319015}.

\bibitem[{\citenamefont{Ishizaki and Tanimura}(2005)}]{Ishizaki2005}
\bibinfo{author}{\bibfnamefont{A.}~\bibnamefont{Ishizaki}} \bibnamefont{and}
  \bibinfo{author}{\bibfnamefont{Y.}~\bibnamefont{Tanimura}},
  \bibinfo{journal}{J. Phys. Soc. Japan} \textbf{\bibinfo{volume}{74}},
  \bibinfo{pages}{3131} (\bibinfo{year}{2005}).

\bibitem[{\citenamefont{Kreisbeck et~al.}(2011)\citenamefont{Kreisbeck, Kramer,
  Rodr{\'{i}}guez, and Hein}}]{Kreisbeck2011}
\bibinfo{author}{\bibfnamefont{C.}~\bibnamefont{Kreisbeck}},
  \bibinfo{author}{\bibfnamefont{T.}~\bibnamefont{Kramer}},
  \bibinfo{author}{\bibfnamefont{M.}~\bibnamefont{Rodr{\'{i}}guez}},
  \bibnamefont{and} \bibinfo{author}{\bibfnamefont{B.}~\bibnamefont{Hein}},
  \bibinfo{journal}{J. Chem. Theory Comput.} \textbf{\bibinfo{volume}{7}},
  \bibinfo{pages}{2166} (\bibinfo{year}{2011}).

\bibitem[{\citenamefont{Shabani et~al.}(2014)\citenamefont{Shabani, Mohseni,
  Rabitz, and Lloyd}}]{Shabani2014}
\bibinfo{author}{\bibfnamefont{A.}~\bibnamefont{Shabani}},
  \bibinfo{author}{\bibfnamefont{M.}~\bibnamefont{Mohseni}},
  \bibinfo{author}{\bibfnamefont{H.}~\bibnamefont{Rabitz}}, \bibnamefont{and}
  \bibinfo{author}{\bibfnamefont{S.}~\bibnamefont{Lloyd}},
  \bibinfo{journal}{Phys. Rev. E} \textbf{\bibinfo{volume}{89}},
  \bibinfo{pages}{042706} (\bibinfo{year}{2014}).

\bibitem[{\citenamefont{Suess et~al.}(2015)\citenamefont{Suess, Strunz, and
  Eisfeld}}]{Suess2015}
\bibinfo{author}{\bibfnamefont{D.}~\bibnamefont{Suess}},
  \bibinfo{author}{\bibfnamefont{W.~T.} \bibnamefont{Strunz}},
  \bibnamefont{and} \bibinfo{author}{\bibfnamefont{A.}~\bibnamefont{Eisfeld}},
  \bibinfo{journal}{J. Stat. Phys.} \textbf{\bibinfo{volume}{159}},
  \bibinfo{pages}{1408} (\bibinfo{year}{2015}).

\bibitem[{\citenamefont{Witt et~al.}(2017)\citenamefont{Witt, Rudnicki,
  Tanimura, and Mintert}}]{Witt2017}
\bibinfo{author}{\bibfnamefont{B.}~\bibnamefont{Witt}},
  \bibinfo{author}{\bibfnamefont{{\L}.}~\bibnamefont{Rudnicki}},
  \bibinfo{author}{\bibfnamefont{Y.}~\bibnamefont{Tanimura}}, \bibnamefont{and}
  \bibinfo{author}{\bibfnamefont{F.}~\bibnamefont{Mintert}},
  \bibinfo{journal}{New J. Phys.} \textbf{\bibinfo{volume}{19}},
  \bibinfo{pages}{013007} (\bibinfo{year}{2017}).

\bibitem[{\citenamefont{Strunz et~al.}(1999)\citenamefont{Strunz, Di{\'{o}}si,
  and Gisin}}]{Strunz1999}
\bibinfo{author}{\bibfnamefont{W.~T.} \bibnamefont{Strunz}},
  \bibinfo{author}{\bibfnamefont{L.}~\bibnamefont{Di{\'{o}}si}},
  \bibnamefont{and} \bibinfo{author}{\bibfnamefont{N.}~\bibnamefont{Gisin}},
  \bibinfo{journal}{Phys. Rev. Lett.} \textbf{\bibinfo{volume}{82}},
  \bibinfo{pages}{1801} (\bibinfo{year}{1999}).

\bibitem[{\citenamefont{Ishizaki and Tanimura}(2006)}]{Ishizaki2006}
\bibinfo{author}{\bibfnamefont{A.}~\bibnamefont{Ishizaki}} \bibnamefont{and}
  \bibinfo{author}{\bibfnamefont{Y.}~\bibnamefont{Tanimura}},
  \bibinfo{journal}{J. Chem. Phys.} \textbf{\bibinfo{volume}{125}},
  \bibinfo{pages}{084501} (\bibinfo{year}{2006}).

\bibitem[{\citenamefont{Tanimura}(2006)}]{Tanimura2006}
\bibinfo{author}{\bibfnamefont{Y.}~\bibnamefont{Tanimura}},
  \bibinfo{journal}{J. Phys. Soc. Japan} \textbf{\bibinfo{volume}{75}},
  \bibinfo{pages}{082001} (\bibinfo{year}{2006}).

\bibitem[{\citenamefont{Uchiyama and Aihara}(2010)}]{Uchiyama2010}
\bibinfo{author}{\bibfnamefont{C.}~\bibnamefont{Uchiyama}} \bibnamefont{and}
  \bibinfo{author}{\bibfnamefont{M.}~\bibnamefont{Aihara}},
  \bibinfo{journal}{Phys. Rev. A} \textbf{\bibinfo{volume}{82}},
  \bibinfo{pages}{044104} (\bibinfo{year}{2010}).

\bibitem[{\citenamefont{Kubo}(1957)}]{kubo1957}
\bibinfo{author}{\bibfnamefont{R.}~\bibnamefont{Kubo}}, \bibinfo{journal}{J.
  Phys. Soc. Japan} \textbf{\bibinfo{volume}{12}}, \bibinfo{pages}{570}
  (\bibinfo{year}{1957}).

\bibitem[{\citenamefont{Mukamel}(1995)}]{Mukamel1995}
\bibinfo{author}{\bibfnamefont{S.}~\bibnamefont{Mukamel}},
  \emph{\bibinfo{title}{{Principles of Nonlinear Optical Spectroscopy}}}
  (\bibinfo{publisher}{Oxford: Oxford University Press}, \bibinfo{year}{1995}).

\bibitem[{\citenamefont{Ban et~al.}(2017)\citenamefont{Ban, Kitajima, Arimitsu,
  and Shibata}}]{Ban2017}
\bibinfo{author}{\bibfnamefont{M.}~\bibnamefont{Ban}},
  \bibinfo{author}{\bibfnamefont{S.}~\bibnamefont{Kitajima}},
  \bibinfo{author}{\bibfnamefont{T.}~\bibnamefont{Arimitsu}}, \bibnamefont{and}
  \bibinfo{author}{\bibfnamefont{F.}~\bibnamefont{Shibata}},
  \bibinfo{journal}{Phys. Rev. A} \textbf{\bibinfo{volume}{95}},
  \bibinfo{pages}{022126} (\bibinfo{year}{2017}).

\bibitem[{\citenamefont{Shen}(2017)}]{Shen2017}
\bibinfo{author}{\bibfnamefont{H.~Z.} \bibnamefont{Shen}},
  \bibinfo{journal}{Phys. Rev. E} \textbf{\bibinfo{volume}{95}},
  \bibinfo{pages}{012156} (\bibinfo{year}{2017}).

\bibitem[{\citenamefont{Novoderezhkin et~al.}(2011)\citenamefont{Novoderezhkin,
  Marin, and van Grondelle}}]{Novoderezhkin2011intra}
\bibinfo{author}{\bibfnamefont{V.}~\bibnamefont{Novoderezhkin}},
  \bibinfo{author}{\bibfnamefont{A.}~\bibnamefont{Marin}}, \bibnamefont{and}
  \bibinfo{author}{\bibfnamefont{R.}~\bibnamefont{van Grondelle}},
  \bibinfo{journal}{Phys. Chem. Chem. Phys.} \textbf{\bibinfo{volume}{13}},
  \bibinfo{pages}{17093} (\bibinfo{year}{2011}).

\bibitem[{\citenamefont{Bennett et~al.}(2013)\citenamefont{Bennett, Amarnath,
  and Fleming}}]{Bennett2013a}
\bibinfo{author}{\bibfnamefont{D.~I.~G.} \bibnamefont{Bennett}},
  \bibinfo{author}{\bibfnamefont{K.}~\bibnamefont{Amarnath}}, \bibnamefont{and}
  \bibinfo{author}{\bibfnamefont{G.~R.} \bibnamefont{Fleming}},
  \bibinfo{journal}{J. Am. Chem. Soc.} \textbf{\bibinfo{volume}{135}},
  \bibinfo{pages}{9164} (\bibinfo{year}{2013}).

\bibitem[{ris()}]{risetimenote}
\bibinfo{note}{We define the rise time here as the 10-90 rise time, i.e., the
  difference between the time values at which $P_e$ is equal to 90 \% and 10 \%
  of its maximum value $P_e^{\text{max}}$.}

\bibitem[{\citenamefont{Chong et~al.}(2010)\citenamefont{Chong, Min, and
  Xie}}]{Chong2010}
\bibinfo{author}{\bibfnamefont{S.}~\bibnamefont{Chong}},
  \bibinfo{author}{\bibfnamefont{W.}~\bibnamefont{Min}}, \bibnamefont{and}
  \bibinfo{author}{\bibfnamefont{X.~S.} \bibnamefont{Xie}},
  \bibinfo{journal}{J. Phys. Chem. Lett.} \textbf{\bibinfo{volume}{1}},
  \bibinfo{pages}{3316} (\bibinfo{year}{2010}).

\bibitem[{\citenamefont{Celebrano et~al.}(2011)\citenamefont{Celebrano, Kukura,
  Renn, and Sandoghdar}}]{Celebrano2011}
\bibinfo{author}{\bibfnamefont{M.}~\bibnamefont{Celebrano}},
  \bibinfo{author}{\bibfnamefont{P.}~\bibnamefont{Kukura}},
  \bibinfo{author}{\bibfnamefont{A.}~\bibnamefont{Renn}}, \bibnamefont{and}
  \bibinfo{author}{\bibfnamefont{V.}~\bibnamefont{Sandoghdar}},
  \bibinfo{journal}{Nat. Photon.} \textbf{\bibinfo{volume}{5}},
  \bibinfo{pages}{95} (\bibinfo{year}{2011}).

\bibitem[{\citenamefont{Scholes and Fleming}(2000)}]{Scholes2000}
\bibinfo{author}{\bibfnamefont{G.~D.} \bibnamefont{Scholes}} \bibnamefont{and}
  \bibinfo{author}{\bibfnamefont{G.~R.} \bibnamefont{Fleming}},
  \bibinfo{journal}{J. Phys. Chem. B} \textbf{\bibinfo{volume}{104}},
  \bibinfo{pages}{1854} (\bibinfo{year}{2000}).

\bibitem[{\citenamefont{Scholes et~al.}(2017)\citenamefont{Scholes, Fleming,
  Chen, Aspuru-Guzik, Buchleitner, Coker, Engel, van Grondelle, Ishizaki, Jonas
  et~al.}}]{Scholes2017}
\bibinfo{author}{\bibfnamefont{G.~D.} \bibnamefont{Scholes}},
  \bibinfo{author}{\bibfnamefont{G.~R.} \bibnamefont{Fleming}},
  \bibinfo{author}{\bibfnamefont{L.~X.} \bibnamefont{Chen}},
  \bibinfo{author}{\bibfnamefont{A.}~\bibnamefont{Aspuru-Guzik}},
  \bibinfo{author}{\bibfnamefont{A.}~\bibnamefont{Buchleitner}},
  \bibinfo{author}{\bibfnamefont{D.~F.} \bibnamefont{Coker}},
  \bibinfo{author}{\bibfnamefont{G.~S.} \bibnamefont{Engel}},
  \bibinfo{author}{\bibfnamefont{R.}~\bibnamefont{van Grondelle}},
  \bibinfo{author}{\bibfnamefont{A.}~\bibnamefont{Ishizaki}},
  \bibinfo{author}{\bibfnamefont{D.~M.} \bibnamefont{Jonas}},
  \bibnamefont{et~al.}, \bibinfo{journal}{Nature}
  \textbf{\bibinfo{volume}{543}}, \bibinfo{pages}{647} (\bibinfo{year}{2017}).

\bibitem[{\citenamefont{Lakowicz}(2006)}]{Lakowicz2006}
\bibinfo{author}{\bibfnamefont{J.~R.} \bibnamefont{Lakowicz}},
  \emph{\bibinfo{title}{Principles of Fluorescence Spectroscopy}}
  (\bibinfo{publisher}{New York: Springer}, \bibinfo{year}{2006}).

\bibitem[{\citenamefont{Kukura et~al.}(2010)\citenamefont{Kukura, Celebrano,
  Renn, and Sandoghdar}}]{Kukura2010}
\bibinfo{author}{\bibfnamefont{P.}~\bibnamefont{Kukura}},
  \bibinfo{author}{\bibfnamefont{M.}~\bibnamefont{Celebrano}},
  \bibinfo{author}{\bibfnamefont{A.}~\bibnamefont{Renn}}, \bibnamefont{and}
  \bibinfo{author}{\bibfnamefont{V.}~\bibnamefont{Sandoghdar}},
  \bibinfo{journal}{J. Phys. Chem. Lett.} \textbf{\bibinfo{volume}{1}},
  \bibinfo{pages}{3323} (\bibinfo{year}{2010}).

\bibitem[{\citenamefont{Gaiduk et~al.}(2010)\citenamefont{Gaiduk, Yorulmaz,
  Ruijgrok, and Orrit}}]{Gaiduk2010}
\bibinfo{author}{\bibfnamefont{A.}~\bibnamefont{Gaiduk}},
  \bibinfo{author}{\bibfnamefont{M.}~\bibnamefont{Yorulmaz}},
  \bibinfo{author}{\bibfnamefont{P.}~\bibnamefont{Ruijgrok}}, \bibnamefont{and}
  \bibinfo{author}{\bibfnamefont{M.}~\bibnamefont{Orrit}},
  \bibinfo{journal}{Science} \textbf{\bibinfo{volume}{330}},
  \bibinfo{pages}{353} (\bibinfo{year}{2010}).

\bibitem[{\citenamefont{Noy et~al.}(2006)\citenamefont{Noy, Moser, and
  Dutton}}]{Noy2006}
\bibinfo{author}{\bibfnamefont{D.}~\bibnamefont{Noy}},
  \bibinfo{author}{\bibfnamefont{C.~C.} \bibnamefont{Moser}}, \bibnamefont{and}
  \bibinfo{author}{\bibfnamefont{P.~L.} \bibnamefont{Dutton}},
  \bibinfo{journal}{Biochim. Biophys. Acta - Bioenerg.}
  \textbf{\bibinfo{volume}{1757}}, \bibinfo{pages}{90} (\bibinfo{year}{2006}).

\bibitem[{\citenamefont{Mandel}(1979)}]{Mandel1979}
\bibinfo{author}{\bibfnamefont{L.}~\bibnamefont{Mandel}}, \bibinfo{journal}{J.
  Opt. Soc. Am.} \textbf{\bibinfo{volume}{69}}, \bibinfo{pages}{1038}
  (\bibinfo{year}{1979}).

\bibitem[{\citenamefont{Agarwal et~al.}(2004)\citenamefont{Agarwal, Gbur, and
  Wolf}}]{Agarwal2004}
\bibinfo{author}{\bibfnamefont{G.~S.} \bibnamefont{Agarwal}},
  \bibinfo{author}{\bibfnamefont{G.}~\bibnamefont{Gbur}}, \bibnamefont{and}
  \bibinfo{author}{\bibfnamefont{E.}~\bibnamefont{Wolf}},
  \bibinfo{journal}{Opt. Lett.} \textbf{\bibinfo{volume}{29}},
  \bibinfo{pages}{459} (\bibinfo{year}{2004}).

\bibitem[{\citenamefont{Kano and Wolf}(1962)}]{Kano1962}
\bibinfo{author}{\bibfnamefont{Y.}~\bibnamefont{Kano}} \bibnamefont{and}
  \bibinfo{author}{\bibfnamefont{E.}~\bibnamefont{Wolf}},
  \bibinfo{journal}{Proc. Phys. Soc.} \textbf{\bibinfo{volume}{80}},
  \bibinfo{pages}{1273} (\bibinfo{year}{1962}).

\bibitem[{\citenamefont{Mehta}(1963)}]{Mehta1963}
\bibinfo{author}{\bibfnamefont{C.}~\bibnamefont{Mehta}}, \bibinfo{journal}{Il
  Nuovo Cimento} \textbf{\bibinfo{volume}{28}}, \bibinfo{pages}{401}
  (\bibinfo{year}{1963}).

\bibitem[{\citenamefont{G03.09}(2012)}]{StandardASTM_2007g173}
\bibinfo{author}{\bibfnamefont{A.}~\bibnamefont{G03.09}},
  \bibinfo{journal}{ASTM International G173} \textbf{\bibinfo{volume}{14.01}}
  (\bibinfo{year}{2012}), \bibinfo{note}{"Standard Tables for Reference Solar
  Spectral Irradiances"}.

\bibitem[{\citenamefont{Amarnath et~al.}(2016)\citenamefont{Amarnath, Bennett,
  Schneider, and Fleming}}]{Amarnath2016}
\bibinfo{author}{\bibfnamefont{K.}~\bibnamefont{Amarnath}},
  \bibinfo{author}{\bibfnamefont{D.~I.} \bibnamefont{Bennett}},
  \bibinfo{author}{\bibfnamefont{A.~R.} \bibnamefont{Schneider}},
  \bibnamefont{and} \bibinfo{author}{\bibfnamefont{G.~R.}
  \bibnamefont{Fleming}}, \bibinfo{journal}{Proc. Natl. Acad. Sci.}
  \textbf{\bibinfo{volume}{113}}, \bibinfo{pages}{1156} (\bibinfo{year}{2016}).

\bibitem[{\citenamefont{Lee and Whitmarsh}(1989)}]{Lee1989}
\bibinfo{author}{\bibfnamefont{W.-J.} \bibnamefont{Lee}} \bibnamefont{and}
  \bibinfo{author}{\bibfnamefont{J.}~\bibnamefont{Whitmarsh}},
  \bibinfo{journal}{Plant Physiol.} \textbf{\bibinfo{volume}{89}},
  \bibinfo{pages}{932} (\bibinfo{year}{1989}).

\bibitem[{\citenamefont{Chylla and Whitmarsh}(1989)}]{Chylla1989}
\bibinfo{author}{\bibfnamefont{R.~A.} \bibnamefont{Chylla}} \bibnamefont{and}
  \bibinfo{author}{\bibfnamefont{J.}~\bibnamefont{Whitmarsh}},
  \bibinfo{journal}{Plant Physiol.} \textbf{\bibinfo{volume}{90}},
  \bibinfo{pages}{765} (\bibinfo{year}{1989}).

\bibitem[{dim()}]{dimerbandwidthnote}
\bibinfo{note}{$\Omega$ is given here by $\pi/\sqrt{ln 2}$ multiplied by the
  sum of the FHWM of the Chla $Q_y$ absorption from [3] and the domain of the
  eigenvalues of $H_{s}^{(2)}$ in the single-excitation manifold.}

\bibitem[{\citenamefont{Roden and Whaley}(2016)}]{Roden2016}
\bibinfo{author}{\bibfnamefont{J.~J.~J.} \bibnamefont{Roden}} \bibnamefont{and}
  \bibinfo{author}{\bibfnamefont{K.~B.} \bibnamefont{Whaley}},
  \bibinfo{journal}{Phys. Rev. E} \textbf{\bibinfo{volume}{93}},
  \bibinfo{pages}{012128} (\bibinfo{year}{2016}).

\bibitem[{\citenamefont{Sarovar et~al.}(2011)\citenamefont{Sarovar, Cheng, and
  Whaley}}]{Sarovar2011}
\bibinfo{author}{\bibfnamefont{M.}~\bibnamefont{Sarovar}},
  \bibinfo{author}{\bibfnamefont{Y.-C.} \bibnamefont{Cheng}}, \bibnamefont{and}
  \bibinfo{author}{\bibfnamefont{K.~B.} \bibnamefont{Whaley}},
  \bibinfo{journal}{Phys. Rev. E} \textbf{\bibinfo{volume}{83}},
  \bibinfo{pages}{011906} (\bibinfo{year}{2011}).

\bibitem[{\citenamefont{Dong and Fleming}(2014)}]{Dong2014}
\bibinfo{author}{\bibfnamefont{H.}~\bibnamefont{Dong}} \bibnamefont{and}
  \bibinfo{author}{\bibfnamefont{G.~R.} \bibnamefont{Fleming}},
  \bibinfo{journal}{J. Phys. Chem. B} \textbf{\bibinfo{volume}{118}},
  \bibinfo{pages}{8956} (\bibinfo{year}{2014}).

\bibitem[{\citenamefont{Butkus et~al.}(2016)\citenamefont{Butkus, Dong,
  Fleming, Abramavicius, and Valkunas}}]{Butkus2016}
\bibinfo{author}{\bibfnamefont{V.}~\bibnamefont{Butkus}},
  \bibinfo{author}{\bibfnamefont{H.}~\bibnamefont{Dong}},
  \bibinfo{author}{\bibfnamefont{G.~R.} \bibnamefont{Fleming}},
  \bibinfo{author}{\bibfnamefont{D.}~\bibnamefont{Abramavicius}},
  \bibnamefont{and} \bibinfo{author}{\bibfnamefont{L.}~\bibnamefont{Valkunas}},
  \bibinfo{journal}{J. Phys. Chem. Lett.} \textbf{\bibinfo{volume}{7}},
  \bibinfo{pages}{277} (\bibinfo{year}{2016}).

\bibitem[{7me()}]{7merpulsechoice}
\bibinfo{note}{This gives a pulse amplitude FWHM equal to $\pi/\sqrt{ln 2}$
  times the sum of the FHWM of the Chla $Q_y$ absorption from [4] and the
  domain of the eigenvalues of $H_{s}^{(7)}$ in the single-excitation manifold,
  thereby ensuring coverage of the full absorption spectrum.}

\bibitem[{\citenamefont{Karasik et~al.}(2008)\citenamefont{Karasik, Marzlin,
  Sanders, and Whaley}}]{Karasik2008}
\bibinfo{author}{\bibfnamefont{R.~I.} \bibnamefont{Karasik}},
  \bibinfo{author}{\bibfnamefont{K.-P.} \bibnamefont{Marzlin}},
  \bibinfo{author}{\bibfnamefont{B.~C.} \bibnamefont{Sanders}},
  \bibnamefont{and} \bibinfo{author}{\bibfnamefont{K.~B.}
  \bibnamefont{Whaley}}, \bibinfo{journal}{Phys. Rev. A}
  \textbf{\bibinfo{volume}{77}}, \bibinfo{pages}{052301}
  (\bibinfo{year}{2008}).

\bibitem[{\citenamefont{Phan et~al.}(2014)\citenamefont{Phan, Cheng, Bessarab,
  and Krivitsky}}]{Phan2014}
\bibinfo{author}{\bibfnamefont{N.~M.} \bibnamefont{Phan}},
  \bibinfo{author}{\bibfnamefont{M.~F.} \bibnamefont{Cheng}},
  \bibinfo{author}{\bibfnamefont{D.~A.} \bibnamefont{Bessarab}},
  \bibnamefont{and} \bibinfo{author}{\bibfnamefont{L.~A.}
  \bibnamefont{Krivitsky}}, \bibinfo{journal}{Phys. Rev. Lett.}
  \textbf{\bibinfo{volume}{112}}, \bibinfo{pages}{213601}
  (\bibinfo{year}{2014}).

\end{thebibliography}

\end{document}